\documentclass[twocolumn,superscriptaddress,showpacs,preprintnumbers,amsmath,amssymb,aps,prd,floatfix]{revtex4-1}
\usepackage{graphicx}% Include figure files
\usepackage{dcolumn}% Align table columns on decimal point
\usepackage{bm}% bold math
\usepackage[%showframe,%Uncomment any one of the following lines to test 
total={7.0in,9.75in}, top=1.2in, left=0.75in, includefoot,
]{geometry}
\usepackage{longtable}
\newcommand{\ra}           {\rightarrow}
\newcommand{\jpsi}         {J/\psi}
\newcommand{\bp}           {B^+}
\newcommand{\bzero}        {B^0}
\newcommand{\bc}           {B_c^+}
\newcommand{\bcst}         {B_c^{\ast+}}
\newcommand{\ppi}          {p\pi}
\newcommand{\bbbar}        {b\bar{b}}
\newcommand{\dphi}         {\Delta \phi}
\newcommand{\jpsihyptrack} {$\jpsi$-track\ }
\newcommand{\jpsisidetrack}{$\jpsi_{\textrm{side}}$-track\ }
\newcommand{\jpsiplustrack}{\jpsi\,\textrm{track}}
\newcommand{\jpsiplusk}    {\jpsi\, K^+}

\newcommand{\jpsiplusmu}   {\jpsi\, \mu^+}
\newcommand{\jpsiplusmunu} {\jpsi\, \mu^+ \nu}
\newcommand{\jpsitomumu}   {\jpsi \to \mu^+\mu^-}
\newcommand{\bctojpsimu}   {\bc \to \jpsi\, \mu^+}

\newcommand{\bctojpsimux}  {\bc \to \jpsi\, \mu^+ X}
\newcommand{\bctojpsimunu} {\bc \to \jpsi\, \mu^+ \nu}

\newcommand{\bptojpsik}    {\bp \to \jpsi\, K^+}
\newcommand{\bptojpsipi}   {\bp \to \jpsi\, \pi^+}
\newcommand{\ltoppi}       {\Lambda \to \ppi}
\newcommand{\fb}           {\ifmmode \textrm{fb}^{-1} \else fb$^{-1}\:$\fi}

\newcommand{\mevc}         {\ifmmode \textrm{MeV}/c \else MeV$/c\:$\fi}
\newcommand{\mevcc}        {\ifmmode \textrm{MeV}/c^2 \else MeV$/c^2\:$\fi}

\newcommand{\gevc}         {\ifmmode \textrm{GeV}/c \else GeV$/c\:$\fi}
\newcommand{\gevcc}        {\ifmmode \textrm{GeV}/c^2 \else GeV$/c^2\:$\fi}
\newcommand{\lowin}        {\textrm{3--4}~\gevcc}
\newcommand{\sigwin}       {\textrm{4--6}~\gevcc}
\newcommand{\hiwin}        {$>$\,6~\gevcc}
\newcommand{\um}           {~\mu\textrm{m}}
\newcommand{\pt}           {p_T}
\newcommand{\lxy}          {L_{xy}}

\newcommand{\ratio}        {\mathcal{R}}
\newcommand{\sigbrbc}      {\sigma(\bc)\mathcal{B}(\bctojpsimunu)}
\newcommand{\sigbrb}       {\sigma(\bp)\mathcal{B}(\bptojpsik)}
\newcommand{\erel}         {\epsilon_{\textrm{rel}}}

\newcommand{\DIF}          {\textrm{decay-in-flight}}
\newcommand{\PT}           {\textrm{punch-through}}
\newcommand{\dedx}         {dE/dx}
\newcommand{\dedxnum}      {dE/dx_{\textrm{meas}} - dE/dx_{\pi}}
\newcommand{\tofnum}       {ToF_{\textrm{meas}} - ToF_{\pi}}

\def\dzero           {D^0}
\def\dst             {D^{*+}}

\begin{document}

%\preprint{Collaboration Draft v2.4}

\title{\textbf{Measurement of the \boldmath{$B_c^{\pm}$} production cross section
 in \boldmath{$p\bar{p}$} collisions at \boldmath{$\sqrt{s}=1.96$}~TeV}}

% Last update: $Date: 2015/12/02 22:50:09 $
\affiliation{Institute of Physics, Academia Sinica, Taipei, Taiwan 11529, Republic of China}
\affiliation{Argonne National Laboratory, Argonne, Illinois 60439, USA}
\affiliation{University of Athens, 157 71 Athens, Greece}
\affiliation{Institut de Fisica d'Altes Energies, ICREA, Universitat Autonoma de Barcelona, E-08193, Bellaterra (Barcelona), Spain}
\affiliation{Baylor University, Waco, Texas 76798, USA}
\affiliation{Istituto Nazionale di Fisica Nucleare Bologna, \ensuremath{^{kk}}University of Bologna, I-40127 Bologna, Italy}
\affiliation{University of California, Davis, Davis, California 95616, USA}
\affiliation{University of California, Los Angeles, Los Angeles, California 90024, USA}
\affiliation{Instituto de Fisica de Cantabria, CSIC-University of Cantabria, 39005 Santander, Spain}
\affiliation{Carnegie Mellon University, Pittsburgh, Pennsylvania 15213, USA}
\affiliation{Enrico Fermi Institute, University of Chicago, Chicago, Illinois 60637, USA}
\affiliation{Comenius University, 842 48 Bratislava, Slovakia; Institute of Experimental Physics, 040 01 Kosice, Slovakia}
\affiliation{Joint Institute for Nuclear Research, RU-141980 Dubna, Russia}
\affiliation{Duke University, Durham, North Carolina 27708, USA}
\affiliation{Fermi National Accelerator Laboratory, Batavia, Illinois 60510, USA}
\affiliation{University of Florida, Gainesville, Florida 32611, USA}
\affiliation{Laboratori Nazionali di Frascati, Istituto Nazionale di Fisica Nucleare, I-00044 Frascati, Italy}
\affiliation{University of Geneva, CH-1211 Geneva 4, Switzerland}
\affiliation{Glasgow University, Glasgow G12 8QQ, United Kingdom}
\affiliation{Harvard University, Cambridge, Massachusetts 02138, USA}
\affiliation{Division of High Energy Physics, Department of Physics, University of Helsinki, FIN-00014, Helsinki, Finland; Helsinki Institute of Physics, FIN-00014, Helsinki, Finland}
\affiliation{University of Illinois, Urbana, Illinois 61801, USA}
\affiliation{The Johns Hopkins University, Baltimore, Maryland 21218, USA}
\affiliation{Institut f\"{u}r Experimentelle Kernphysik, Karlsruhe Institute of Technology, D-76131 Karlsruhe, Germany}
\affiliation{Center for High Energy Physics: Kyungpook National University, Daegu 702-701, Korea; Seoul National University, Seoul 151-742, Korea; Sungkyunkwan University, Suwon 440-746, Korea; Korea Institute of Science and Technology Information, Daejeon 305-806, Korea; Chonnam National University, Gwangju 500-757, Korea; Chonbuk National University, Jeonju 561-756, Korea; Ewha Womans University, Seoul, 120-750, Korea}
\affiliation{Ernest Orlando Lawrence Berkeley National Laboratory, Berkeley, California 94720, USA}
\affiliation{University of Liverpool, Liverpool L69 7ZE, United Kingdom}
\affiliation{University College London, London WC1E 6BT, United Kingdom}
\affiliation{Centro de Investigaciones Energeticas Medioambientales y Tecnologicas, E-28040 Madrid, Spain}
\affiliation{Massachusetts Institute of Technology, Cambridge, Massachusetts 02139, USA}
\affiliation{University of Michigan, Ann Arbor, Michigan 48109, USA}
\affiliation{Michigan State University, East Lansing, Michigan 48824, USA}
\affiliation{Institution for Theoretical and Experimental Physics, ITEP, Moscow 117259, Russia}
\affiliation{University of New Mexico, Albuquerque, New Mexico 87131, USA}
\affiliation{The Ohio State University, Columbus, Ohio 43210, USA}
\affiliation{Okayama University, Okayama 700-8530, Japan}
\affiliation{Osaka City University, Osaka 558-8585, Japan}
\affiliation{University of Oxford, Oxford OX1 3RH, United Kingdom}
\affiliation{Istituto Nazionale di Fisica Nucleare, Sezione di Padova, \ensuremath{^{ll}}University of Padova, I-35131 Padova, Italy}
\affiliation{University of Pennsylvania, Philadelphia, Pennsylvania 19104, USA}
\affiliation{Istituto Nazionale di Fisica Nucleare Pisa, \ensuremath{^{mm}}University of Pisa, \ensuremath{^{nn}}University of Siena, \ensuremath{^{oo}}Scuola Normale Superiore, I-56127 Pisa, Italy, \ensuremath{^{pp}}INFN Pavia, I-27100 Pavia, Italy, \ensuremath{^{qq}}University of Pavia, I-27100 Pavia, Italy}
\affiliation{University of Pittsburgh, Pittsburgh, Pennsylvania 15260, USA}
\affiliation{Purdue University, West Lafayette, Indiana 47907, USA}
\affiliation{University of Rochester, Rochester, New York 14627, USA}
\affiliation{The Rockefeller University, New York, New York 10065, USA}
\affiliation{Istituto Nazionale di Fisica Nucleare, Sezione di Roma 1, \ensuremath{^{rr}}Sapienza Universit\`{a} di Roma, I-00185 Roma, Italy}
\affiliation{Mitchell Institute for Fundamental Physics and Astronomy, Texas A\&M University, College Station, Texas 77843, USA}
\affiliation{Istituto Nazionale di Fisica Nucleare Trieste, \ensuremath{^{ss}}Gruppo Collegato di Udine, \ensuremath{^{tt}}University of Udine, I-33100 Udine, Italy, \ensuremath{^{uu}}University of Trieste, I-34127 Trieste, Italy}
\affiliation{University of Tsukuba, Tsukuba, Ibaraki 305, Japan}
\affiliation{Tufts University, Medford, Massachusetts 02155, USA}
\affiliation{Waseda University, Tokyo 169, Japan}
\affiliation{Wayne State University, Detroit, Michigan 48201, USA}
\affiliation{University of Wisconsin-Madison, Madison, Wisconsin 53706, USA}
\affiliation{Yale University, New Haven, Connecticut 06520, USA}

\author{T.~Aaltonen}
\affiliation{Division of High Energy Physics, Department of Physics, University of Helsinki, FIN-00014, Helsinki, Finland; Helsinki Institute of Physics, FIN-00014, Helsinki, Finland}
\author{S.~Amerio\ensuremath{^{ll}}}
\affiliation{Istituto Nazionale di Fisica Nucleare, Sezione di Padova, \ensuremath{^{ll}}University of Padova, I-35131 Padova, Italy}
\author{D.~Amidei}
\affiliation{University of Michigan, Ann Arbor, Michigan 48109, USA}
\author{A.~Anastassov\ensuremath{^{w}}}
\affiliation{Fermi National Accelerator Laboratory, Batavia, Illinois 60510, USA}
\author{A.~Annovi}
\affiliation{Laboratori Nazionali di Frascati, Istituto Nazionale di Fisica Nucleare, I-00044 Frascati, Italy}
\author{J.~Antos}
\affiliation{Comenius University, 842 48 Bratislava, Slovakia; Institute of Experimental Physics, 040 01 Kosice, Slovakia}
\author{G.~Apollinari}
\affiliation{Fermi National Accelerator Laboratory, Batavia, Illinois 60510, USA}
\author{J.A.~Appel}
\affiliation{Fermi National Accelerator Laboratory, Batavia, Illinois 60510, USA}
\author{T.~Arisawa}
\affiliation{Waseda University, Tokyo 169, Japan}
\author{A.~Artikov}
\affiliation{Joint Institute for Nuclear Research, RU-141980 Dubna, Russia}
% ...

\collaboration{CDF Collaboration}
\altaffiliation[With visitors from]{
\ensuremath{^{a}}University of British Columbia, Vancouver, BC V6T 1Z1, Canada,
\ensuremath{^{b}}Istituto Nazionale di Fisica Nucleare, Sezione di Cagliari, 09042 Monserrato (Cagliari), Italy,
\ensuremath{^{c}}University of California Irvine, Irvine, CA 92697, USA,
\ensuremath{^{d}}Institute of Physics, Academy of Sciences of the Czech Republic, 182~21, Czech Republic,
\ensuremath{^{e}}CERN, CH-1211 Geneva, Switzerland,
\ensuremath{^{f}}Cornell University, Ithaca, NY 14853, USA,
\ensuremath{^{g}}University of Cyprus, Nicosia CY-1678, Cyprus,
\ensuremath{^{h}}Office of Science, U.S. Department of Energy, Washington, DC 20585, USA,
\ensuremath{^{i}}University College Dublin, Dublin 4, Ireland,
\ensuremath{^{j}}ETH, 8092 Z\"{u}rich, Switzerland,
\ensuremath{^{k}}University of Fukui, Fukui City, Fukui Prefecture, Japan 910-0017,
\ensuremath{^{l}}Universidad Iberoamericana, Lomas de Santa Fe, M\'{e}xico, C.P. 01219, Distrito Federal,
\ensuremath{^{m}}University of Iowa, Iowa City, IA 52242, USA,
\ensuremath{^{n}}Kinki University, Higashi-Osaka City, Japan 577-8502,
\ensuremath{^{o}}Kansas State University, Manhattan, KS 66506, USA,
\ensuremath{^{p}}Brookhaven National Laboratory, Upton, NY 11973, USA,
\ensuremath{^{q}}Istituto Nazionale di Fisica Nucleare, Sezione di Lecce, Via Arnesano, I-73100 Lecce, Italy,
\ensuremath{^{r}}Queen Mary, University of London, London, E1 4NS, United Kingdom,
\ensuremath{^{s}}University of Melbourne, Victoria 3010, Australia,
\ensuremath{^{t}}Muons, Inc., Batavia, IL 60510, USA,
\ensuremath{^{u}}Nagasaki Institute of Applied Science, Nagasaki 851-0193, Japan,
\ensuremath{^{v}}National Research Nuclear University, Moscow 115409, Russia,
\ensuremath{^{w}}Northwestern University, Evanston, IL 60208, USA,
\ensuremath{^{x}}University of Notre Dame, Notre Dame, IN 46556, USA,
\ensuremath{^{y}}Universidad de Oviedo, E-33007 Oviedo, Spain,
\ensuremath{^{z}}CNRS-IN2P3, Paris, F-75205 France,
\ensuremath{^{aa}}Universidad Tecnica Federico Santa Maria, 110v Valparaiso, Chile,
\ensuremath{^{bb}}Sejong University, Seoul 143-747, Korea,
\ensuremath{^{cc}}The University of Jordan, Amman 11942, Jordan,
\ensuremath{^{dd}}Universite catholique de Louvain, 1348 Louvain-La-Neuve, Belgium,
\ensuremath{^{ee}}University of Z\"{u}rich, 8006 Z\"{u}rich, Switzerland,
\ensuremath{^{ff}}Massachusetts General Hospital, Boston, MA 02114 USA,
\ensuremath{^{gg}}Harvard Medical School, Boston, MA 02114 USA,
\ensuremath{^{hh}}Hampton University, Hampton, VA 23668, USA,
\ensuremath{^{ii}}Los Alamos National Laboratory, Los Alamos, NM 87544, USA,
\ensuremath{^{jj}}Universit\`{a} degli Studi di Napoli Federico I, I-80138 Napoli, Italy
}
\noaffiliation
% Last update: $Date: 2015/12/02 22:50:09 $

%The CDF author list to be inserted here.

\date{March 1, 2016} 

\begin{abstract}
%\linenumbers\relax

We describe a measurement of the ratio of the cross sections times branching fractions of the $\bc$ meson in the decay mode $\bctojpsimunu$ to the $\bp$ meson in the decay mode $\bptojpsik$ in proton-antiproton collisions at center-of-mass energy $\sqrt{s}=1.96$~TeV. 
The measurement is based on the complete CDF Run\,II data set, which %has 
comes from an integrated luminosity of $8.7\,\textrm{fb}^{-1}$. 
%We select a sample of 1370 events with a $\jpsi\ra\mu^+\mu^-$ candidate 
%matched to a high-quality third muon.
The ratio of the 
production cross sections times branching fractions for $\bc$ and $\bp$ mesons with momentum transverse to the beam greater than $6~\gevc$ and rapidity magnitude smaller than 0.6 is 
$0.211\pm 0.012~\textrm{(stat)}^{+0.021}_{-0.020}~\textrm{(syst)}$.  Using the known $\bptojpsik$ branching fraction, the known $\bp$ production cross section, and a selection of the predicted $\bctojpsimunu$ branching fractions, %With additional assumptions,
the range for the total $\bc$ production cross section is estimated.
% to be in the range 25 $\pm$ 4 nb to 52 $\pm$ 8 nb.

\end{abstract}

% activate the following line for publication

\maketitle

%\newpage
%\tableofcontents

%\linenumbers\relax
%%%%%%%%%%%%%%%%%%%%%%%%%%%%%%%%%%%%%%%%%%%%%%%%%%%%%%%%%%%%
\section{Introduction} \label{sec:intro}
%%%%%%%%%%%%%%%%%%%%%%%%%%%%%%%%%%%%%%%%%%%%%%%%%%%%%%%%%%%%   
   We report a measurement of the ratio of the production cross sections times 
branching fractions (BF) 
\begin{equation}
{\cal R}=\frac{\sigbrbc}{\sigbrb}
\label{eq:ratio}
\end{equation}
in proton-antiproton ($p\bar{p}$) collisions at a center-of-mass energy of 1.96 TeV measured using the full CDF data set collected from February of 2001 through September of 2011 (Run\,II), which %has
comes from an integrated luminosity of 8.7~$\fb$. 
%We require $\pt(B)>6$ $\gevc$  to allow comparison with the Run\,I measurements~\cite{Ref:RunI} and theoretical predictions.

   The $\bc$-meson~\footnote{Charge-conjugate states are implied throughout the paper
unless otherwise specified.} production cross section is predicted to be three orders
of magnitude smaller than the $\bp$-meson production cross
section~\cite{Ref:Chang_ExcitedBc,Ref:FONLL}.  The branching fraction
of the $\bptojpsik$ decay is $(1.027\pm0.031)\times10^{-3}$~\cite{Ref:PDG3}, while the
branching fraction of the $\bctojpsimunu$ is predicted to be
approximately 2\%~\cite{Ref:Kiselev,Ref:Ivanov}.  Thus, we expect ${\cal R}$ 
%the ratio of production cross sections times branching fractions 
to be ${\cal O}(10^{-2})$. 
     
   The $\bc$ meson, with a mass of 6.2756 $\pm$ 0.0011 GeV/$c^2$~\cite{Ref:PDG3}, 
is the most massive meson involving 
unlike-quark flavors, with a ground state consisting of a
$\bar{b}$ and a $c$ quark.  Both the
$b$ and $c$ quarks decay through the weak interaction and, unlike in $c\bar{c}$ and $b\bar{b}$ 
quarkonia, %they 
cannot annihilate into gluons.  Consequently, there
are many possible final states to explore new aspects of heavy-quark dynamics.  Studies of strong-interaction $\bc$ production have been
possible only at hadron colliders because of the low center-of-mass energy at $e^+e^-$ colliders operating at the $\Upsilon(4S)$ resonance and the small $q\bar{q}$ cross section in $e^+e^-$ collisions at the $Z$ resonance.  %the limited $B$ hadron data samples available to the LEP experiments.  
%Compared to the Run\,I detectors at the Tevatron~\cite{Ref:RunI}, 
The CDF II detector features significant
improvements in the system for reconstructing charged-particle trajectories (tracking) that increase the acceptance and facilitate the detection and
precise measurement of the kinematic properties of $b$ hadrons and their decay products. Together with the increased luminosity, this makes it possible to measure more precisely the properties of the $\bc$ meson with the significantly larger samples of $B$ hadrons collected in Run\,II.

%     One of the goals in the study of particles containing $b$ quarks is to measure the most basic properties, such as lifetimes and production cross sections.  
     Since the production cross section of the
$\bp$-meson and its branching fraction in the decay channel $\bptojpsik$ are
well measured, it is convenient to
measure the $\bc$ production cross section with the $\bctojpsimunu$ channel using the kinematically similar $\bptojpsik$ channel as a reference.  Many systematic effects related to detector and online-event-selection (trigger) efficiencies are expected to cancel in the ratio ${\cal R}$, given that the event topologies are similar and all $J/\psi$ candidates in either the $\bctojpsimunu$ or the $\bptojpsik$ final state are reconstructed using a common set of trigger criteria. %for finding the $J/\psi$.
   
    Both the $\bp$ and $\bc$ production cross sections include production
from excited $B$ states that subsequently decay into $\bp$ or $\bc$ mesons.
Excited $\bp$ states that contribute to the $\bp$ ground state
include the radiative decay $B^{\ast+}\ra\bp\gamma$, as well as
orbital excitations of the $\bp$ and $\bzero$ mesons, e.g.,
$B^{\ast\ast0}\ra B^{+(\ast)}\pi^-$.  In the case of the $\bc$ meson, besides direct production of the
ground state, contributions are only allowed from excited states of the
$\bc$ meson itself because of flavor conservation.  Therefore, any
excited $\bc$ state whose mass is smaller than the sum of the bottom and charm meson masses cascades into the $\bc$ ground state, primarily through
radiative decay.  For example, the production cross section of the
$\bcst$ meson~\cite{Ref:Chang_ExcitedBc} is estimated to be approximately 2.5 times
the cross section to the ground state $\bc$, and the $\bcst$ meson reaches the
ground state through the radiative decay $\bcst\ra\bc\gamma$, where the
mass splitting between the $\bcst$ and $\bc$ mesons is estimated to be within
the range 40--76~$\mevcc$~\cite{Ref:BcStar_mass}. Less
important are the $P$-wave excited $B^{*+}_{cJ,L=1}$ states whose
total cross section is estimated to be about 1/2 of that of direct production
to the ground state $\bc$~\cite{Ref:Chang_PwaveBc}.
  
The ratio ${\cal R}$ can be measured using the formula
%of the production cross sections times branching fractions
%is given by the formula
\begin{equation}
{\cal R} = \frac{N_{\bc}}{N_{\bp}}\frac{\epsilon_{\bp}}{\epsilon_{\bc}}\frac{1}{\epsilon_{\mu}}\,,
\label{relcross}
\end{equation} 
where $N_{\bc}$ and $N_{\bp}$ are the numbers of reconstructed $\bctojpsimunu$ and $\bptojpsik$ events estimated in experimental data after all background subtractions and other corrections, respectively; $\epsilon_{\bp}$ and $\epsilon_{\bc}$ are the total efficiencies for selecting and reconstructing the decays $\bptojpsik$ and $\bctojpsimunu$, respectively; and $\epsilon_{\mu}$ is the muon identification efficiency.  On the right side of Eq.~(\ref{relcross}), the first factor is the relative yield for the two decays, the second term gives the scaling for the relative geometrical acceptance and detection efficiency, and the third term is a correction for the muon efficiency relative to kaons. The overall relative efficiency $\erel$ is defined by $\erel=\epsilon_{\bp}/(\epsilon_{\bc}\times\epsilon_{\mu})$.  The selection criteria for both $\bc$ and $\bp$ events are made as nearly identical as possible to minimize systematic uncertainties in both the relative yields and in determining $\erel$.
 
  The number of $\bptojpsik$ decays is determined from a fit to the invariant-mass spectrum around the known $\bp$ mass value, which includes a background component, a signal component, and a correction for the Cabibbo-suppressed $J/\psi\,\pi^+$ final state. Since the $\bc$ decay is only partially reconstructed, the number of $\bctojpsimunu$ candidates is determined by counting the total number of $\jpsiplusmu$ events in the invariant-mass window $4~\gevcc< M(\jpsiplusmu)< 6~\gevcc$ and subtracting the contributions of known
backgrounds.  The quantity $M(\jpsiplusmu)$ is the invariant mass of the trimuon partial reconstruction of the $\jpsiplusmu X$ final state, where $X$ represents any undetected particles. Because the signal events are spread over a 2 $\gevcc$ invariant-mass interval, the background cannot be determined by a simple sideband subtraction.  A large fraction of this paper is devoted to describing the methods used to determine the various backgrounds included in the $\bctojpsimunu$ candidate sample.  The principal classes of background events are the following: a wrongly identified or misidentified-$\jpsi$ candidate with a real third muon, a real $\jpsi$ meson with a wrongly identified or misidentified third muon, and a real $\jpsi$ meson with a real muon that originated from different $b$ quarks in the same event.  These
backgrounds are determined quantitatively from independent data
samples wherever possible and from Monte Carlo~(MC) simulation otherwise.  
We correct for misidentified-$\jpsi$ candidates with misidentified muons that are contained in two of the major backgrounds above and for backgrounds from other $\bc$ decay modes that yield a $\jpsiplusmu X$ final state (for examples see Table~\ref{tab:branch_frac} in Sec.~\ref{sec:odm}).  % and should be subtracted only once, but not twice.  
The analysis demonstrates that about half of the inclusive $\jpsiplusmu X$ sample is $\bctojpsimunu$ events, and the remainder is background with a small contribution from other $\bc$ decay modes.
    
Because the signal events are confined to a 2 $\gevcc$ mass region between 4 and 6 $\gevcc$, we use the events at masses between 3 and 4 $\gevcc$ and greater than 6 $\gevcc$ as a control sample to check the predictions for the major backgrounds in the signal region. 
%Uncertainties arise from several sources, including knowledge of the $\pt$ production spectrum of the $\bc$ and $\bp$, uncertainties in the estimation of the misidentified backgrounds listed above, and branching fractions of $\bc$ to intermediate decay modes which contribute to the $\bctojpsimux$ sample.
    
     The elements of the CDF\,II detector most relevant to this analysis are discussed in Sec.~\ref{sec:detector_des}.  The selection of $\bc$ and $\bp$ candidates is described in Sec.~\ref{sec:selec}.  Backgrounds are described in Sec.~\ref{sec:bc_bkg}.  Contributions from other $\bc$ decays are estimated in Sec.~\ref{sec:odm}, and the final corrected $\bctojpsimunu$ signal is discussed in
Sec.~\ref{sec:bc_excess}.  Since the measurement of $\bctojpsimunu$ 
is made relative to the decay $\bptojpsik$, the relative
reconstruction efficiency of the two decay modes in the CDF\,II
detector is estimated using MC simulation, which is
described in Sec.~\ref{sec:rel_eff}.  Systematic uncertainties
assigned to the measurement are described throughout the paper.  
Final results are presented in Sec.~\ref{sec:results}.

%%%%%%%%%%%%%%%%%%%%%%%%%%%%%%%%%%%%%%%%%%%%%%%%%%%%%%%%%%%%%%%%
\section{\label{sec:detector_des}CDF\,II Detector Description}
%%%%%%%%%%%%%%%%%%%%%%%%%%%%%%%%%%%%%%%%%%%%%%%%%%%%%%%%%%%%%%%%

     The CDF\,II detector is a multipurpose, nearly cylindrically symmetric detector consisting of a collection of silicon-strip detectors, a %open-cell multi-wire proportional 
drift chamber, and a time-of-flight (ToF) detector immersed in a 1.4 T solenoidal magnetic field, surrounded by electromagnetic and hadronic calorimeters with a projective-tower geometry, and followed by absorber and wire-chamber muon detectors. The apparatus is described in more detail in Refs.~\cite{Ref:cot, Ref:jpsi_inclusive}.
  
     Because the CDF\,II detector has a nearly azimuthally symmetric geometry that extends along the $p\bar{p}$ beam axis, the detector is described with a cylindrical coordinate system in which $\phi$ is the azimuthal angle, $r$ is the
radial distance from the nominal beam line, and $z$ points in the proton-beam
direction with the origin at the center of the detector.  The
transverse $r$-$\phi$ or $x$-$y$ plane is the plane perpendicular to the $z$ axis. The pseudorapidity $\eta$ is defined in
terms of the polar angle $\theta$ by $\eta=-\textrm{ln}[\textrm{tan}
(\theta/2)]$, where $\theta=0$ corresponds to the proton
direction. The transverse momentum $\pt$ of a particle is given by
$p_T = p\,\textrm{sin}(\theta)$ where $p$ is the magnitude of the particle momentum.
  
\subsection{\label{sec:tracking}Charged-particle trajectories}
  
  Charged-particle trajectories (tracks) are measured in the CDF\,II detector by a combination of silicon-strip detectors and a %open-cell multi-wire proportional 
drift chamber called the central outer tracker 
(COT).  The two innermost components of the charged-particle-tracking system used in this analysis are %layer 00 (L00)~\cite{Hill:2004lz,Aaltonen:2013} located at $r$ between 1.35 and 1.6 cm,
the silicon vertex detector (SVX\,II)~\cite{Sill:2000zz,Aaltonen:2013} with five double-sided layers with $r$ between 2.5 and 10.6~cm, and the intermediate silicon layers (ISL)~\cite{Affolder:2000tj,Aaltonen:2013} with three double-sided partial layers with $r$ between 20 and 29~cm. %This analysis uses the SVX\,II and the ISL silicon detectors.  
%In order to avoid bias in the measurement of candidate decay times, there is no selection requirement on the position of the secondary vertex to eliminate prompt events; also, L00 hits are explicitly removed in the estimation of the charged particle trajectories.  Including L00 would improve the secondary vertex resolution, but its tracking efficiency is only 65\%.  Thus, it was decided to eliminate L00 from the analysis in order to avoid any systematic uncertainties arising from having a signal sample composed of tracks with different efficiencies.  
%Thus, the improvement in vertex resolution close to the beam provided by the L00 detector is not particularly useful.  It is not used and the loss of efficiency entailed with its use is also avoided.
  
\begin{figure}[Htbp]
\centerline{
\includegraphics[scale=0.45]{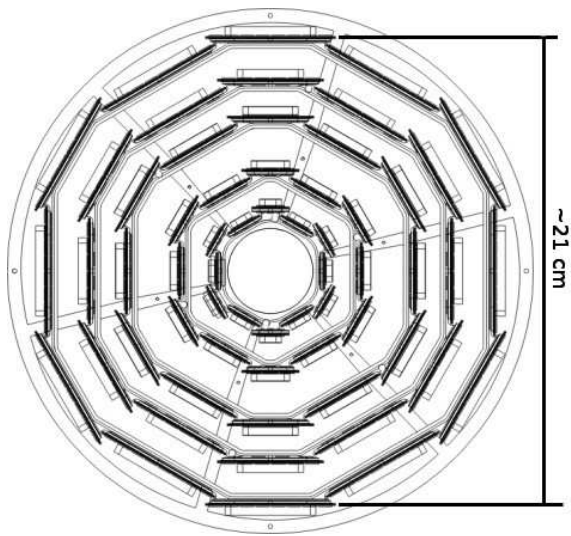}}
\caption{Arrangement of sensors in the five SVX\,II layers in an $r$-$\phi$ slice.}
\label{svx_end}
\end{figure}

    The five layers of the SVX\,II are arranged in five cylindrical shells and divided into three identical sections (barrels) along the beam axis for a total $z$ coverage of 90~cm excluding gaps.  Each barrel is divided into 12 azimuthal wedges of $30^{\circ}$ as illustrated in Fig.~\ref{svx_end}, which shows an $r$-$\phi$ slice of the SVX\,II.  
%%%%%%%%%%%%%%
%%%%%%%%%%%%%%%
%Each of the five layers in a wedge is further divided into two electrically independent modules called ladders.  There are a total of 360 ladders in the SVX\,II. They are read out independently and in parallel, which makes possible the use of the SVX\,II data in the level-2 trigger discussed in Sec.~\ref{sec:xft}.  
The sensors have strip pitches ranging from 60 to 140~$\mu$m depending on the radius.  They have strips on both sides of the silicon to allow for two position measurements at each layer.  All layers have axial strips parallel 
to the beam direction  for $\phi$ measurements.  Three layers have strips perpendicular to the beam direction
to measure $z$ position, while the remaining two layers have %small-angle stereo 
strips that are tilted by 1.2$^{\circ}$ 
relative to the axial strips.  %The small-angle stereo sensors help remove the ambiguity involved in matching $\phi$ and $z$ information when there is more than one particle producing ionization in a given sensor.
 
     The ISL detector serves to extend the precision of the SVX\,II to larger radius and allows for better matching of tracking information
between the silicon detectors and the COT.   %As with the SVX\,II, the ISL is approximately symmetric for rotations in $\phi$.  
%The arrangement of ISL layers relative to the SVX\,II can be seen in Figure~\ref{isl_rad}.  
The ISL sensors are 
double sided with axial and 1.2$^{\circ}$ %small-angle stereo 
strips spaced with a pitch of 112~$\mu$m.
  
   The silicon detectors provide a precise measurement of the azimuth of tracks and of their transverse impact parameter, the distance by which trajectories extrapolated back in the $r$-$\phi$ plane miss the beam line.  For particles with $p_T=$ 2~GeV/$c$, the transverse-impact-parameter resolution %relative to the $p\bar{p}$ interaction point 
given by the SVX\,II is about 50~$\mu$m; this includes a contribution of approximately 30~$\mu$m %added in quadrature 
due to the transverse beam-spot size~\cite{Aaltonen:2013}.  In this analysis the silicon detectors provide precise measurements of the decay vertices for $\bc$ and $\bp$ candidates.
 
\begin{figure*}[tbp]
\centerline{
\includegraphics[scale=0.7]{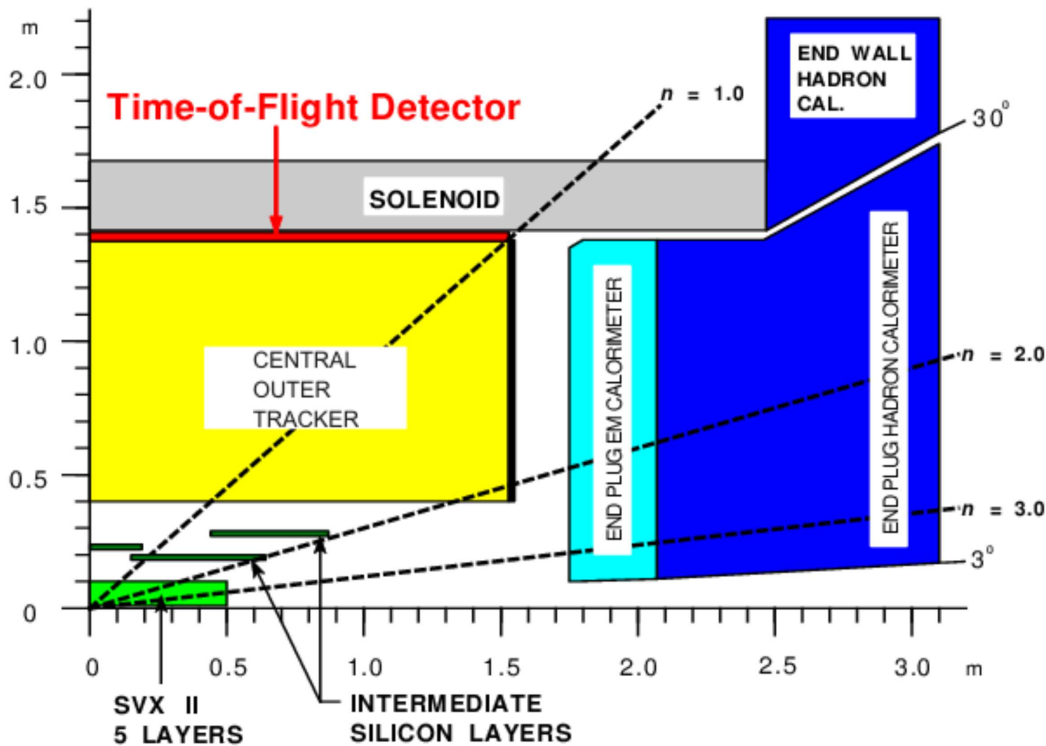}}
\caption{One quarter $r$-$z$ side view of the COT showing its position relative to other detectors.}
\label{cot_side}
\end{figure*}

The 310~cm long COT~\cite{Affolder:2003ep} is an open-cell multiwire proportional drift chamber consisting of 96 sense-wire layers from $r=$ 40~cm to $r=$ 137~cm.  The layers are grouped into alternating axial and $\pm 2^{\circ}$ stereo superlayers.  
%The COT together with the silicon systems are located inside of a 1.4~T solenoidal magnetic field pointing along the negative $z$ axis. 
The relative positions of the silicon and COT tracking systems are shown in Fig.~\ref{cot_side}.
%%%%%%%%%%%%%%%%%%%
%%%%%%%%%%%%%%%%%%%%%%%%%%
The COT alone provides excellent %charged-particle-
track reconstruction and momentum resolution.  For the combined COT, ISL, and SVX II tracking system, the asymptotic transverse
momentum resolution $\delta p_T/p_T$ has a $p_T$ dependence  given by 
%\begin{equation}
$\delta p_T/p_T = 0.0007 p_T \textrm{ (GeV/$c$)}$.
%\end{equation}
In addition the COT provides sampling of the specific-ionization-energy loss $dE/dx$ along a track, 
%as given by the Bethe-Bloch formula~\cite{Blum:1993nw}, 
which provides particle-type identification~\cite{DedxStudies}.  %The COT gives a 1.25$\sigma$ separation between pions and kaons at a  momentum of 2~GeV/$c$.
   
    Following the COT in radius, but located inside the solenoid coil, is a ToF detector~\cite{Acosta:2004kc} consisting of scintillator bars with photomultiplier tubes at both ends.  The ToF system has a resolution of approximately 110~ps~\cite{Cabrera:2004nw} that corresponds to a separation of 0.6$\sigma$ between pions and
kaons at $p=$ 3~$\gevc$. Both the ToF and $dE/dx$ measurements are important in determining the particle fractions in the analysis of the misidentified-muon background discussed in Sec.~\ref{sec:fake_mu}.
\subsection{\label{sec:muon_det}Muon detectors}
 
     The central muon detector (CMU)~\cite{Ascoli:1987av} consists of %a set of 
single-wire drift cells located outside of each calorimeter wedge, covering $|\eta|<$~0.6, starting at $r=347$~cm.  For particle trajectories at 90$^{\circ}$, there are approximately 5.5 interaction lengths for hadron attenuation before the wire drift cells.  
%The complete detector is formed from 48 wedges with 24 covering positive $z$ and 24 covering negative $z$.  
%The wedges are 12.6$^{\circ}$ in $\phi$ and 226~cm long.  There are 2.4$^{\circ}$ gaps between drift cell arrays, resulting in an overall $\phi$ coverage of 84\%.  
The drift cell arrays sample the trajectories in up to four positions in the $r$-$\phi$ plane that are used to form straight track segments.  The track segments are matched to extrapolated COT tracks to form muon candidates using both position and slope.

     The central muon upgrade detector (CMP) covers the same $|\eta|<$~0.6 range as the CMU.  Arranged in a box 
that surrounds the central region of the detector, the CMP consists of single-wire drift cells stacked in four
layers similar to the CMU.  Since the CMP is located behind an additional 60~cm of steel (approximately 3.3 interaction lengths), there are considerably fewer kaons and pions that penetrate to the CMP compared to the CMU.  Muon candidates associated with track segments in both the CMU and CMP are called CMUP muons.
 
     The central muon extension detector (CMX) extends the %CDF\,II 
muon coverage to the kinematic region $0.6<|\eta|<1.0$.  The
CMX consists of eight layers of single-wire drift tubes. %arranged in 15$^{\circ}$ $\phi$ wedges. 
%There is no additional shielding added between the CMX and the interaction point as the long path through the calorimetry and the intervening detector supports is adequate to remove the hadronic component of the event.
The calorimeter, together with detector supports, provides approximately 6 (at $\eta=0.6$) to 10 (at $\eta=1.1$) interaction lengths of absorber in front of the CMX for hadron attenuation~\cite{Blair:1996kx}.

     This analysis uses the CMU and CMX to identify the muon candidates  for reconstructing $\jpsi$ mesons,  but requires the CMUP %muons 
for the third muon in the semileptonic decay $B_c^+ \rightarrow J/\psi\, \mu^+ X$.
\subsection{\label{sec:xft}Online event-selection system%The CDF\,II trigger sys
}

    The Tevatron average beam crossing rate is %provides $p\bar{p}$ collisions at a rate of 
1.7~MHz, and the typical CDF\,II triggered event size is 
about 300~kB.  Since the data-acquisition system can write about 20~MB/s to permanent storage, it is necessary to reject 99.996$\%$
of the $p\bar{p}$ collisions.  This is accomplished by a three-level online event-selection system (trigger).  The first two levels use
custom electronic logic circuits to choose or reject events and the third level uses a parallel array of commodity personal computers. %commercial computers.

%\subsubsection{The level 1 trigger}
    The level-1 trigger makes decisions using information from the COT, calorimeters, and muon detectors.  The extremely fast tracker (XFT)~\cite{Thompson:2002}, a pattern-recognition system for fast
COT track reconstruction, %using dedicated hardware
provides the tracks for the level-1 trigger~\cite{Holm:1999uf}.  
The decision time is fixed at 5.5~$\mu$s and this requires a pipeline buffer with a depth of 42 events for the storage of event data while decisions
are made. The typical level-1 rate of event acceptance is approximately 20~kHz. %and the maximum accept rate is 50 kHz.  
For this analysis events are collected by one of two level-1 triggers: two XFT tracks corresponding to charged particles with $p_T>1.5$~GeV/$c$
are matched with track segments in the CMU detector, or one XFT track corresponding to a particle with $p_T>1.5$~GeV/$c$ is matched with a CMU track segment,
while another with $p_T>2.0$~GeV/$c$ is matched with a CMX track segment.

%\subsubsection{The level 2 trigger}
     After an event is accepted by the level-1 trigger, it is passed to the level-2 trigger~\cite{Anikeev:2006kw}.  The level-2 trigger uses the
same information as the level-1 trigger with additional track position information from the 
silicon vertex trigger (SVT).  The SVT applies pattern recognition to SVX\,II silicon hits (a positive detector response to the passage of a charged particle) that are matched to XFT tracks
and calculates impact parameters for the tracks~\cite{Ashmanskas:2004uv}.  Events with track vertices (two or more tracks originating from a common point) displaced from the beam line, i.e., likely to contain the decay of a long-lived particle such as a $B$ or $D$ meson, are chosen by 
requiring two SVT tracks with nonzero impact parameters.  For the case of the dimuon
triggers used to collect signal candidates for this analysis, the SVT is not used, but SVT-triggered events are used to reconstruct control samples used in the analysis, such as $D^{*+}\rightarrow D^0\pi^+$ followed by $D^0\rightarrow K^- \pi^+$.  These decays are used to define cleanly identified samples of pions and kaons to measure the probabilities that such hadrons are misidentified as muons.  The level-2 trigger typically has a total output rate of 200--800~Hz.  

%\subsubsection{Level 3 Trigger}  
  The level-3 trigger system~\cite{Chung:2005iz} %runs on a computer cluster and 
uses information from 
all parts of the CDF\,II detector to reconstruct and select events.  The typical output rate for level 3 is approximately 100~Hz.  For the level-3-$J/\psi$ trigger %part of this rate 
used in this analysis, there is a selection on the $J/\psi$ that requires the invariant mass of the muon pair used in the reconstruction to fall in the range 2.7--4.0~$\gevcc$. 
%%%%%%%%%%%%%%%%%%%%%%%%%%%%%%%%%%%%%%%%%%%%%%%%%%%%%%%%%%%%
\section{Event Selection}  \label{sec:selec}
%%%%%%%%%%%%%%%%%%%%%%%%%%%%%%%%%%%%%%%%%%%%%%%%%%%%%%%%%%%%

The high spatial resolution provided by the silicon-tracking system in the plane transverse to the beam line makes it ideal for the reconstruction of $B$ hadrons.  Because tracks curve in the
transverse plane, the transverse momentum is
well measured.  Additionally, the small transverse $p\bar{p}$
interaction region %(about 30~$\mu$m) 
constrains the
location of the $p\bar{p}$ collision space point (primary vertex) in this plane.  Consequently,
we use the transverse momentum $\pt$ of the reconstructed hadron and
transverse decay length $\lxy$, which is the decay length of the
reconstructed three-track system projected into the transverse plane,
when selecting $\bc$ and $\bp$ candidates and when discriminating
against backgrounds.  Unless otherwise noted, $\lxy$ is measured from
the primary vertex of the event to the candidate $B$-meson decay point (decay vertex).  
%The decay length $ct$ of the $\bc$ or $\bp$ candidate is
%constructed from $\lxy$ and $\pt$ of the three-track system: $ct
%\equiv m \lxy/\pt$, where $m$ is the known mass of the
%corresponding $B$ candidate~\cite{Ref:PDG3}.

%Our selection requirements closely follow those of the $\bc$ lifetime 
%measurement~\cite{Ref:MarkThesis}.  
     We use similar selection requirements for both 
the $\bctojpsimux$ and $\bptojpsik$ decays to minimize possible systematic uncertainties in the relative efficiency between 
the two modes.
%%%%%%%%%%%%%%%%%%%%%%%%%%%%%%%%%%%%%%%%%%%%%%%%%%%%%%%%%%%%
\subsection[$\jpsitomumu$ selection]{\boldmath{$\jpsitomumu$} selection}
\label{sec:jpsi_selec}
%%%%%%%%%%%%%%%%%%%%%%%%%%%%%%%%%%%%%%%%%%%%%%%%%%%%%%%%%%%%

\begin{figure}[tbp]
\centerline{
\makebox{\includegraphics[width=1.0\hsize]{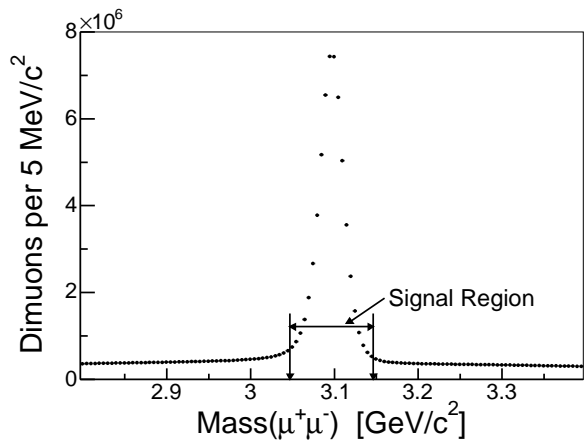}}}
\caption{Dimuon invariant-mass distribution for oppositely charged muon pairs near the $\jpsi$-meson mass.  The signal region for selecting a $\jpsi$ meson is shown.}
\label{dimuon_mass}
\end{figure}

The data are collected with a dimuon trigger
that requires two oppositely charged muon candidates (see Sec.~\ref{sec:xft}).
%, both of which have a transverse momentum of $\pt>1.5~\gevc$.  The trigger requires that the muons be identified at the first level of the CDF\,II trigger system by the XFT track processor, which reconstructs tracks in the COT and determines the transverse momentum.  The trigger also requires that at least one muon candidate track extrapolates to a signal in the CMU detector and that the second muon candidate track either extrapolates to a signal in the CMU detector, which has $|\eta| <0.6$ or to a signal in the CMX detector, with $|\eta| < 1.0$ but with a $\pt>2.0~\gevc$. 
The trigger requirements are confirmed in our offline 
analysis using track variables reconstructed from track fits for track candidates passing our selection criteria.  To guarantee good track quality, each track is required to have at least three $r$-$\phi$ hits in the silicon detector
and hits in at least ten axial and ten stereo layers in the
COT.  We define a likelihood ratio ${\cal
LR}(\mu)$ that incorporates information from the muon detectors,
calorimeters, and tracking detectors to optimize the separation of real
muons from hadrons~\cite{Ref:GiurgiuThesis}.  %The ratio is defined to take on values from zero (hadrons misidentified as muons) to one (real muons); in this analysis we require ${\cal LR}(\mu) >0.06$ for both muons. 
This muon likelihood selection is determined from an
optimization study carried out on the signal and sideband regions of
the $\mu^+\mu^-$ invariant-mass
distribution~\cite{Ref:MarkThesis}. The dimuon invariant mass distribution near the $J/\psi$-meson mass with muon candidates that satisfy the muon likelihood selection is shown in Fig.~\ref{dimuon_mass}. In the $\jpsi$ signal region, there are $6.1\times 10^7$ dimuon events.
% ++++++++++++++++++++++++++++++++++++++++++++++++++++++++++++++++++++++
% ++++++++++++++++++++++++++++++++++++++++++++++++++++++++++++++++++++++
Selection of the $\jpsi$ meson requires the two muons to come from a common decay point and have an invariant mass that lies within $50~\mevcc$~of the known $\jpsi$-meson mass~\cite{Ref:PDG3}.
The selection requirements
applied to the $\jpsitomumu$ candidates are listed in
Table~\ref{tab:jpsi_cuts}.
% ++++++++++++++++++++++++++++++++++++++++++++++++++++++++++++++++++++++
\begin{table*}[Htbp]
\caption{
Selection requirements applied to the muons of $\jpsi$ candidates
and to the two-particle $\jpsi$ candidates.  The two muons are
labeled $\mu_1$ and $\mu_2$ to identify the two tracks of the trigger.}
\begin{center} 
\begin{tabular}{ll}
\hline\hline
Selection requirement             & $\,\,\,\,$Value                     \\
\hline
$\mu_1$                & $|\eta|<0.6$ and $\pt>1.5~\gevc$                \\
$\mu_2$                & ($|\eta|<0.6$ and $\pt>1.5~\gevc$) \\
                       & or ($0.6\le|\eta|<1.0$ and $\pt>2.0~\gevc$)  \\
COT hits/track         & Hits in ten axial and ten stereo layers         \\ 
$r$-$\phi$ silicon hits/track  & $\ge 3$                                \\
Muon likelihood/muon   & Optimized using likelihood ratio %${\cal LR}(\mu)>0.06$ 
\\
$|M(\mu_1 \mu_2)-M_{\jpsi}|$ & $< 50.0~\mevcc$   \\
\hline\hline
\end{tabular}
\end{center}
\label{tab:jpsi_cuts}
\end{table*}
% ++++++++++++++++++++++++++++++++++++++++++++++++++++++++++++++++++++++

%%%%%%%%%%%%%%%%%%%%%%%%%%%%%%%%%%%%%%%%%%%%%%%%%%%%%%%%%%%%
\subsection{Three-track-system selection} 
\label{sec:three_trk}
%%%%%%%%%%%%%%%%%%%%%%%%%%%%%%%%%%%%%%%%%%%%%%%%%%%%%%%%%%%%

   The three-track event candidates used in this analysis are chosen by matching a third track to a $\jpsi$ candidate in three dimensions, 
%by requiring the third track 
%to form a three-track vertex with the $\jpsi$ candidate, 
where the $\chi^2$ probability for the kinematic fit to a common vertex is greater than 0.001 with the dimuons from the $\jpsi$ decay constrained to the known invariant mass of the $\jpsi$ meson~\cite{Ref:PDG3}. The selection requirements used to choose the sample of three tracks consistent with a common origin are listed in Table~\ref{tab:thirdTrk_cuts}.  The three-track sample is also called the \jpsihyptrack sample and is the sample from which decays $\bctojpsimux$ and $\bptojpsik$ are reconstructed.  Candidates for the $\bctojpsimunu$ decay are chosen by requiring the third track to be identified as a muon in both the CMU and CMP detectors (CMUP) as described in Sec.~\ref{sec:muon_det}.  In addition to the continuum background that contributes to the $\bptojpsik$ decay candidates, there is the Cabibbo-suppressed decay $\bp\to\jpsi\pi^+$.  Background to $\bctojpsimux$ decays arises when a $\pi^+$, $K^+$, or $p$ is misidentified as a muon (misidentified-muon background).  Another background is contributed when a real muon from one $B$-hadron decay combined with a real $\jpsi$ candidate from a different $B$-hadron decay passes the three-track vertex-selection requirements ($\bbbar$ background).  The \jpsihyptrack sample is used extensively to determine the rate of hadrons producing muon signatures in the detector (see Sec.~\ref{sec:fake_mu}). The third, fourth, and fifth columns in Table~\ref{tab:thirdTrk_cuts} identify which selection
criteria are applied to the $\bc$, $\bp$, and \jpsihyptrack candidates, respectively.
% ++++++++++++++++++++++++++++++++++++++++++++++++++++++++++++++++++++++
\begin{table*}[Htbp]
\caption{Selection requirements applied to the third track and the three-particle \jpsihyptrack system and samples selected from the \jpsihyptrack system.}
\begin{center} 
\begin{tabular}{llccc}
\hline\hline
Selection requirement & Value            & $\bc$ & $\bp$ & \jpsihyptrack \\
\hline%\hline		           				                    
\multicolumn{5}{c}{Third track}         \\
\hline
Muon type              & CMUP                                        & X  &   &            \\
CMUP boundary          & Track extrapolates to CMU and CMP detectors & X  & X & X          \\
Match with XFT         & Track is required to trigger XFT            & X  & X & X          \\
Isolation at CMU       & No other extrapolated track within          & X  & X & X          \\
                       & 40 cm at CMU   \\
$p_T$                  & $> 3.0~\gevc$                               & X  & X & X          \\
$r$-$\phi$ silicon hits/track  & $\ge 3$                             & X  & X & X          \\
COT hits/track         & Ten stereo and ten axial hits                 & X  & X  & X  \\
$\dedx$ hits/track     & $\ge 43$ hits                               & X  & X & X          \\
\hline%\hline
\multicolumn{5}{c}{\jpsihyptrack system} \\
\hline
Kinematic-fit probability     & $>0.001$                             & X  & X & X          \\
$\Delta\phi$           & $<\pi/2$                                    & X  & X & X          \\
$\sigma_{L_{xy}}$      & $<200$ $\mu m$                                & X  & X & X          \\
$\lxy$/$\sigma_{\lxy}$ & $> 3$                                       & X  & X & X          \\
$\bc$ mass region      & $|M(\jpsiplustrack)-5.0~\gevcc|<1.0~\gevcc$ & X  &   & X          \\
$\jpsiplusk$ mass Veto & $|M(\jpsiplusk)-5.279~\gevcc|>0.05~\gevcc$  & X  &   & X          \\
\hline\hline
\end{tabular}
\end{center}
\label{tab:thirdTrk_cuts}
\end{table*}
% ++++++++++++++++++++++++++++++++++++++++++++++++++++++++++++++++++++++
%In addition to the requirements on the muons from the $\jpsi$ decay, we require the event to have a third track that is matched to the same vertex as the $\jpsi$ candidate.  This third track might originate from one of three sources: the $\bctojpsimux$ decay; the $\bptojpsik$ decay (including its related combinatorial and $\bp\to\jpsi\pi^+$ backgrounds); or a $\pi^+$, $K^+$ or $p$ that mimics a third muon.  This third category of tracks represents an irreducible background to our measurement and must be accounted for in the calculation of the trimuon backgrounds.  Consequently, we select a sample of events that we name the $\jpsiplustrack$ sample.  This sample is used to determine the rate that a hadronic track produces a muon-like signature in the detector.  The selection requirements for the third track and the three-track system are listed in Table~\ref{tab:thirdTrk_cuts}. The third, fourth, and fifth right-most columns identify which selection criteria are applied to the $\bc$, $\bp$, and $\jpsiplustrack$ candidates, respectively.

    %To reduce the contribution of hadrons misidentified as muons in the $\bc$ sample, we require the third muon candidate to be identified as a CMUP muon. %extrapolate to track segments in both the CMU and CMP detectors. 
    The CMUP requirement %muon track segment 
is not made
for the $\bp$ or \jpsihyptrack samples.  However, to ensure that the acceptance is consistent across samples, the third track is required to extrapolate to the same region of the CMU and CMP detectors as the third-muon candidates and to satisfy the isolation cut applied to third-muon candidates.  In all three samples the third track is required to meet the XFT criteria because the events of the control sample used to determine the probabilities that pions and kaons are misidentified as muons (see Sec.~\ref{sec:fake_mu}) are selected with the XFT trigger.
The muon selection also requires that no other track with $\pt>1.45~\gevc$ extrapolates to within a transverse distance of
40~cm in the $r$-$\phi$ plane at the front face of the CMU element relative to the track candidate observed.  This ``track isolation requirement'' ensures that the
estimation of the misidentified-muon background is consistent across
the various data samples used in the analysis and does not require a
correction for local track density.

To penetrate the additional absorber between the CMU and CMP
detectors, a muon must have a minimum initial transverse momentum greater than 3~\gevc.  Consequently, the third track in all three samples is
required to have a transverse momentum greater than 3~\gevc. To ensure
good-quality track reconstruction in all samples, standard criteria (see Table~\ref{tab:thirdTrk_cuts}) for good track and vertex reconstruction and reliable $dE/dx$ information are imposed.  %we require the third track to have at least three $r$-$\phi$ hits in the silicon detector and five hits in each of at least two COT super-layers.  Since $\dedx$ is used to determine the particle composition of the samples that measure the misidentified-muon probabilities, the third track is also required to have information from at least 43 wires in the COT that provide $\dedx$ information.  The number of required wires is determined from a $\dedx$ calibration study.
 
%We require the three-track system to form a vertex in three dimensions. In the kinematic fit, the tracks are constrained to come from a common point with a $\chi^2$ probability greater than 0.1\%.  
     The azimuthal opening angle $\phi$ in the lab frame
between the $\jpsi$ and third track is required to be less than
$\pi/2$ in all samples because no signal events are expected to contribute outside of this azimuthal aperture.  The uncertainty $\sigma_{\lxy}$ on $\lxy$ %the displacement between the measured beam line position and the reconstructed decay point 
is required to be less than 200~$\mu$m in the transverse plane.  Simulation studies indicate that this requirement removes primarily
background events and a negligible number of signal events.  The selection criterion $\lxy/\sigma_{\lxy}>
3$ is chosen to eliminate the prompt $\jpsi$ background that arises
from $\jpsi$ mesons produced directly in the $p\bar{p}$ interaction.  The
invariant masses of events in the $\jpsi\,\mu^+$ and \jpsihyptrack samples are
reconstructed with the mass of the third charged particle assigned as a pion, kaon, or muon mass, depending on how the event is used in the analysis.  The signal region for $\bctojpsimunu$ candidates is set between 4 and 6~$\gevcc$. In the $\jpsi\,\mu^+$ sample the mass of the third charged particle is normally assumed to be that of a muon, but to eliminate residual $\bptojpsik$ background, we remove all events with an invariant mass within 50~$\mevcc$ of the known value of the $\bp$ mass~\cite{Ref:PDG3} assuming the mass of the third particle to be that of a kaon.  %The criterion used to remove residual $\bptojpsik$ events is also applied to the evaluation of all backgrounds in the \sigwin signal region.  
%These invariant-mass requirements are not applied to the $\bp$ candidate sample.

%Since the luminosity increases over the time of taking data, one might expect 
%that our $\bc$ and $\bp$ yields will vary versus the calendar time due to the 
%$\jpsi$ dimuon trigger prescale modification. The question arises as to how 
%stable the ratio of the $\bc$ to $\bp$ yields are as a function of time. 
%Fig.~\ref{BcBpYiels_vsTime} shows the ratio of $\bc$ to $\bp$ yields versus 
%the dataset section number.
% ++++++++++++++++++++++++++++++++++++++++++++++++++++++++++++++++++++++
%\begin{figure}[htbp]
%\centerline{
%\makebox{\includegraphics[width=0.5\hsize]{feps/prd_NbcToNbpRatio.eps}}}
%\caption{The ratio of $\bc$ to $\bp$ yields versus the dataset section number.}
%\label{BcBpYiels_vsTime}
%\end{figure}
% ++++++++++++++++++++++++++++++++++++++++++++++++++++++++++++++++++++++
%A fit to the ratio $N(B_c^+)/N(B^+)$ shown in Fig.~\ref{BcBpYiels_vsTime} 
%yields the slope value 0.00003$\pm$0.00032 consistent with zero.

\begin{figure}[tbp]
  \centerline{
  \makebox{\includegraphics[width=1.0\hsize]{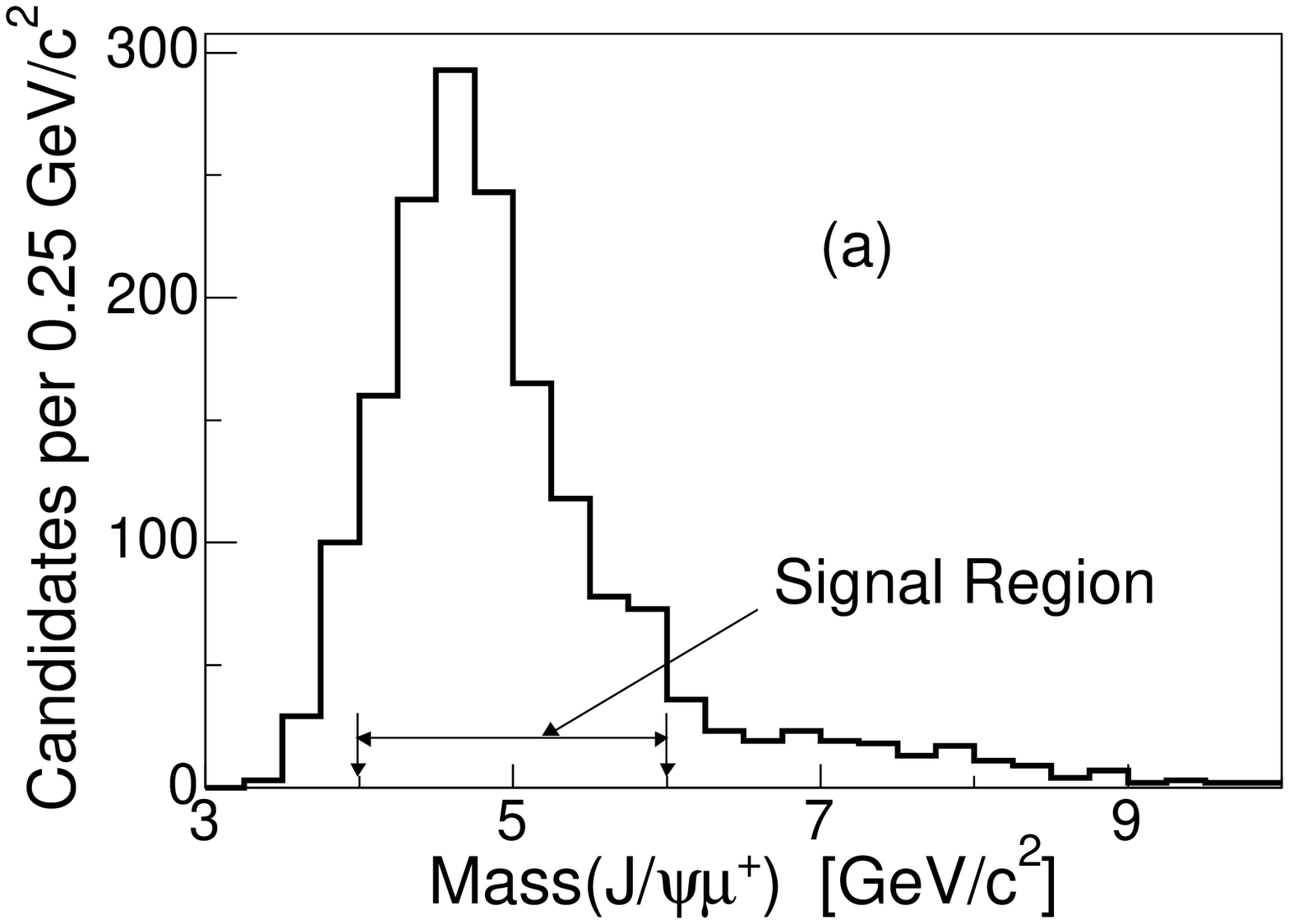}}}
  \centerline{
  \makebox{\includegraphics[width=1.0\hsize]{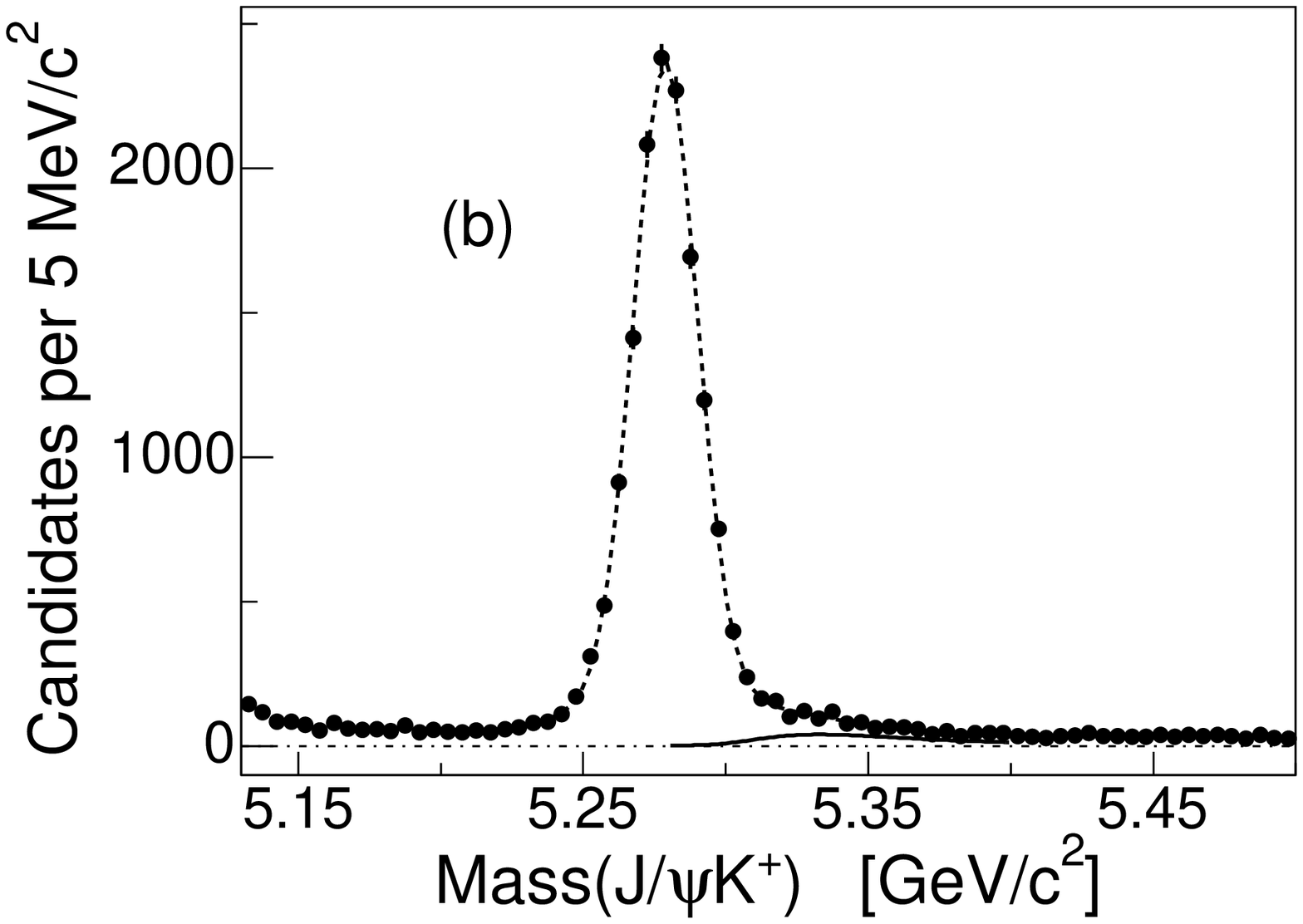}}}
  \caption{(a) Distribution of invariant mass for $\jpsi\mu^+$ candidates with transverse momentum of the $\jpsi\mu^+$ system greater than 6~$\gevc$ and (b) invariant-mass
  distribution of the $\jpsiplusk$ candidates for $\bp$ decay. 
  The Cabibbo-suppressed $\bptojpsipi$ contribution is shown as a solid 
  curve in (b). %N($\bptojpsipi$) is fixed to 3.83\% of $N(\bptojpsik)$ 
  %following Ref.~\cite{Ref:LHCbBR_BpJpsiPi}. 
}
  \label{fig:threeTrk_mass}
\end{figure}

Using the $\jpsitomumu$ selection requirements from Table~\ref{tab:jpsi_cuts} and the $\bc$ and $\bp$ selection requirements from Table~\ref{tab:thirdTrk_cuts}, the invariant-mass distributions of the $\jpsiplusmu$ and $\jpsiplusk$ candidates are constructed.  These are shown in Fig.~\ref{fig:threeTrk_mass}. Both samples are subsets of the \jpsihyptrack sample and must pass a minimum
$p_T > 6~\gevc$ requirement applied to the three-track system, where the third track is assumed to be either a muon or kaon, depending on the sample.

We select $1370\pm 37$ $\jpsiplusmu$ candidate events within a \sigwin 
signal mass window.  % for $\pt(\jpsiplusmu)> 6~\gevc$. 
To extract the number of $\bptojpsik$ events, the $\jpsiplusk$ invariant-mass distribution is fit with a function that consists of a double Gaussian for $\bptojpsik$ decays, a template for the invariant-mass distribution generated by MC simulation for the Cabibbo-suppressed $\bptojpsipi$ contribution within the 
mass range 5.28--5.4 $\gevcc$, and a second-order polynomial for the continuum background. The Cabibbo-suppressed 
$\bptojpsipi$ contribution is fixed to 3.83$\%$ of the number of $\bptojpsik$ decays following
Ref.~\cite{Ref:LHCbBR_BpJpsiPi}.  The fit determines a yield of 14 338$\pm 125$ $\bptojpsik$ decays.% for $\pt(\jpsiplusk)> 6~\gevc$.
% ++++++++++++++++++++++++++++++++++++++++++++++++++++++++++++++++++++++
% ++++++++++++++++++++++++++++++++++++++++++++++++++++++++++++++++++++++

%%%%%%%%%%%%%%%%%%%%%%%%%%%%%%%%%%%%%%%%%%%%%%%%%%%%%%%%%%%%
\section{\boldmath{$\protect\bc$} Backgrounds} 
\label{sec:bc_bkg}
%%%%%%%%%%%%%%%%%%%%%%%%%%%%%%%%%%%%%%%%%%%%%%%%%%%%%%%%%%%%

We consider contributions to the $\bc$ backgrounds from events in which a $\jpsi$ candidate is misidentified, a third muon is misidentified, or $\bbbar$ pairs decay in which one of the $b$
quarks produces the $\jpsi$ meson and the other produces the third
muon.  The misidentified-$\jpsi$-meson background is due to the
reconstruction of a $\jpsitomumu$ candidate that does not consist of
real muons originating from a $\jpsi$ meson, but from hadrons incorrectly identified as muons that produce a mass consistent
with that of the $\jpsi$ meson.  This background is estimated from the sidebands of the $\mu^+\mu^-$ invariant-mass 
distribution and is discussed in Sec.~\ref{sec:fake_jpsi}.  The
misidentified-muon background is due to a third track
that satisfies the vertex requirement and mimics a muon in the
CDF\,II detector but is a hadron.  This mistaken identification
can arise either because a kaon or pion decays in flight to a muon and
produces a muon signature in the detector,  
%because a hadron ``punches through'' to the muon detectors.  Hadron punch through can occur when 
a hadron passes through the calorimeter, %and is detected in both the CMU and CMP muon chambers, 
or a hadron shower yields a track segment in the CMU and CMP chambers.
%or when a hadron interacts strongly in the calorimeter and produces a track in the shower that is matched to the direction of the initial hadron and that can reach the muon system.  
The estimation of the misidentified-muon background directly from the data is discussed in Sec.~\ref{sec:fake_mu}.  Finally, the
$\bbbar$ background is estimated from a parametrization of the
azimuthal opening angle between the reconstructed $\jpsi$ meson and
the third muon trajectory using  MC simulation.  This is
discussed in Sec.~\ref{sec:b_bbar}.

%%%%%%%%%%%%%%%%%%%%%%%%%%%%%%%%%%%%%%%%%%%%%%%%%%%%%%%%%%%%
\subsection[Misidentified-$\jpsi$-meson background]{Misidentified-\boldmath{$\jpsi$}-meson background} 
\label{sec:fake_jpsi}
%%%%%%%%%%%%%%%%%%%%%%%%%%%%%%%%%%%%%%%%%%%%%%%%%%%%%%%%%%%%

\begin{figure}[tbp]
\centerline{
\makebox{\includegraphics[width=1.0\hsize]{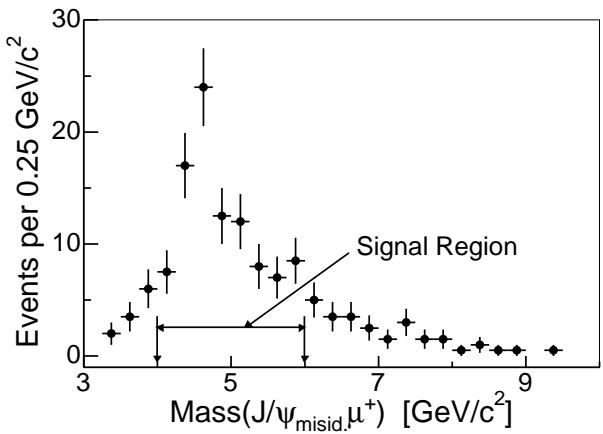}}}
\caption{Invariant-mass distribution of the $\jpsi_{\textrm{misid}}\mu^+$ 
system.}
\label{jpsiSideMu_mass}
\end{figure}

The misidentified-$\jpsi$-meson background is estimated using the
track pairs from the sideband regions of the $\mu^+\mu^-$ invariant-mass distribution, $M(\mu^+\mu^-)$.  These dimuon pairs are required to share a common vertex with 
the third muon. The signal dimuon mass region is defined to be within 
50~MeV/$c^2$ of the known value of the $\jpsi$-meson mass,
$M_{\jpsi}=3.0969~\gevcc$~\cite{Ref:PDG3}.  The sideband regions
are defined as $|(M_{\jpsi}\pm 0.150)-M(\mu^+\mu^-)| < 0.050~\gevcc$. 
%Since the width of each sideband is the same as the signal region, we weight each event observed by half when counting the contribution of these events to the total background. 
The resulting $\jpsiplusmu$ invariant-mass 
distribution based on misidentified-$\jpsi$ mesons, $\jpsi_{\textrm{misid}}$, is presented in Fig.~\ref{jpsiSideMu_mass}. We find 11.5$\pm$2.4 events within 3--4~$\gevcc$, 96.5$\pm$6.9 events within the 4--6~$\gevcc$ signal region, and 25$\pm$3.5 events at masses greater than 6~$\gevcc$.
% ++++++++++++++++++++++++++++++++++++++++++++++++++++++++++++++++++++++
% ++++++++++++++++++++++++++++++++++++++++++++++++++++++++++++++++++++++

%%%%%%%%%%%%%%%%%%%%%%%%%%%%%%%%%%%%%%%%%%%%%%%%%%%%%%%%%%%%
\subsection{Misidentified-muon background} 
\label{sec:fake_mu}
%%%%%%%%%%%%%%%%%%%%%%%%%%%%%%%%%%%%%%%%%%%%%%%%%%%%%%%%%%%%

\begin{figure}[tbp]
\centerline{
\makebox{\includegraphics[width=1.0\hsize]{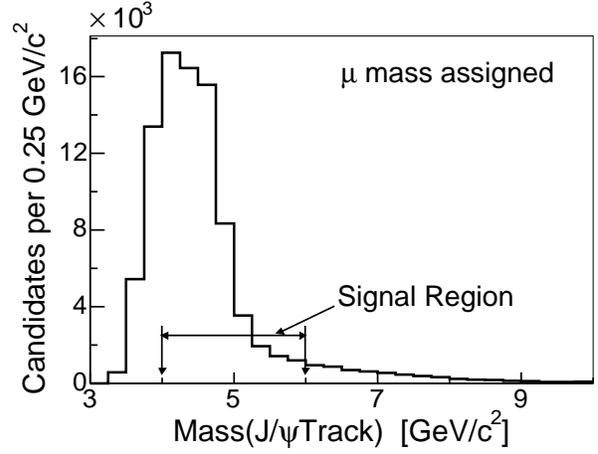}}
}
\caption{Invariant-mass distribution of the \jpsihyptrack
system. This sample is used in the misidentified-muon calculation.}
\label{fig:jpsiTrkMass}
\end{figure}

     The misidentified-muon background arises from real $\jpsi$ decays that form a good three-track vertex with a hadron that is misidentified as a muon. We determine this background from the data as a function of the momentum of the third charged particle by using the \jpsihyptrack sample combined with knowledge of the fraction of pions, kaons, and protons in the \jpsihyptrack sample and the probability of each hadron type to
be misidentfied as a muon.  Equation~(\ref{fakeMu_W_formula}) gives the total probability $W$ that the third track in an event in the \jpsihyptrack sample is misidentified as a muon:
% ++++++++++++++++++++++++++++++++++++++++++++++++++++++++++++++++++++++
\begin{eqnarray}
W&=&\epsilon_{\pi}(1+F^{\textrm{out}}_{\pi})F_{\pi}\nonumber \\ 
& &+ \epsilon_{K}(1+\alpha F^{\textrm{out}}_{K})F_{K} + \epsilon_{p}F_{p}\,,
\label{fakeMu_W_formula}
\end{eqnarray}
% ++++++++++++++++++++++++++++++++++++++++++++++++++++++++++++++++++++++
where $\epsilon_{\pi,K,p}$ are the probabilities for the relevant particle type to 
be misidentified as a muon, %$F^{out}_{\pi,K}$ is the fraction of misidentified 
%events outside of the $\dzero$ mass peak for a given particle type, $\alpha$=1
%for $K^-$, $\alpha=\epsilon_{K^-}/\epsilon_{K^+}$ for $K^+$, 
and $F_{\pi,K,p}$ are the fractions of the relevant particle types within the \jpsihyptrack sample. The $\epsilon_{\pi,K,p}$ are determined as functions of the $\pt$ of the third particle, and the $F_{\pi,K,p}$ are determined as functions of the momentum of the third particle. The terms $1+F^{\textrm{out}}_{\pi}$ and $1+\alpha F^{\textrm{out}}_{K}$ are corrections to the probabilities for pions and kaons, respectively, to be misidentified as muons and are discussed in Sec.~\ref{sec:fakeMu_fracOffPeak}.  For each event in the \jpsihyptrack sample, reconstructed assuming that the third track is a muon, we determine $W$ and sum these weights as functions of the $\jpsiplusmu$ invariant mass of the events.  The result is a measurement of the misidentified-muon background in the $\jpsiplusmu$-event sample as a function of the $\jpsiplusmu$ invariant mass.  The invariant-mass distribution of the \jpsihyptrack system is shown in Fig.~\ref{fig:jpsiTrkMass}.
% ++++++++++++++++++++++++++++++++++++++++++++++++++++++++++++++++++++++
%      The determination of the probability for a $\pi^{\pm}$ or $K^{\pm}$ to be misidentified as a muon is presented below.
%     In the next section we calculate the probabilities $\epsilon_{p,\pi,K}$ that protons, pions, and kaons, respectively, are misidentified as muons.

%%%%%%%%%%%%%%%%%%%%%%%%%%%%%%%%%%%%%%%%%%%%%%%%%%%%%%%%%%%%
\subsubsection{Probability for a $p$, $\pi^{\pm}$, or $K^{\pm}$ to be misidentified as a muon}
\label{sec:fakeMu_probability}
%%%%%%%%%%%%%%%%%%%%%%%%%%%%%%%%%%%%%%%%%%%%%%%%%%%%%%%%%%%%
% ++++++++++++++++++++++++++++++++++++++++++++++++++++++++++++++++++++++
\begin{table*}[Htbp]
\caption{Pion and kaon particle selection requirements.}
\begin{center}
\begin{tabular}{lcc}
\hline\hline
Selection requirement           & Value         & Comments   \\
\hline
$q(\pi) q(\pi)$ & 1             & Same sign             \\
$\pt$ of $\pi$ or $K$    & $>$3 $\gevc$  & Same as in $\bctojpsimunu$ \\
$\pt(K^{-}\pi^{+})$        & $>$3 $\gevc$  & $\dzero$                   \\
$\Delta\phi(K^{-}\pi^{+})$ & 0.035 - 2.36 rad. &                        \\
Vertex $\chi^{2}$ prob     & $>$0.001      & $\dzero$ and $\dst$        \\
$L_{xy}$                   & $>$100 $\um$  & $\dzero$                   \\
$|M(K\pi\pi)-M(K\pi)-145.7~\mevcc|$ & $<$2 $\mevcc$ & $\dst\to\dzero\pi^{+}$ tagging\\
CMUP boundary             & Inside boundary   & Same as in $\bctojpsimunu$ \\
Match with XFT             & Is XFT        & Same as in $\bctojpsimunu$ \\
Isolation at CMU           & No tracks $<$40 cm & Same as in $\bctojpsimunu$ \\
$dE/dx$ hits               & $\ge$43 hits  & Same as in $\bctojpsimunu$ \\
\hline\hline
\end{tabular}
\end{center}
\label{piK_selection}
\end{table*}
% ++++++++++++++++++++++++++++++++++++++++++++++++++++++++++++++++++++++
% ++++++++++++++++++++++++++++++++++++++++++++++++++++++++++++++++++++++
\begin{figure*}[tbp]
\centerline{
\makebox{\includegraphics[width=0.33\hsize]{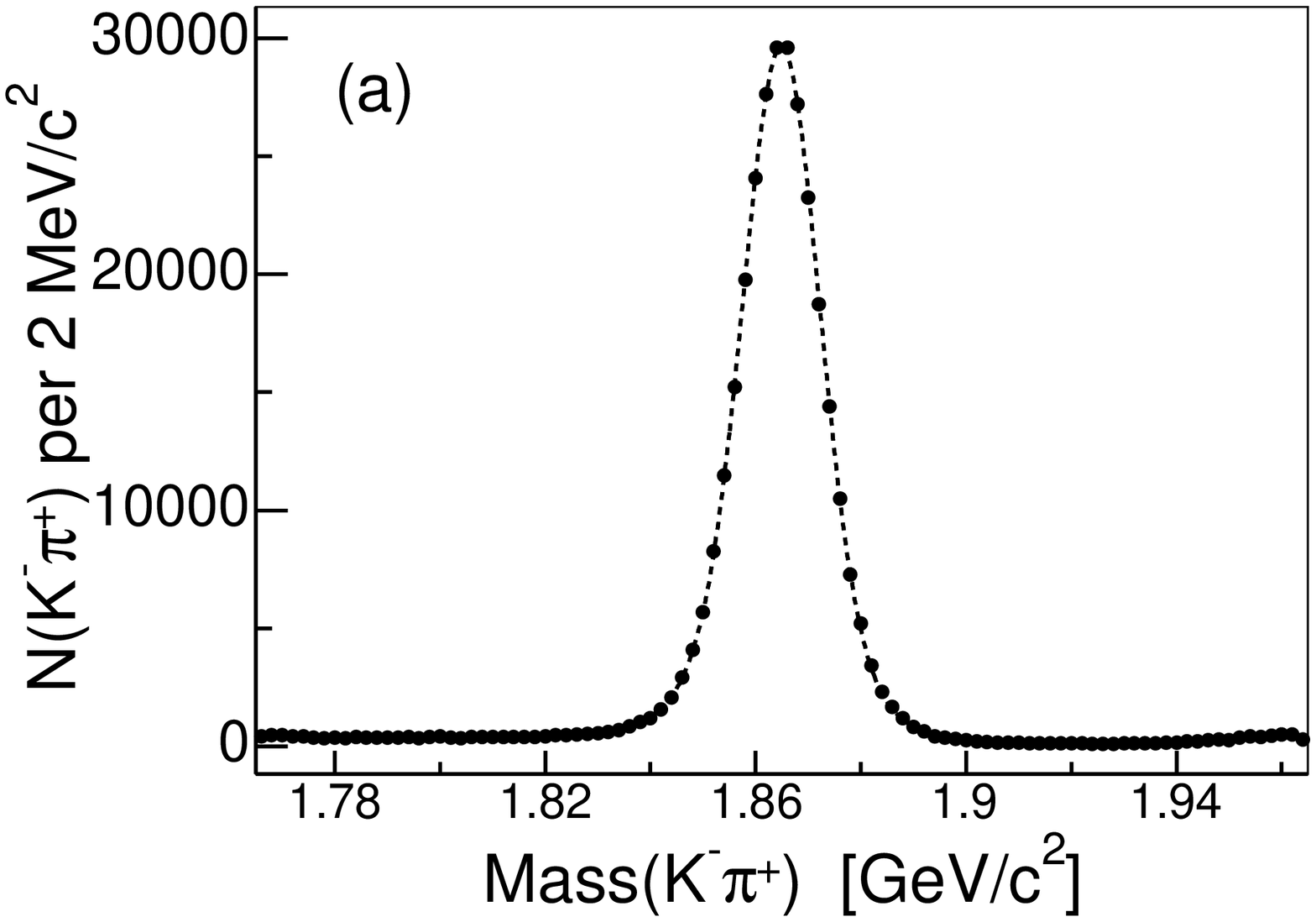}}
\makebox{\includegraphics[width=0.33\hsize]{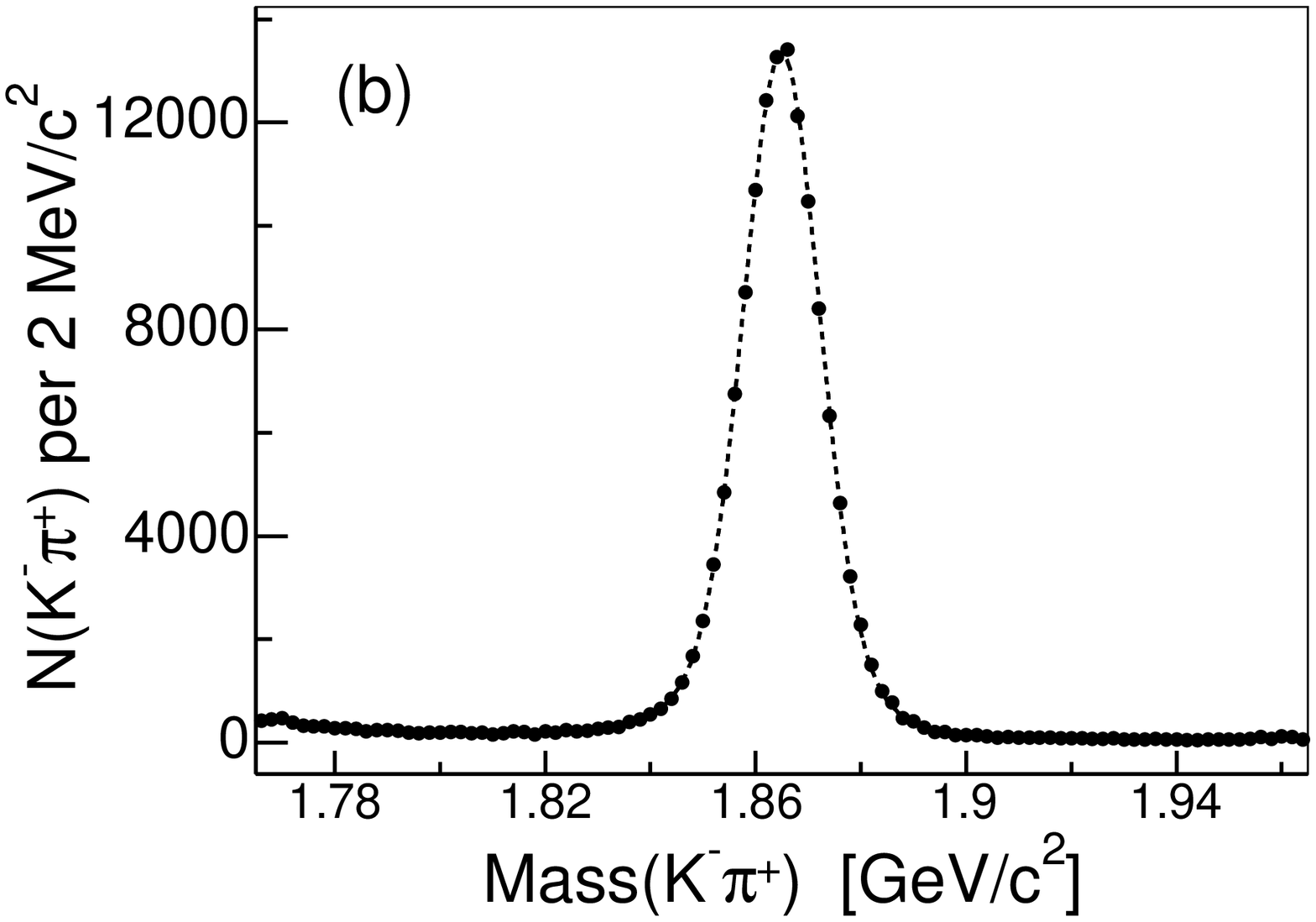}}
\makebox{\includegraphics[width=0.33\hsize]{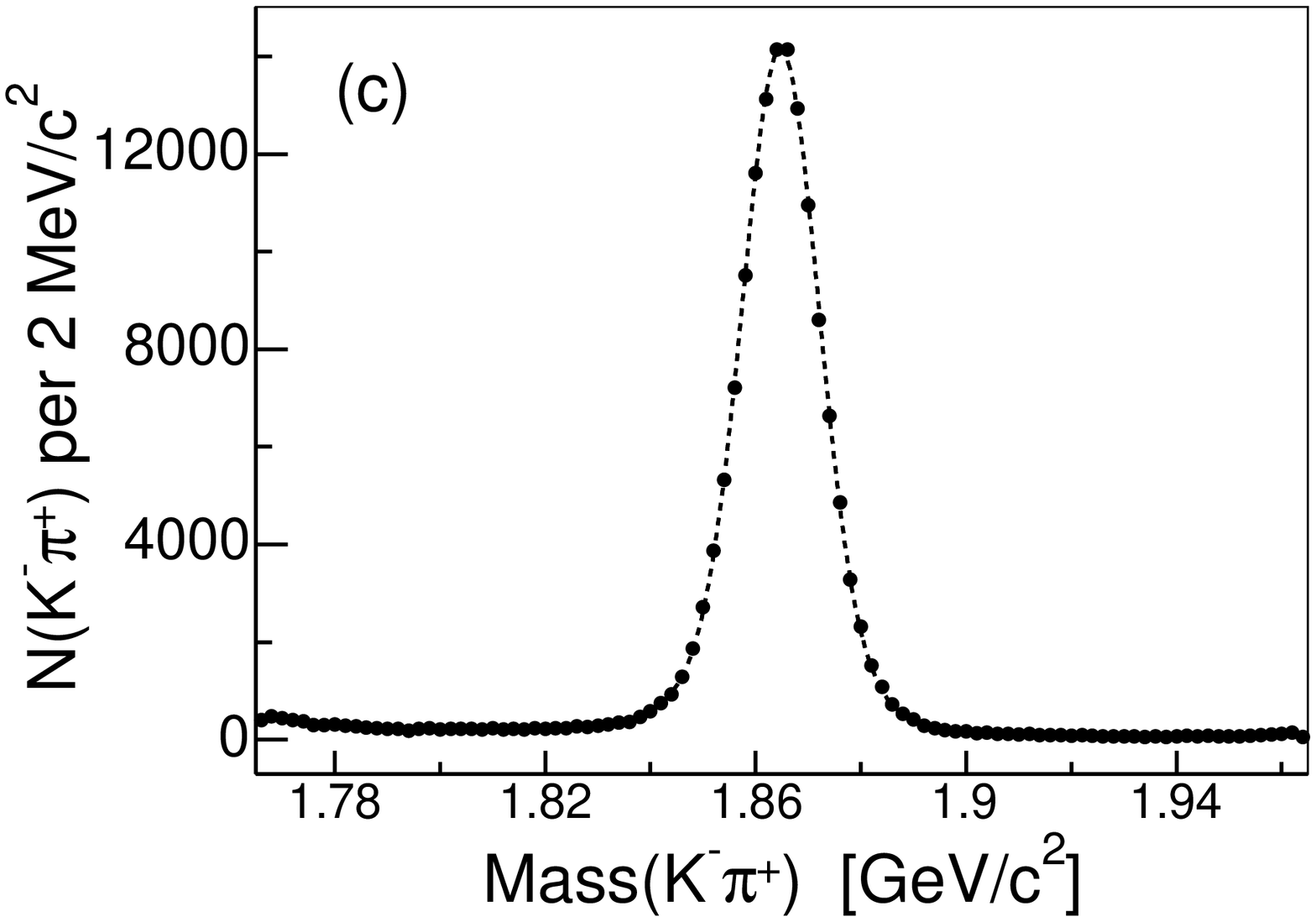}}}
\centerline{
\makebox{\includegraphics[width=0.33\hsize]{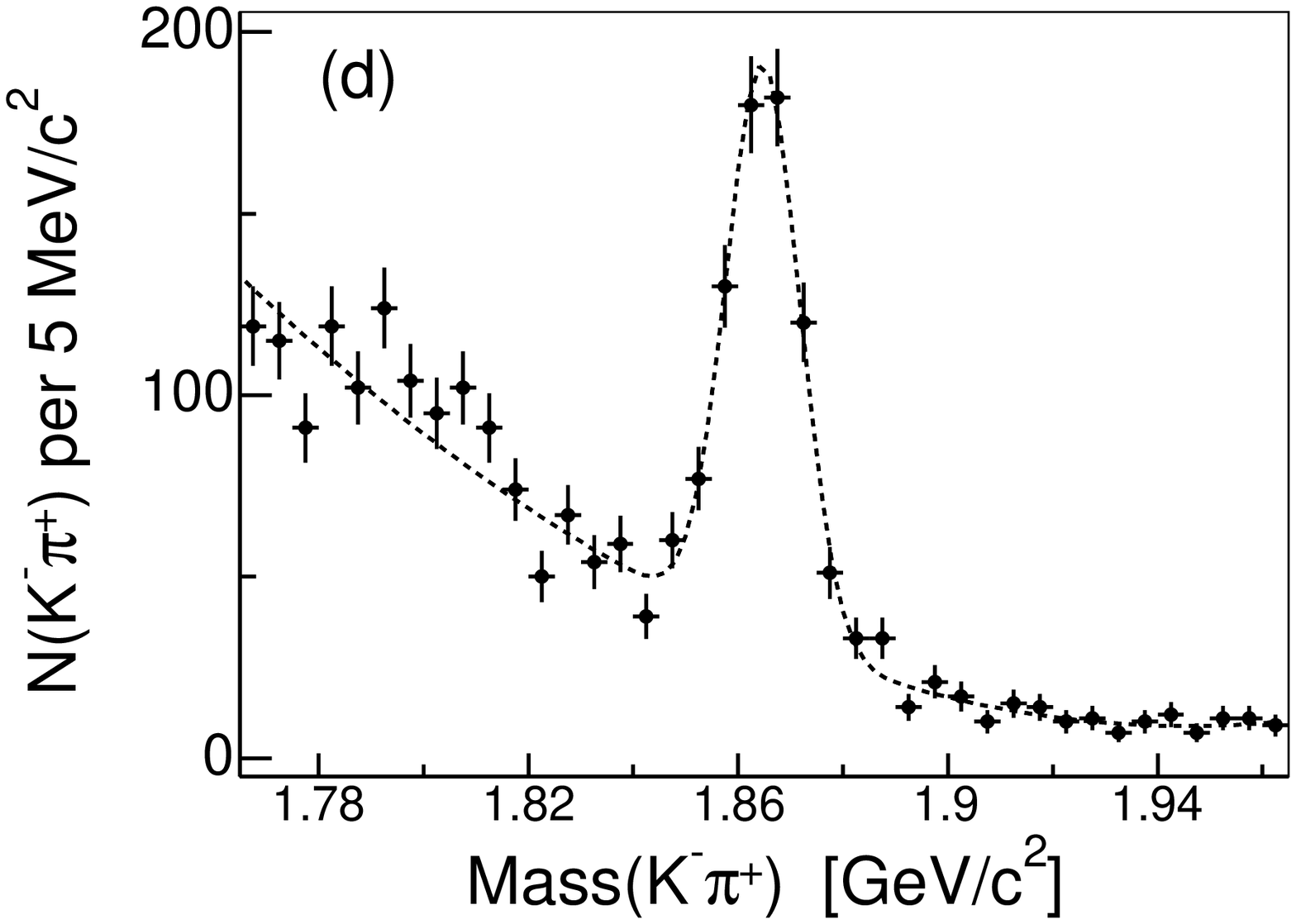}}
\makebox{\includegraphics[width=0.33\hsize]{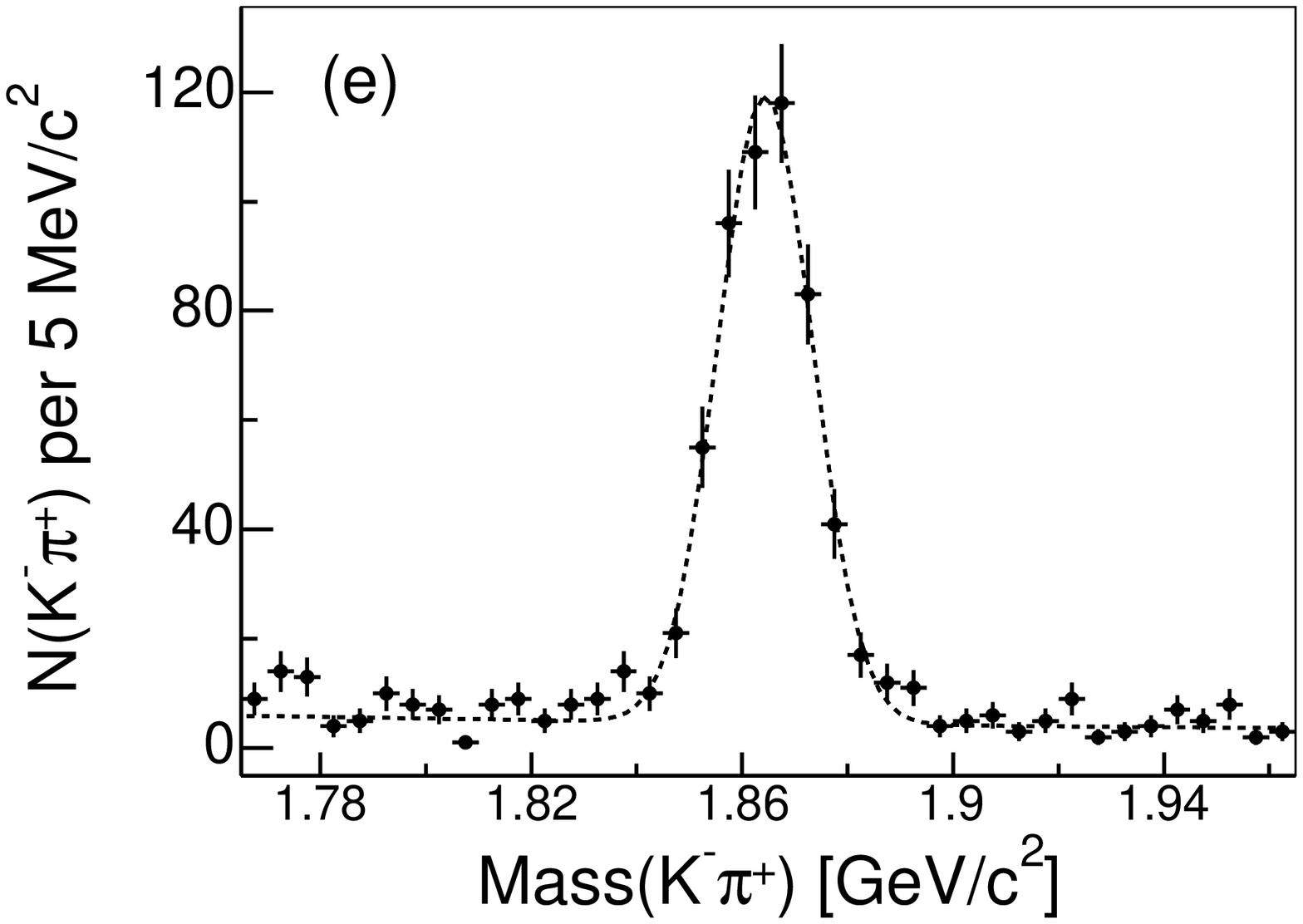}}
\makebox{\includegraphics[width=0.33\hsize]{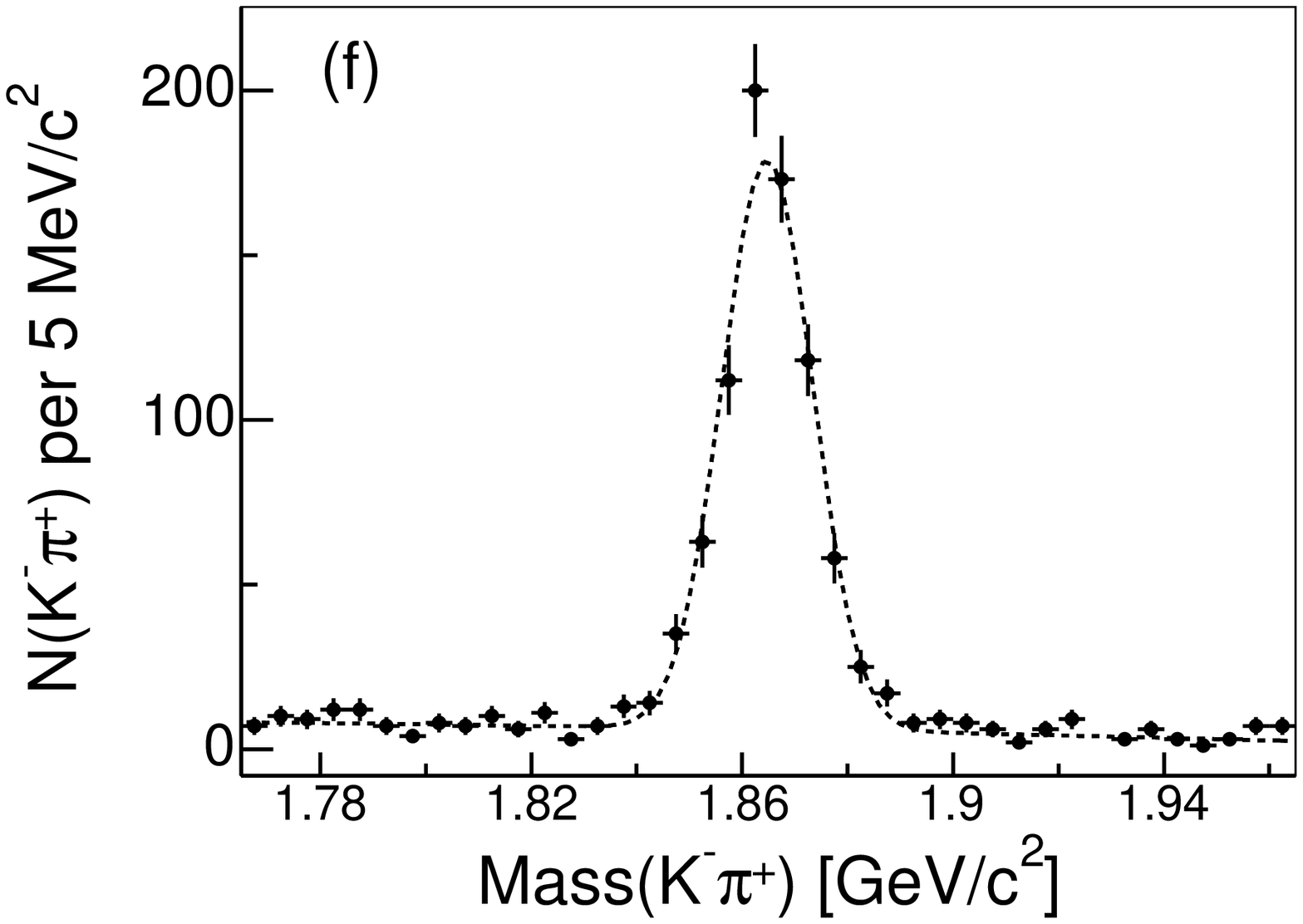}}}
\caption{Invariant-mass distributions of $K^{-}\pi^{+}$ pairs from $\dzero$ decays where the hadron of interest is not matched (a)--(c), and where it is matched
(d)--(f) with a muon that satisfies the third-muon selection requirements. Figures (a)--(f) are paired vertically with (a) and (d), (b) and (e), and (c) and (f) corresponding to $\pi^{\pm}$, $K^{-}$, and $K^{+}$, respectively. Examples are shown for the $\pt$ region 
3.0--3.3 $\gevc$. The fit function consists of a double Gaussian (a)--(c), or a single Gaussian (d)--(f), plus a second-order polynomial. %The enhancement in the low-mass sideband observed in the sample in which pions are misidentified as CMUP muons results from $\dzero\to K^{-}\mu^{+}\nu$ semileptonic decays.
}
\label{fig:hadron_d0Ms}
\end{figure*}
% ++++++++++++++++++++++++++++++++++++++++++++++++++++++++++++++++++++++
        
   The calculation of the probability for a proton to be misidentified as a muon 
is done using protons from reconstructed $\ltoppi$ 
%~\cite{Ref:MarkThesis}. 
decays.  In selecting the proton candidates we use the selection requirements for the third charged-particle from the $\bctojpsimunu$ candidates to be a muon.  To determine an appropriate $\Lambda$ mass range, we reconstruct the $p\pi^-$ final state for candidates with no muon match requirement.  Based on the mass resolution of the $p\pi^-$ final state fit to a single Gaussian, we search in a mass range that is six standard deviations wide and centered at the known $\Lambda$ mass.  We find no evidence for the proton $\PT$ 
process.  Therefore, using the uniform distribution of the invariant mass of $p\pi^{-}$ 
pairs in the $\Lambda$ mass region for a data sample with matched CMUP muons, 
we establish an upper limit at the 95\% confidence level that $\epsilon_{p}$ is less than 3.4$\times$10$^{-4}$. 
This upper limit applies to antiprotons as well.

     To measure the probability for charged pions and kaons to be misidentified as muons, we use samples of well-identified  pions and kaons obtained from  a $\dst$ sample collected using the SVT trigger as discussed in Sec.~\ref{sec:xft}.  We reconstruct the decay chain $\dst\to\dzero(K^-\pi^+)\pi^+$.  The pions and kaons are selected using the requirements listed in Table~\ref{piK_selection}. 
%requirements are the same as described in Ref.~\cite{Ref:MarkThesis}. 
We also require that in a $\dzero$ decay, the track being examined for a misidentified muon meets the same selection requirements as the third track in the \jpsihyptrack sample.  %corresponds to a track found with the XFT (see Sec.~\ref{sec:xft}), is isolated in the same way as the third track in the $\bctojpsimunu$ analysis, and has the same minimal number of $dE/dx$ hits as the third track in the $\jpsihyptrack$ sample. 
%In the previous analysis~\cite{Ref:MarkThesis} with 1 $\fb$ of 
%data, the misidentified-muon probabilities were presented in 4 $\pt$ bins. 
%In the current analysis the complete CDF dataset allows us to make smaller 
%$\pt$ bins, now we have 11 bins. 
Figure~\ref{fig:hadron_d0Ms} shows the 
invariant-mass distributions of $K^{-}\pi^{+}$ pairs from $\dzero$ decays 
where the hadron under test is not matched [Figs.~\ref{fig:hadron_d0Ms}(a)--(c)], and where it is matched 
[Figs.~\ref{fig:hadron_d0Ms}(d)--(f)] with a muon that satisfies the third-muon selection requirements. %Figures~\ref{fig:hadron_d0Ms}a-f are paired vertically with (a,d), (b,e), and (c,f) corresponding to $\pi^{\pm}$, $K^{-}$, and $K^{+}$, respectively. Examples are shown for the $\pt$ region 3.0-3.3 $\gevc$. 
The fit function consists of a double Gaussian [Figs.~\ref{fig:hadron_d0Ms}(a)--(c)], or a single Gaussian [Figs.~\ref{fig:hadron_d0Ms}(d)--(f)], plus a second-order polynomial. Simulation shows that the enhancement in the low-mass sideband of the sample in which pions are misidentified as CMUP muons results from $\dzero\to K^{-}\mu^{+}\nu$ semileptonic decays.
   
     We consider two options to fit the $\dzero$ peak shown in 
Figs.~\ref{fig:hadron_d0Ms}(d)--(f): first, using a double-Gaussian template 
derived from the fit of the data sample where no matched muons are present, and second, with a 
single Gaussian. We choose the single-Gaussian fit because the matched sample has poor statistics and the unmatched and matched samples are not expected to have the same widths because additional broadening may occur as 
a result of the $\DIF$ phenomenon discussed in Sec.~\ref{sec:fakeMu_fracOffPeak}.  We compare the results from the double-Gaussian template with the single-Gaussian fit in order to estimate the 
systematic uncertainty associated with the fit model.

\begin{figure}[tbp]
\centerline{
\makebox{\includegraphics[width=1.0\hsize]{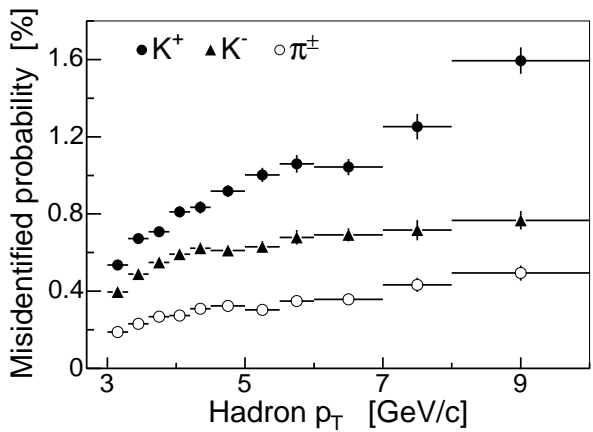}}
}
\caption{Muon-misidentification probabilities for $K^+$, $K^-$, and $\pi^{\pm}$  as functions of hadron $\pt$.}
\label{fig:KpiToMuRates}
\end{figure}

     The muon-misidentification probability $\epsilon_{\pi^{\pm},K^-,K^+}$ is given by Eq.~(\ref{fakeMu_fake_formula}),
% ++++++++++++++++++++++++++++++++++++++++++++++++++++++++++++++++++++++
\begin{equation}
\epsilon_{h} = \frac{N^{\textrm{with}\,\mu}_{h}}{N^{\textrm{no}\,\mu}_{h} + N^{\textrm{with}\,\mu}_{h}}\,,
\label{fakeMu_fake_formula}
\end{equation}
% ++++++++++++++++++++++++++++++++++++++++++++++++++++++++++++++++++++++
where $h$ is a $\pi^{\pm}$, $K^-$, or $K^+$; $N^{\textrm{no}\,\mu}_{h}$ represents the number of candidates where $h$ is not matched with a CMUP muon; and $N^{\textrm{with}\,\mu}_{h}$ is the number of candidates where $h$ is matched with a CMUP muon. The $N^{\textrm{no}\,\mu}_{h}$ values are determined by the integrals under the fitted double Gaussian within a 100~$\mevcc$ range, and the $N^{\textrm{with}\,\mu}_{h}$ are determined by the corresponding single-Gaussian integrals also within a 100~$\mevcc$ range.  The muon-misidentification probabilities as functions of hadron $\pt$ are %summarized in Table~\ref{tab:piK_toMuFakeRates} and 
shown in Fig.~\ref{fig:KpiToMuRates}. The uncertainties shown are statistical only. 
% ++++++++++++++++++++++++++++++++++++++++++++++++++++++++++++++++++++++

The muon-misidentification probabilities for $K^{+}$ hadrons are significantly 
higher than for $K^{-}$. %The $\DIF$ properties for the same particle with 
%opposite charge are the same. 
The observed difference results from the different interaction cross sections for $K^+$ and $K^-$ hadrons with matter, which leads to 
different $\PT$ probabilities.  These effects are discussed further in Sec.~\ref{sec:fakeMu_fracOffPeak}.  We find no significant differences in the misidentification probabilities of $\pi^+$ and $\pi^-$ mesons.
%One might ask if there are any differences for the positive and negative pions in the muon misidentification results? In this study we find that the misidentification probabilities for $\pi^{+}$ and for $\pi^{-}$ mesons are the same. 

%%%%%%%%%%%%%%%%%%%%%%%%%%%%%%%%%%%%%%%%%%%%%%%%%%%%%%%%%%%%
\subsubsection{Corrections to $\pi^{\pm}$, $K^{-}$, and $K^{+}$ probabilities to be misidentified as a muon}%Fraction of muon tagged events outside of the $\dzero$ mass peak}
\label{sec:fakeMu_fracOffPeak}
%%%%%%%%%%%%%%%%%%%%%%%%%%%%%%%%%%%%%%%%%%%%%%%%%%%%%%%%%%%%
     In Eq.~(\ref{fakeMu_W_formula}) the terms $1+F^{\textrm{out}}_{\pi}$ and $1+\alpha F^{\textrm{out}}_{K}$ are corrections to the probabilities for pions and kaons, respectively, to be misidentified as muons.  They arise because of mass resolution effects associated with the decay in flight of pions and kaons where the decay muon is ultimately matched with a third track and results in the event contributing to the misidentified-muon background.  The misidentified-muon probabilities determined above are derived under the assumption that the pion and kaon tracks, even after a possible kink resulting from a decay in flight, yield a two-body invariant mass that remains within 50~MeV/$c^2$ of the known $\dzero$ mass.  %This is not always the case.  
However, the mass resolution can be spoiled because of a kink, while the pion or kaon track is still matched to a CMUP muon.  Because the signal region for $\bctojpsimunu$ decays has a width of 2~$\gevcc$, background events from decays in flight may contribute to the signal region but remain excluded from the measurement of the probability that a pion or kaon is misidentified as a muon using the decay $\dzero\to K^{-}\pi^{+}$.  We correct for this effect by determining the fraction $F^{\textrm{out}}_h$ ($h$ is a pion or kaon) of misidentified events that fall outside of the $\dzero$ mass peak for a given particle type through a MC simulation.
    
     The term $1+\alpha F^{\textrm{out}}_{K}$ involves an additional correction factor $\alpha$ that is set to 1 for $K^-$ mesons and to $\alpha=\epsilon_{K^-}/\epsilon_{K^+}$, which is less than 1, for $K^+$ mesons.  The rationale is as follows: Fig.~\ref{fig:KpiToMuRates} shows that the muon-misidentification probabilities for $K^{+}$ mesons are significantly higher than for $K^{-}$ mesons. This difference arises because $K^+$ mesons have an additional $\PT$ component, which is not present for $K^-$ mesons because the strong-interaction cross section in matter for $K^-$ mesons is larger than that for $K^+$ mesons. The $\PT$ component does not produce any kink in the track and for this component of $\epsilon_{K^+}$ the outside-of-peak correction should not be applied. The outside-of-peak correction is needed only for the $\DIF$ fraction of $\epsilon_{K^+}$, which is modeled as the ratio $\alpha=\epsilon_{K^-}/\epsilon_{K^+}$. 

\begin{figure*}[Htbp]
\centerline{
\makebox{\includegraphics[width=0.33\hsize]{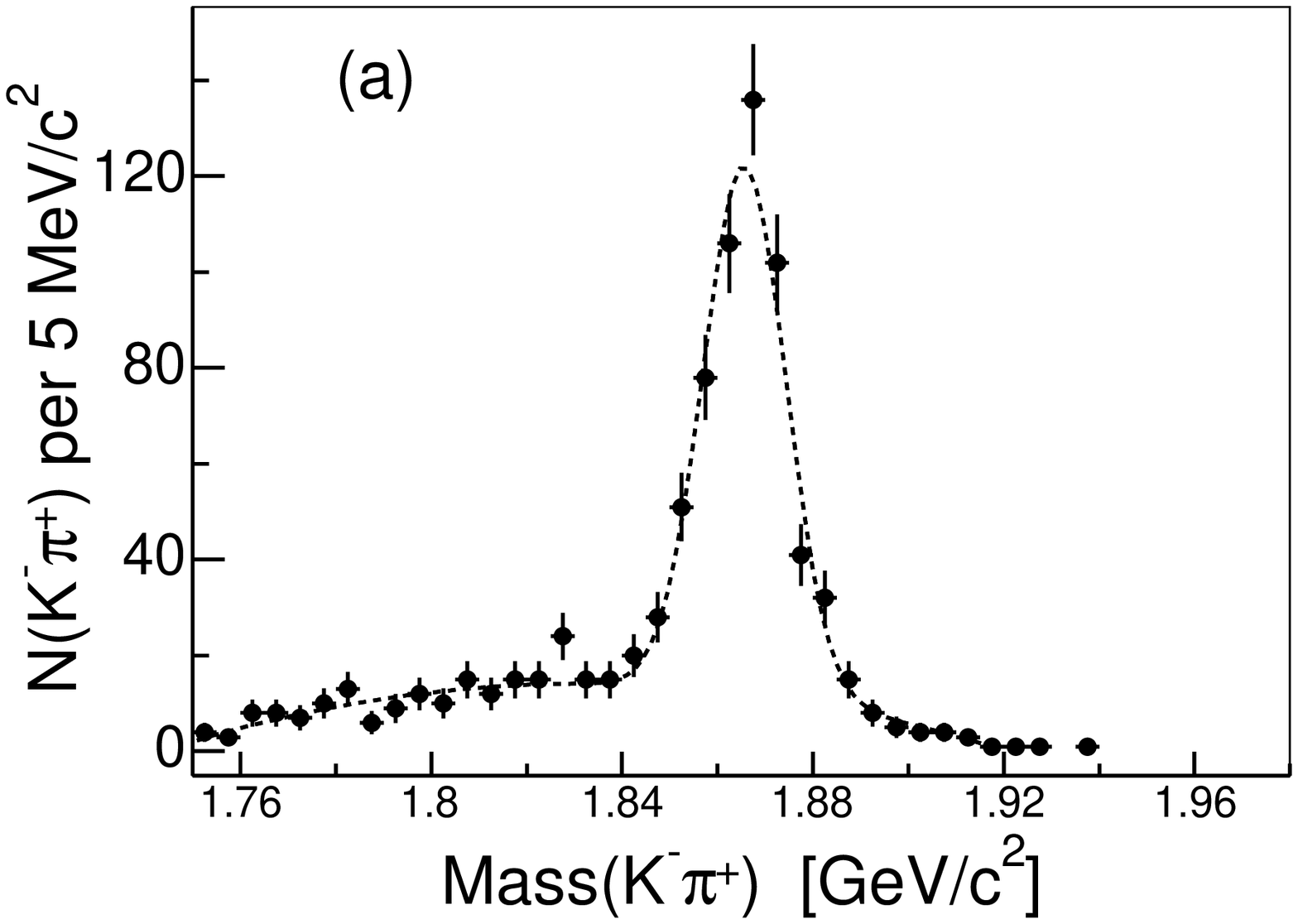}}
\makebox{\includegraphics[width=0.33\hsize]{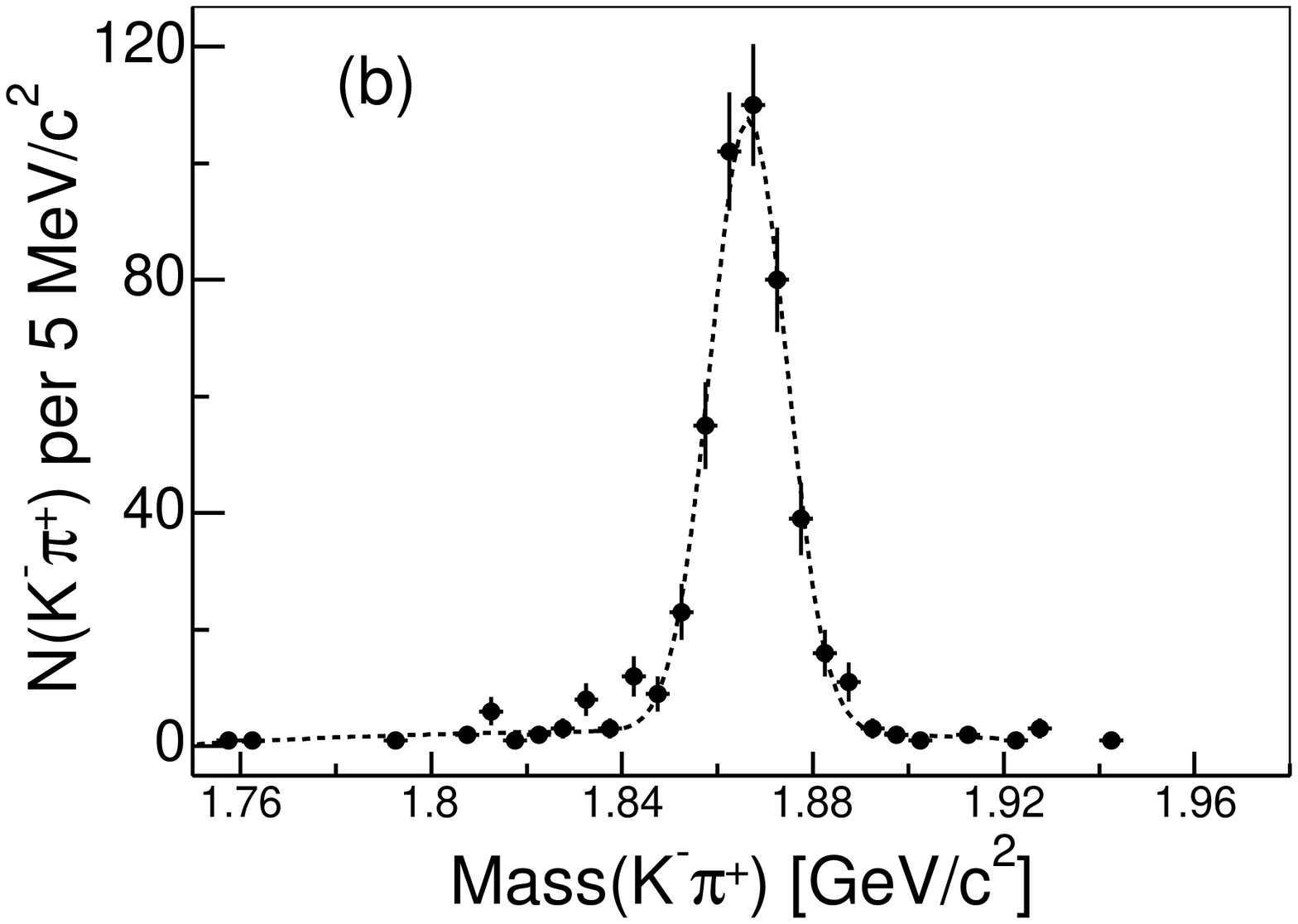}}
\makebox{\includegraphics[width=0.33\hsize]{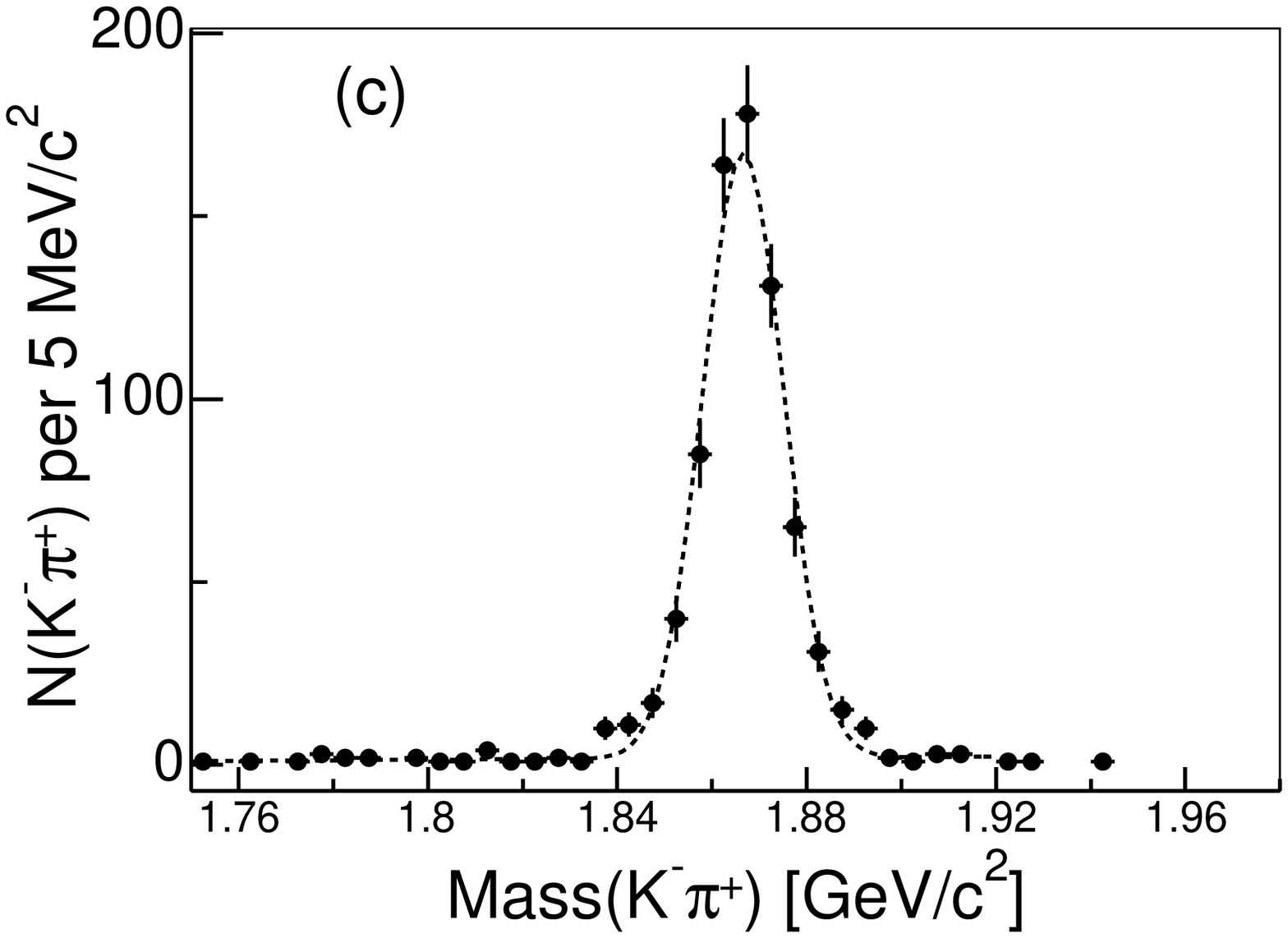}}}
\caption{Invariant-mass distributions of 
$K^{-}\pi^{+}$ pairs from simulated $\dzero\to K^{-}\pi^{+}$ decays for (a) $\pi^{\pm}$, 
(b) $K^{-}$, and (c) $K^{+}$ mesons passing the selection requirements 
for a CMUP muon. Example distributions 
are given for the $\pt$ range 3.0--3.3 $\gevc$. The fit function 
consists of a single Gaussian plus a second-order polynomial.}
\label{fig:MCpionKaonToMu}
\end{figure*}

We determine the fractions $F^{\textrm{out}}_h$ as functions of pion and 
kaon $\pt$ by using simulated $\dst\to\dzero(\to K^-\pi^+)\pi^+$ decays 
selected as the corresponding control sample of data.
Figure~\ref{fig:MCpionKaonToMu} shows simulated invariant-mass 
distributions of $K^{-}\pi^{+}$ pairs from $\dzero$ decays for $\pi^{\pm}$, 
$K^{-}$, and $K^{+}$ mesons passing the selection requirements for a CMUP muon 
(see Sec.~\ref{sec:muon_det}). Example distributions are given for the $\pt$ 
range 3.0--3.3~$\gevc$.
% ++++++++++++++++++++++++++++++++++++++++++++++++++++++++++++++++++++++
% ++++++++++++++++++++++++++++++++++++++++++++++++++++++++++++++++++++++

     The simulated data shown in Fig.~\ref{fig:MCpionKaonToMu} are fit with 
a single Gaussian plus a second-order polynomial. The fraction 
of the muon misidentifications outside of the $\dzero$ mass peak for each $\pt$ interval is calculated as follows:
% ++++++++++++++++++++++++++++++++++++++++++++++++++++++++++++++++++++++
\begin{equation}
F^{\textrm{out}}_h = \frac{N_{h} - N^{\textrm{peak}}_{h}}{N^{\textrm{peak}}_{h}}\,,
\label{offPeakFr_form}
\end{equation}
% ++++++++++++++++++++++++++++++++++++++++++++++++++++++++++++++++++++++
where $N_{h}$ represents, in each $\pt$ bin of the relevant final-state hadron $h$, the number of events that pass the requirements for $h$ to match a CMUP muon %matched to hadron $h$ in the invariant-mass distribution of $K^{-}\pi^{+}$ pairs, and  
%from 1.745 $\gevcc$ through 1.985 $\gevcc$, or $\pm$120 $\mevcc$ with respect 
%to the $\dzero$ mass value;
and $N^{\textrm{peak}}_{h}$ represents the integral under the single-Gaussian component of 
the fit to the distribution within 50~$\mevcc$ of the peak of the 
Gaussian. %The width of the distributions is about 9 $\mevcc$ for the lowest 
% and 13 $\mevcc$ for the
%highest $\pt$ bin. For the $K^{-}$ and $K^{+}$ mesons the $\sigma$ values are 
%7-9 $\mevcc$. This calculation is based entirely on Monte Carlo simulation.  
In order to estimate the systematic uncertainty, we fit the above distributions
with width values derived from the experimental data analyzed in 
Sec.~\ref{sec:fakeMu_probability}.
  
\begin{figure}[tbp]
\centerline{
\makebox{\includegraphics[width=1.0\hsize]{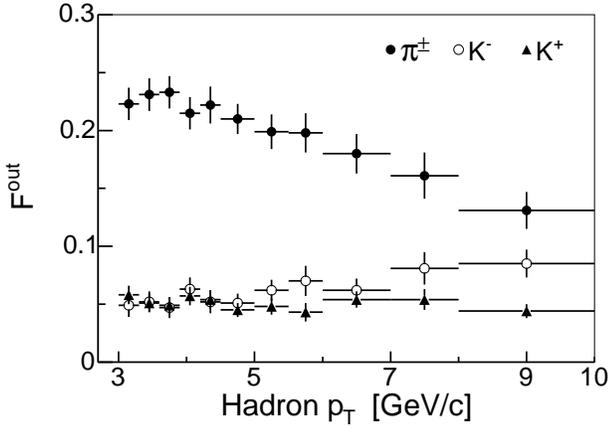}}}
\caption{Fraction of events with a CMUP muon whose $K \pi$ invariant mass falls
outside of the $\dzero$ mass peak due to decay in flight:  
$\bullet$ for the $F^{\textrm{out}}_{\pi^{\pm}}$, $\circ$ for the $F^{\textrm{out}}_{K^{-}}$, 
and $\blacktriangle$ for the $F^{\textrm{out}}_{K^{+}}$, respectively.} 
%from the $\dst\to\dzero(K^-\pi^+)\pi^+$ decay chain as a function 
%of hadron $\pt$.}
\label{fig:muOutOffPeak}
\end{figure}

    Figure~\ref{fig:muOutOffPeak} shows the fraction of events with a CMUP muon whose $K\pi$ invariant mass falls outside of the $\dzero$ mass peak due to decay in flight for $\pi^{\pm}$, $K^{-}$, and 
$K^{+}$ mesons from the $\dst\to\dzero(K^-\pi^+)\pi^+$ decay chain 
as a function of hadron $\pt$. 
% ++++++++++++++++++++++++++++++++++++++++++++++++++++++++++++++++++++++
% ++++++++++++++++++++++++++++++++++++++++++++++++++++++++++++++++++++++

%%%%%%%%%%%%%%%%%%%%%%%%%%%%%%%%%%%%%%%%%%%%%%%%%%%%%%%%%%%%
\subsubsection{Hadron fractions within the \jpsihyptrack sample}
\label{sec:fakeMu_particleFrac}
%%%%%%%%%%%%%%%%%%%%%%%%%%%%%%%%%%%%%%%%%%%%%%%%%%%%%%%%%%%%
     The proton, pion, and kaon fractions in the \jpsihyptrack sample comprise the other essential component required to complete the data-driven calculation of the misidentified-muon background.

   The pion fraction $F_{\pi}$ of the tracks in the \jpsihyptrack sample as a function of particle momentum is determined using $\dedx$ measured in the COT.  %~\cite{Ref:MarkThesis}. 
The remaining fraction $F_{K+p}$ of tracks in the \jpsihyptrack sample is a combination of kaons and protons because the kaon and proton $\dedx$ distributions in the COT are indistinguishable at 
momentum greater than $3~\gevc$. %Consequently, the kaon and proton fractions are combined as $F_{K+p}$.  
The proton fraction $F_{p}$ is measured within the 
2.0--3.3~$\gevc$ momentum range using a simultaneous fit of the $\dedx$ and 
time-of-flight data. %However, 
Also available is the predicted $F_{p}$ from MC simulation for momenta greater than 3.0~$\gevc$.  Using the two fractions $F_p$ in the 3.0--3.3 $\gevc$ momentum range, one from the 
experimental data and a second from simulation, $F_p$ from simulation is scaled to agree with the experimental data in the momentum range 3.0--3.3 $\gevc$.  Thus, $F_p$ is taken from the scaled simulation for particle momentum greater than 3.0~$\gevc$.  Then the kaon particle fraction $F_K$ in the \jpsihyptrack sample for particle momentum greater than 3.0~$\gevc$ is given by $1-F_{\pi}-F_p$.%allows use of a single scale factor for the simulated events so they match the experimental data at low momentum. In the higher momentum range ($p>$ 3.3 $\gevc$), we follow the corrected predictions for $F_{p}$ from the MC simulation. 
   
   To estimate the $F_{\pi}$ and $F_{K+p}$ fractions, we use the $\dedx$ information contained in the separation-significance quantity $\mathcal{S}$,
% ++++++++++++++++++++++++++++++++++++++++++++++++++++++++++++++++++++++
\begin{equation}
{\mathcal{S}}=\frac{\dedxnum}{\sigma_{\dedx}}\,,
\label{dedx_equation}
\end{equation}
% ++++++++++++++++++++++++++++++++++++++++++++++++++++++++++++++++++++++
where $\dedx_{\textrm{meas}}$ is the measured energy loss 
%on a track-by-track basis 
for a given third track from the \jpsihyptrack sample, 
$\dedx_{\pi}$ is the predicted energy loss for the $\pi$ hypothesis, 
and $\sigma_{\dedx}$ is the estimated uncertainty of the measurement. 
%The distribution of ${\mathcal{S}}$ for pion tracks is expected to be 
% Gaussian. with the mean at zero and $\sigma$=1. 
In this analysis the 
third track in the \jpsihyptrack sample has contributions not 
only from pions, but also kaons and protons. The predicted mean value is about 
$-1.5$ for kaons and protons and about zero for pions.  %The ${\mathcal{S}}$ 
%distributions for pions and kaons combined with protons are both asymmetric.  
%Fitting the above quantity with two Gaussians, where the mean for one is zero 
%and the mean for the other is about -1.5, allows an estimate of the $\pi$ and 
%$K+p$ fractions. The sum of two single Gaussians does not describe well the 
%pion or kaon positive tails. Therefore, we use the sum of two gamma 
%distributions to estimate $F_{\pi}$ and $F_{K+p}$ 
%because the gamma distribution describes better the experimental data when 
%there is some asymmetry present. The difference 
%between the results derived with Gaussian fits compared with those from gamma 
%distributions is used to estimate the systematic uncertainties.
%, but the central values are based on the fits using two gamma distributions.
Because the ${\mathcal{S}}$ distribution for each component is asymmetric, we 
model it empirically with the sum of two gamma distributions and use the 
results from a simpler Gaussian fit to evaluate the systematic uncertainty 
associated with the fit model. The probability density written in terms of 
${\mathcal{S}}$ is defined by 
\begin{equation}
G(\gamma,\beta,\mu;\mathcal{S})=\frac{(\frac{\mathcal{S}-\mu}{\beta})^{\gamma-1}\exp(-(\frac{\mathcal{S}-\mu}{\beta}))}
{\beta\Gamma(\gamma)}
\end{equation}
for $\mathcal{S}>\mu$ and zero otherwise  %The free parameters in the expression for $G$ are $\gamma$, $\beta$, and $\mu$; and 
where $\Gamma$ is the Euler gamma 
function. The distribution has a mean $\gamma\beta$ + $\mu$  and variance 
$\gamma\beta^2$.  In the limit of large $\gamma$ this asymmetric 
distribution approaches a Gaussian distribution.  The parameters $\gamma$ 
and $\beta$ are positive real numbers that control the shape, mean, and 
variance of the distribution, and $\mu$ is the location parameter.
     
\begin{figure}[tbp]
\centerline{
\makebox{\includegraphics[width=1.0\hsize]{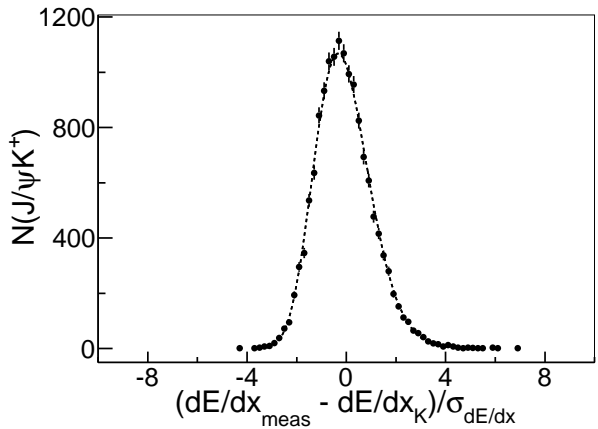}}}
\caption{Distribution of ${\mathcal{S}}$ from Eq.~(\ref{dedx_equation}) 
for the $K$ hypothesis using the $K^+$ tracks from the $\bptojpsik$ decays 
[see Fig.~\ref{fig:threeTrk_mass}(b)].
%Comparison of the $\dedx$ response with the predicted energy loss for 
%the $K$ hypothesis using the $K^+$ tracks from the $\bptojpsik$ decays (see 
%Fig.~\ref{fig:threeTrk_mass}b). 
The distribution is fit with a gamma 
function, where all three parameters ($\gamma$, $\beta$, and $\mu$) are 
allowed to float.}
\label{fig:gammaPdf_koans}
\end{figure}

     In order to find parameters to use for the kaon gamma distribution 
and gain guidance for pions, we use kaons from the 
$\bptojpsik$ decays [see Fig.~\ref{fig:threeTrk_mass}(b)]. 
The kaon tracks are identified by requiring the $\jpsiplusk$ mass to be 
within 40~$\mevcc$ (approximately $3\sigma$) of the known $\bp$ mass~\cite{Ref:PDG3}. 
Figure~\ref{fig:gammaPdf_koans} shows the distribution of a quantity similar 
to ${\mathcal{S}}$ from Eq.~(\ref{dedx_equation}), but where the kaon hypothesis for the predicted energy loss $\dedx_K$ is used. 
% ++++++++++++++++++++++++++++++++++++++++++++++++++++++++++++++++++++++
% ++++++++++++++++++++++++++++++++++++++++++++++++++++++++++++++++++++++
Figure~\ref{fig:gammaPdf_koans} illustrates that the $\dedx$ distribution 
for kaons is not Gaussian. 
%The asymmetrical deviation from a Gaussian distribution arises because 
%of the stochastic nature of the energy losses. 
A least-squares fit returns the following values: $\gamma$ = 23.5$\pm$2.5, 
$\beta$ = 0.230$\pm$0.013, and $\mu$ = -5.49$\pm$0.28. Using these  
parameters we calculate the width $\sigma$ = 
$\beta\sqrt{\gamma} = 1.11\pm0.13$ and mean 
$\gamma\beta + \mu = -0.09\pm0.01$.  
%The $\gamma$ parameter with a value of about 23 
%is found to be consistent across the momentum spectrum of the kaons in 
%the selected $\bptojpsik$ decays. Thus, for the fit for the $K+p$ fraction in 
%the \jpsihyptrack sample, we use the same $\gamma$ value. 
A value of 23 for the $\gamma$ parameter models accurately the kaons across 
their full momentum spectrum and is used for the fit of the $K+p$ fraction in 
the \jpsihyptrack sample.
For the width $\sigma$, we choose a higher 
value of 1.15 to take into account the contributions from protons. 
As a fit function for the pions we also use the gamma distribution.  
%The $\gamma$ value for the pions was set to 6.8 to account for the asymmetric positive pion tail, but the resulting particle fraction are not sensitive to this quantity. The $\sigma$ value for pions was set to 1.11 and the mean to zero. 
Finally, the $\dedx$ data are fit with the following formula:
% ++++++++++++++++++++++++++++++++++++++++++++++++++++++++++++++++++++++
\begin{eqnarray}
N_{\textrm{ev}}(\mathcal{S})&=&N_{\textrm{fit}}[F_{\pi} G(\gamma_{\pi},\beta_{\pi},\mu_{\pi};\mathcal{S})+(1-F_{\pi})\nonumber \\ 
& &\times G(\gamma_{K+p},\beta_{K+p},\mu_{K+p};\mathcal{S})]\,,
\label{fit_function}
\end{eqnarray}
% ++++++++++++++++++++++++++++++++++++++++++++++++++++++++++++++++++++++
where $N_{\textrm{ev}}(\mathcal{S})$ is the prediction as a function of the
quantity in Eq.~(\ref{dedx_equation}), $N_{\textrm{fit}}$ is the number of events, 
$F_{\pi}$ is the pion fraction, 
%$F_{K+p} = 1-F_{\pi}$, 
and $G$ is the probability density function of the 
gamma distribution. There are only two free parameters in this least-squares 
fit: $N_{\textrm{fit}}$ and $F_{\pi}$.  
%The rest of the quantities are fixed: $\gamma_{\pi} = 6.8$, $\beta_{\pi} = \frac{1.11}{\sqrt{\gamma_{\pi}}}$, and $\mu_{\pi} = -\gamma_{\pi}\beta_{\pi}$ for pions; $\gamma_{K+p} = 23$, $\beta_{K+p} = \frac{1.15}{\sqrt{\gamma_{K+p}}}$ for kaons. 
The parameter $\mu_{K+p}$ is adjusted as a function of pion and kaon momentum because the $K+p$ $\dedx$ distribution changes slowly with respect to that of the pion as the particle momentum changes.

\begin{figure*}[tbp]
\centerline{
\makebox{\includegraphics[width=0.33\hsize]{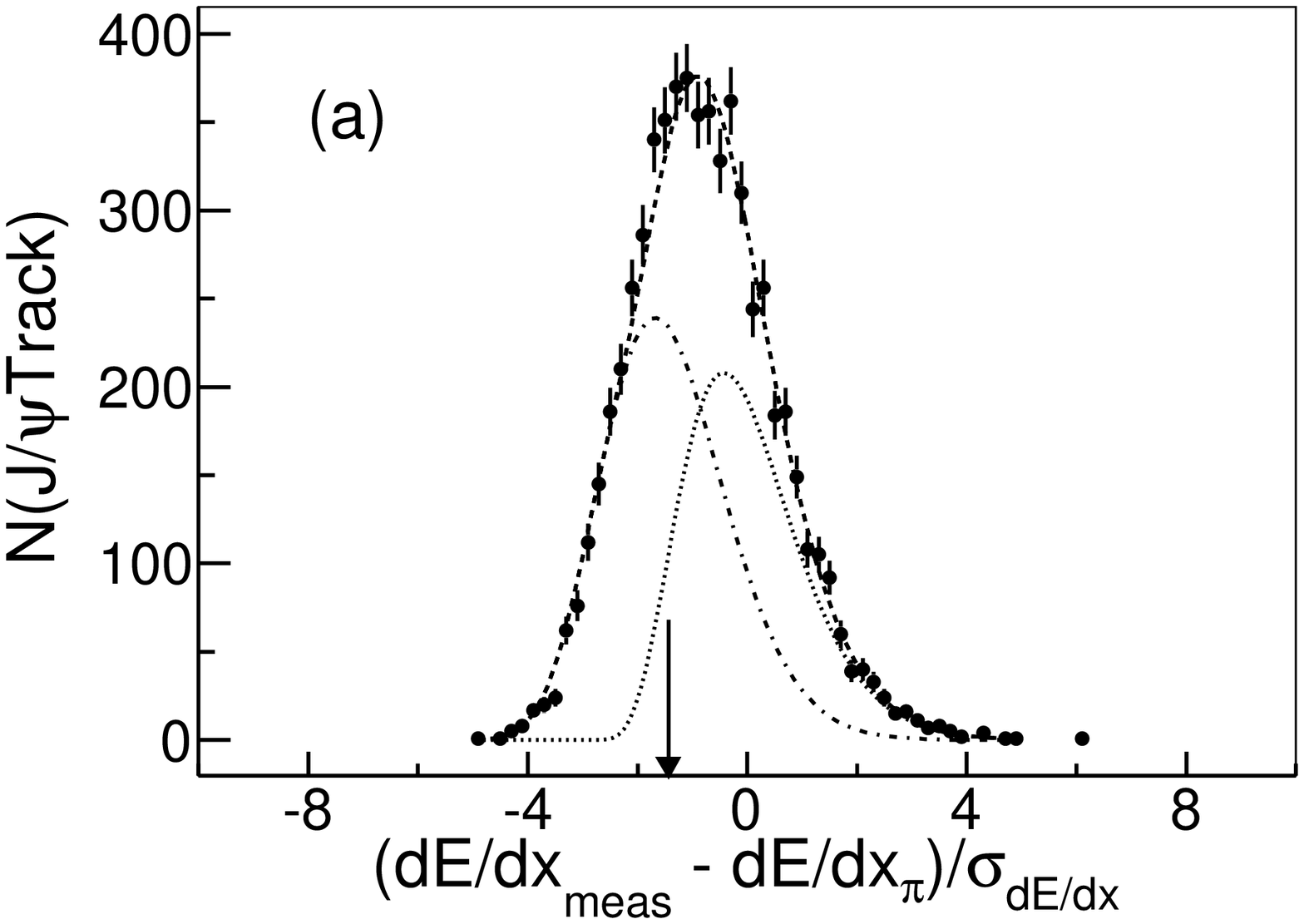}}
%\centerline{
\makebox{\includegraphics[width=0.33\hsize]{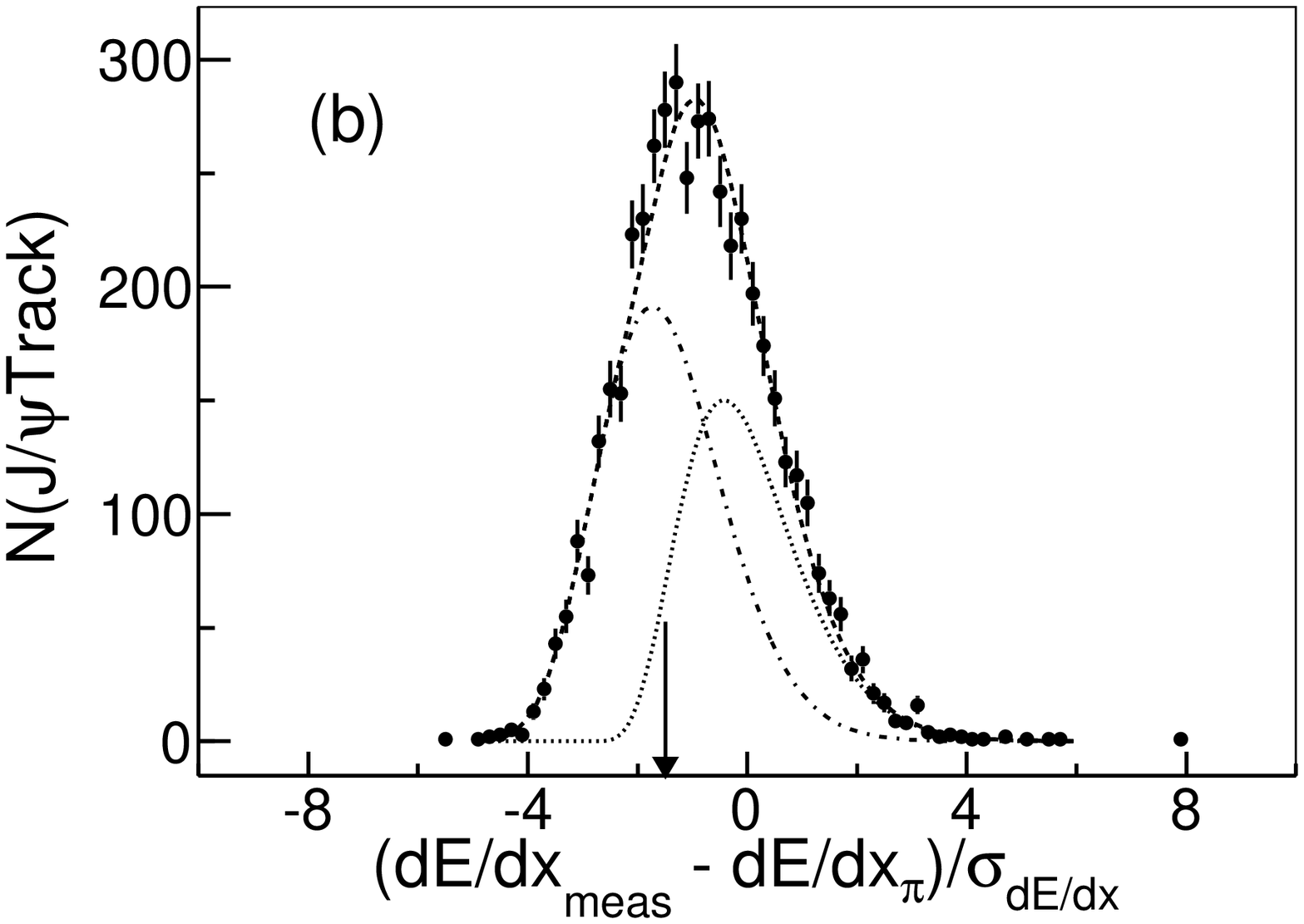}}
%\centerline{
\makebox{\includegraphics[width=0.33\hsize]{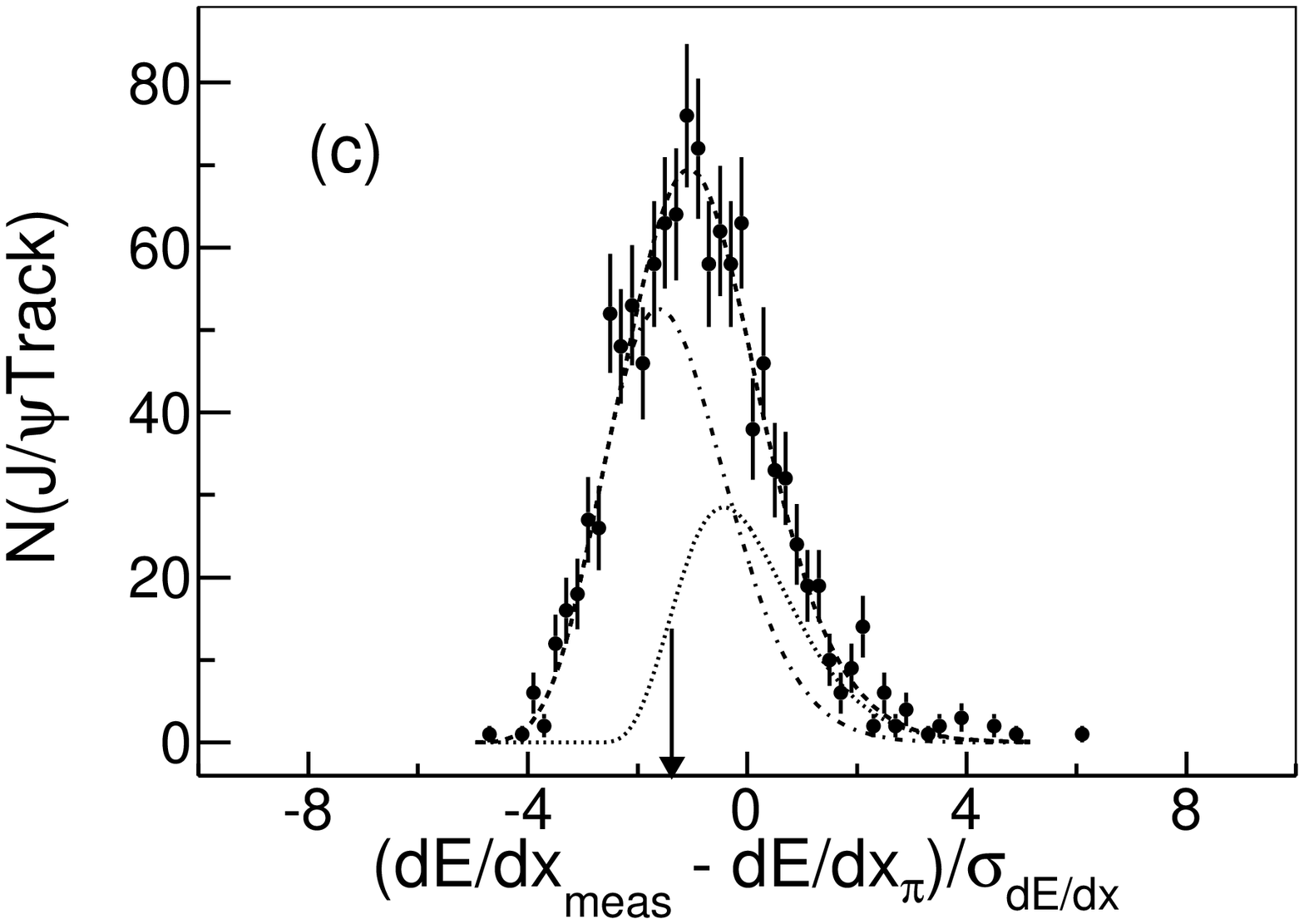}}}
\caption{Distributions of ${\mathcal{S}}$ %from Eq.~(\ref{dedx_equation}) 
for the positively charged third tracks in three momentum ranges: (a) 3.0--3.3 
$\gevc$, (b) 4.2--4.5 $\gevc$, and (c) 6.0--7.0 $\gevc$. The fit
function consists of the sum of two gamma distributions, one for pions (dotted curve), and 
a second one for $K+p$ (dash-dot curve). The total fit function is
shown as a dashed curve. Details of the fit are discussed in the text.%There are only two free parameters in the fit: $N_{fit}$ 
%and $F_{\pi}$.  The rest of the quantities are fixed: $\gamma_{\pi}$ = 6.8, 
%$\beta_{\pi}$ = $\frac{1.11}{\sqrt{\gamma_{\pi}}}$, and $\mu_{\pi}$ = 
%-$\gamma_{\pi}\beta_{\pi}$ for pions; $\gamma_{K+p}$ = 23, $\beta_{K+p}$ 
%= $\frac{1.15}{\sqrt{\gamma_{K+p}}}$ for kaons. The parameter $\mu_{K+p}$ has to be adjusted as a function of the $\pi/K$ momentum because the $K+p$ $\dedx$ distribution changes slowly with respect to the pion as the particle momentum changes.
}
\label{fig:dedxPlot_Gpdf_Qplus}
\end{figure*}

  Figure~\ref{fig:dedxPlot_Gpdf_Qplus} shows the distributions 
of ${\mathcal{S}}$ 
%from Eq.~(\ref{dedx_equation}) 
for the positively charged 
third tracks in three momentum ranges fit with a sum of two gamma 
distributions as described in Eq.~(\ref{fit_function}). 
% ++++++++++++++++++++++++++++++++++++++++++++++++++++++++++++++++++++++
% ++++++++++++++++++++++++++++++++++++++++++++++++++++++++++++++++++++++
%Table~\ref{tab:dedx_piKpFrac} summarizes the fractions of $F_{\pi}$ and 
%$F_{K+p}$ particles in the \jpsihyptrack sample determined using the 
%$\dedx$ method, where the fitted function consists of two gamma 
%distributions as shown in Fig.~\ref{fig:dedxPlot_Gpdf_Qplus}.
% ++++++++++++++++++++++++++++++++++++++++++++++++++++++++++++++++++++++
%\begin{table}[tbp]
%\caption{Fractions for $F_{\pi}$ 
%and $F_{K+p}$ particles 
%in the 
%\jpsihyptrack sample determined using the $\dedx$ method. The fitted 
%functions, one for each particle-charge sign, consists of two gamma 
%distributions as shown in Fig.~\ref{fig:dedxPlot_Gpdf_Qplus}.}
%\begin{center}
%\begin{tabular}{lcccc}
%\hline\hline
%$p$ $\gevc$ & $F_{\pi^{+}}$ %& $F_{K^{+}+p}$  
%& $F_{\pi^{-}}$ %& $F_{K^{-}+\bar{p}}$ 
%\\
%\hline         
% 3.0-3.3 & 0.445$\pm$0.011 & 0.441$\pm$0.011 \\
% 3.3-3.6 & 0.410$\pm$0.009 & 0.445$\pm$0.010 \\
% 3.6-3.9 & 0.422$\pm$0.011 & 0.443$\pm$0.011 \\
% 3.9-4.2 & 0.416$\pm$0.012 & 0.390$\pm$0.012 \\
% 4.2-4.5 & 0.407$\pm$0.014 & 0.390$\pm$0.014 \\
% 4.5-5.0 & 0.420$\pm$0.013 & 0.383$\pm$0.013 \\
% 5.0-5.5 & 0.349$\pm$0.016 & 0.378$\pm$0.016 \\
% 5.5-6.0 & 0.335$\pm$0.020 & 0.328$\pm$0.020 \\
% 6.0-7.0 & 0.317$\pm$0.019 & 0.325$\pm$0.019 \\
% 7.0-8.0 & 0.333$\pm$0.028 & 0.265$\pm$0.027 \\
% 8.0-10.0& 0.293$\pm$0.033 & 0.278$\pm$0.031 \\
% $>$ 10.0 & 0.294$\pm$0.052 & 0.203$\pm$0.051 \\
%\hline\hline
%\end{tabular}
%\end{center}
%\label{tab:dedx_piKpFrac}
%\end{table}
% ++++++++++++++++++++++++++++++++++++++++++++++++++++++++++++++++++++++
    
\begin{figure}[tbp]
\centerline{
\makebox{\includegraphics[width=1.0\hsize]{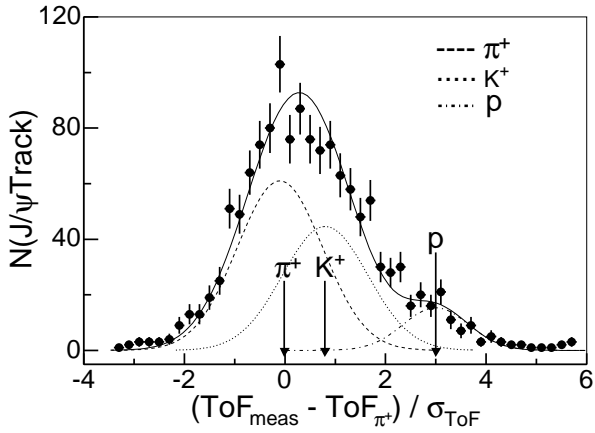}}}
\caption{Distribution of the quantity 
$\frac{\tofnum}{\sigma_{\textrm{ToF}}}$
using the momentum range 2.0--2.2 $\gevc$ for the events restricted to the 
subset with $-1.7<\mathcal{S}<-1.5$. Arrows 
show the central positions of the $\pi$, $K$, and $p$ hadrons.}
\label{fig:ToF_Slice}
\end{figure}

\begin{figure}[tbp]
\centerline{
\makebox{\includegraphics[width=1.0\hsize]{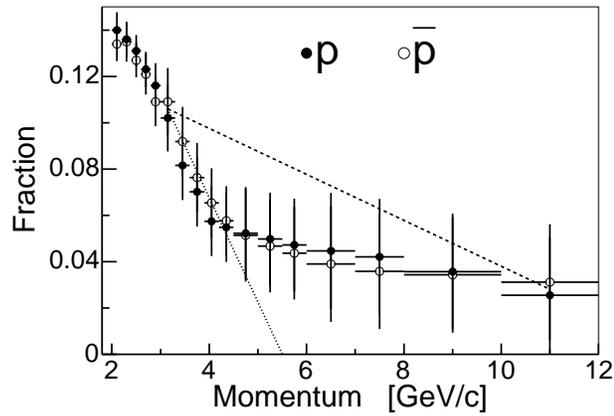}}}
\caption{Third-track $p$ and $\bar{p}$ fractions in the \jpsihyptrack sample. 
%The details are described in the text.
%momentum range 2.0-3.3 $\gevc$ shows the results from the $\dedx$ and ToF 
%simultaneous fit using the data, while the region $>$ $3~\gevc$ 
%represents the prediction from a {\sc pythia} simulation that is set to be in 
%agreement with the experimental data in the momentum range 3.0-3.3 $\gevc$.
The systematic uncertainty for the simulation prediction is bounded from 
above and below using the dashed and dotted lines in the figure.}
\label{fig:pbar_fractions}
\end{figure}

To calculate the proton fraction, we first calibrate the ToF performance using 
the kaon tracks from $\bptojpsik$ decays in the momentum range 2.0--3.3 
$\gevc$.
Then we perform a simultaneous two-dimensional likelihood fit of the ToF and 
the $\dedx$ data for the third track in the \jpsihyptrack sample.
As an example of the ToF standalone information, Fig.~\ref{fig:ToF_Slice} 
shows the distribution of the quantity $\frac{\tofnum}{\sigma_{\textrm{ToF}}}$ 
using 
the momentum range 2.0--2.2 $\gevc$ for events restricted to the subset with
$-1.7<\mathcal{S}<-1.5$.  Here, $ToF_{\textrm{meas}}$ is the
measured time, $ToF_{\pi}$ is the predicted time for the pion hypothesis, and $\sigma_{\textrm{ToF}}$ is the uncertainty in the measured time. %and the $S$ is from Eq.~(\ref{dedx_equation}).
% ++++++++++++++++++++++++++++++++++++++++++++++++++++++++++++++++++++++
To make use of the determination of $F_p$ in the momentum range 2.0--3.3~$\gevc$, we simulate $F_p$ for momentum in the range greater than 3.0~$\gevc$.  The MC procedure generates realistic $\bbbar$ quark events using the \textsc{pythia}~\cite{Ref:PYTHIA} simulation package with all $2\rightarrow 2$ QCD processes and initial- and final-state radiation.  The CTEQ5L~\cite{Ref:CTEQ} parton distributions for protons are used, and fragmentation of the $b$ quarks employs the Lund string model~\cite{Lund:pythia,Lund:proton}. The decay of $B$ mesons and baryons utilizes EVTGEN and the CDF\,II detector simulation is based on GEANT3~\cite{Brun:1987ma}. 
Studies show that the PYTHIA 
simulation predictions are lower than the experimental measurements in the 
momentum range 3.0--3.3~$\gevc$. To achieve 
consistency between the simulation and the experimental data, 
we scale the PYTHIA predictions for the whole range of momenta greater 
than 3~$\gevc$ so that the simulation and the experimental measurement of $F_p$ agree in the momentum range 3.0--3.3~$\gevc$. Both the experimental measurements for the $p$ and $\bar{p}$ 
fractions in the momentum range 2.0--3.3 $\gevc$ and the scaled PYTHIA
predictions in the range of momenta greater than 3 $\gevc$ are shown in 
Fig.~\ref{fig:pbar_fractions}.

% ++++++++++++++++++++++++++++++++++++++++++++++++++++++++++++++++++++++
% ++++++++++++++++++++++++++++++++++++++++++++++++++++++++++++++++++++++

For the study of the systematic uncertainty in the proton fractions we 
consider two options.  The first is to follow the slope of the simulation in 
the region 3.0--4.2~$\gevc$ assuming that $F_p$ = 0 at momenta higher than 
5.5~$\gevc$, and the second is to assume a straight line connecting the lowest 
and highest momentum points in the simulation (see the dotted and dashed lines 
in Fig.~\ref{fig:pbar_fractions}).
   
\begin{figure}[tbp]
\centerline{
\makebox{\includegraphics[width=1.0\hsize]{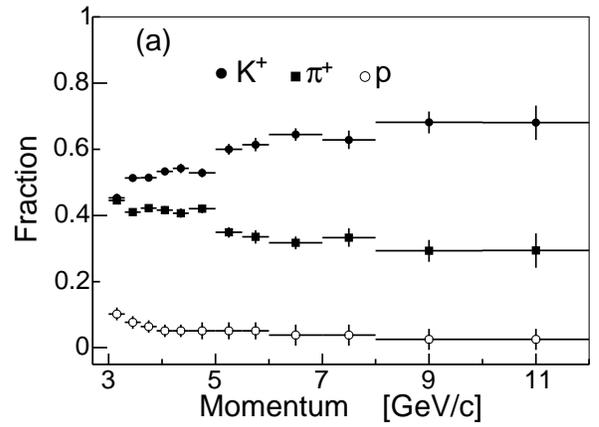}}}
\centerline{
\makebox{\includegraphics[width=1.0\hsize]{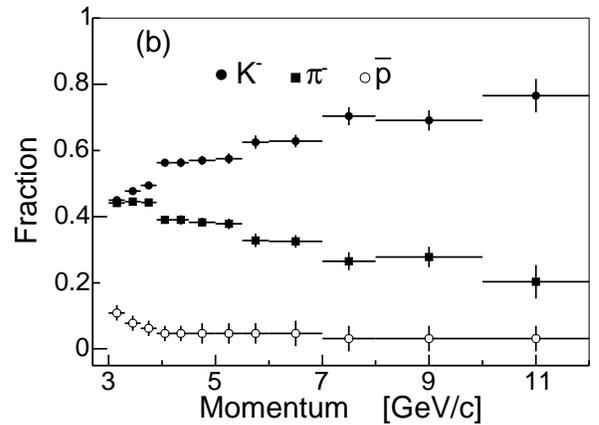}}}
\caption{Fractions $F_{\pi}$, $F_{K}$, and $F_{p}$ for (a) positively
and (b) negatively charged particles with momenta greater than 3~$\gevc$ corresponding to the third tracks in the \jpsihyptrack 
system.}
\label{fig:piKaPr_FractionRes}
\end{figure}

   Using the combined fraction $F_{K+p}$, determined from the fit illustrated 
in Fig.~\ref{fig:dedxPlot_Gpdf_Qplus},
and the standalone fraction $F_p$ illustrated in Fig.~\ref{fig:pbar_fractions},
the fraction $F_{K}$ is determined.  Figure~\ref{fig:piKaPr_FractionRes} 
shows the $F_{\pi}$, $F_{K}$, and $F_{p}$ fractions for the (a) positively 
and (b) negatively charged particles with momenta greater than 3 $\gevc$
corresponding to the third tracks in the \jpsihyptrack system.
% ++++++++++++++++++++++++++++++++++++++++++++++++++++++++++++++++++++++
% ++++++++++++++++++++++++++++++++++++++++++++++++++++++++++++++++++++++

%%%%%%%%%%%%%%%%%%%%%%%%%%%%%%%%%%%%%%%%%%%%%%%%%%%%%%%%%%%%
\subsubsection{Results and systematic uncertainties for the misidentified-muon background}
\label{sec:fakeMu_results}
%%%%%%%%%%%%%%%%%%%%%%%%%%%%%%%%%%%%%%%%%%%%%%%%%%%%%%%%%%%%

To complete the misidentified-muon background calculation, for each third track in the \jpsihyptrack sample, we assign a weight $W$ according to Eq.~(\ref{fakeMu_W_formula}) using the kaon and pion misidentification probabilities %from Table~\ref{tab:piK_toMuFakeRates} and 
shown in Fig.~\ref{fig:KpiToMuRates}; 
the fraction of the muon events outside of the $\dzero$ mass peak shown in 
Fig.~\ref{fig:muOutOffPeak}; and the pion, kaon, and proton fractions 
%in the \jpsihyptrack sample 
shown in Fig.~\ref{fig:piKaPr_FractionRes}.  %We histogram the weights versus the invariant masses of the $\jpsihyptrack$ candidates.  
The weighted mass distribution corresponds to the distribution of the 
misidentified-muon background as a function of the $\jpsiplusmu$ invariant mass.
               
     An additional small misidentified-muon component is produced if a 
misidentified $\jpsi$ makes a three-track vertex with a 
misidentified muon (``doubly misidentified'').  Since this background is in both
 the misidentified-$\jpsi$ and the misidentified-muon backgrounds, it must be 
determined to avoid double counting it. The 
doubly misidentified correction is calculated using the invariant-mass 
distribution of the sideband dimuons in the \jpsisidetrack system following 
procedures the same as those discussed in this section of the paper. 

     Because of the large size of the \jpsihyptrack sample, the statistical 
uncertainties in the calculation of the misidentified-muon background are 
negligible compared with the systematic uncertainties.  For the 
misidentified-muon uncertainties, the following procedures are used to estimate
 the various components of the systematic uncertainty:
\begin{enumerate}
%\item For the muon misidentification probabilities of pions or kaons, the  
%      distributions associated with a CMUP muon (shown in 
%      Figs.~\ref{fig:hadron_d0Ms}d-f) are fitted with the double-Gaussian 
%      templates derived from the non-muon sample and compared with 
%      the results from the single-Gaussian fits.
\item For the muon-misidentification probabilities of pions or kaons,
      a comparison is made of results from two fit functions applied to the 
      same distributions associated with a CMUP muon: the single-Gaussian 
      function versus the double-Gaussian templates derived from the nonmuon 
      sample.

\item For the fraction of the muon-matched events outside of the $\dzero$ mass 
      peak, 
%      the MC invariant-mass distributions fitted with a single Gaussian 
%      forced to have the same width as the experimental data are compared with
%      a single-Gaussian fit where the MC simulation was allowed to set the 
%      width of the fit. 
      fits to simulated mass distributions based on single-Gaussian functions 
      with widths fixed to those observed in data are compared with fits in 
      which widths are free to float.
      The resulting differences are used to estimate the systematic 
      uncertainty for this part of the misidentified-muon calculation.

\item For the particle fractions in the \jpsihyptrack
      system, fits of the $\dedx$ data with a sum of two Gaussian 
      distributions are compared to the fits with the sum of two gamma 
      distributions to determine the systematic uncertainty in the fitting 
      procedure.

\item For the proton fractions, variation bounds are obtained from
      the data-normalized PYTHIA simulation. 
%      is bounded conservatively from above and below to determine the 
%      systematic uncertainty. 
      For the lower bound we follow the slope of the PYTHIA simulation 
      in the region 3.0--4.2 GeV/c assuming that $F_p$ = 0 beyond 5.5 GeV/c
      (Fig.~\ref{fig:pbar_fractions}, dotted line).
      For the upper bound we assume a straight line connecting the lowest and 
      highest momentum points in the PYTHIA simulation (see the dashed 
      line in Fig.~\ref{fig:pbar_fractions}).
\end{enumerate}

\begin{figure}[tbp]
\centerline{
\makebox{\includegraphics[width=1.0\hsize]{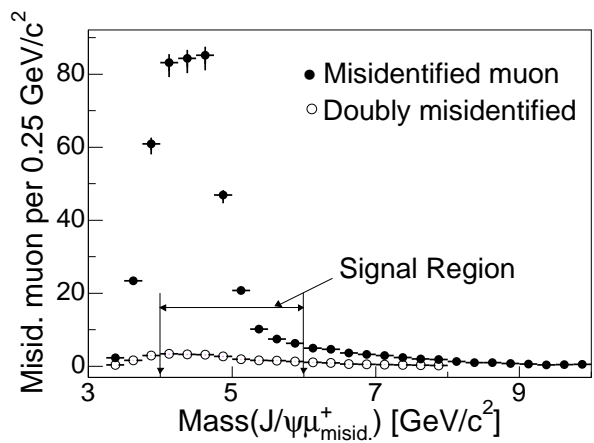}}}
\caption{Weighted invariant-mass distribution of the \jpsihyptrack 
system showing the misidentified (filled circles) and the doubly misidentified
(open circles) muon backgrounds to the $\bctojpsimux$ decays. The error 
bars represent the estimated systematic uncertainties. Because of the large 
size of the \jpsihyptrack sample, the statistical errors in the 
misidentified-muon calculation are negligible.}
\label{fig:fakeMuon_mass}
\end{figure}

The systematic uncertainties for the misidentified and doubly misidentified-muon backgrounds are shown in Table~\ref{tab:fakeMuon_system}.
% ++++++++++++++++++++++++++++++++++++++++++++++++++++++++++++++++++++++
\begin{table*}[Htbp]
\caption{Systematic uncertainties in the number of events involving misidentified muons and doubly misidentified muons.}
\begin{center}
\begin{tabular}{lcc}
\hline \hline
Source                              & Misidentified & Doubly misidentified \\
%of systematic uncertainty           & muons         & muons               \\ 
\hline
Misidentification probability                      & $\pm7.3$ & $\pm0.4$  \\
Fraction of events outside of the $\dzero$ mass peak  & $\pm1.2$ & $\pm0.1$  \\
Particle fractions in the \jpsihyptrack system     & $\pm4.7$ & $\pm0.3$  \\
Proton fractions           & $^{+4.0}_{-14.0}$ & $^{+0.2}_{-0.7}$  \\
\hline
Total                & $^{+9.6}_{-16.5}$ & $^{+0.5}_{-0.9}$ \\
\hline \hline
\end{tabular}
\end{center}
\label{tab:fakeMuon_system}
\end{table*}
% ++++++++++++++++++++++++++++++++++++++++++++++++++++++++++++++++++++++
The misidentified and doubly misidentified-muon backgrounds as functions of the
 invariant mass of the \jpsihyptrack system and their associated 
systematic uncertainties are shown in Fig.~\ref{fig:fakeMuon_mass}. 
Numerical results are given in Table~\ref{tab:fakeMuon_results}.  

% ++++++++++++++++++++++++++++++++++++++++++++++++++++++++++++++++++++++
% ++++++++++++++++++++++++++++++++++++++++++++++++++++++++++++++++++++++
% ++++++++++++++++++++++++++++++++++++++++++++++++++++++++++++++++++++++
\begin{table*}[tbp]
\caption{Number of events involving misidentified-muon and 
doubly misidentified-muon backgrounds within the signal and sideband mass 
ranges and associated systematic uncertainties.}
\label{tab:fakeMuon_results}
\begin{center}
\begin{tabular}{lccc}
\hline\hline
Mass range ($\gevcc$) & $3$-$4$              & $4$-$6$                & $>6$ \\
\hline
Misidentified muons   & 86.7$^{+2.4}_{-4.2}$ & 344.4$^{+9.6}_{-16.5}$ & 32.1$^{+0.9}_{-1.5}$  \\
Doubly misidentified muons & 5.1$^{+0.1}_{-0.2}$  & 19.0$^{+0.5}_{-0.9}$   & 5.2$^{+0.1}_{-0.3}$ \\
\hline\hline
\end{tabular}
\end{center}
\end{table*}
% ++++++++++++++++++++++++++++++++++++++++++++++++++++++++++++++++++++++

%%%%%%%%%%%%%%%%%%%%%%%%%%%%%%%%%%%%%%%%%%%%%%%%%%%%%%%%%%%%
\subsection[$\bbbar$ background]{\boldmath{$\bbbar$} background} 
\label{sec:b_bbar}
%%%%%%%%%%%%%%%%%%%%%%%%%%%%%%%%%%%%%%%%%%%%%%%%%%%%%%%%%%%%

The $\bbbar$ background arises from the combination of a $\jpsi$ meson produced by the decay of a $b$ quark with the third muon produced from the decay of the $\bar{b}$ quark in the same event, or vice
versa.  The production of $\bbbar$ pairs in $p\bar{p}$ collisions 
is dominated by the leading-order flavor-creation (FC) process and the
next-to-leading-order flavor-excitation (FE) and gluon-splitting (GS)
processes~\cite{Ref:Field2002}. Flavor creation corresponds to the production of a $\bbbar$ pair by gluon fusion or by the annihilation of light quarks via two 2-to-2 parton subprocesses $gg \to \bbbar$ and $q\bar{q}\to \bbbar$.  Flavor excitation refers to the QCD hard 2-to-2 reaction corresponding to the scattering of a $b$ quark out of the initial state into the final state by a gluon or a light quark or light antiquark via the subprocesses $gb \to gb$, $qb \to qb$, and $\bar{q} b \to \bar{q} b$.  The $\bar{b}$ partner from the original initial state $\bbbar$ pair will also appear in the final state.  There are three more processes corresponding to the scattering of the $\bar{b}$ quarks in the high $Q^2$ sea of gluons and heavy-quark pairs that define the $p$ and $\bar{p}$ structure functions.  Gluon splitting occurs when only gluons and light quarks and light antiquarks participate in the 2-to-2 hard parton scattering subprocess, but one of the final-state gluons fragments into a $\bbbar$ pair, e.g., $gg \to g(g\to\bbbar)$ or $qg \to q(g\to \bbbar)$.  Flavor creation is expected to produce the largest opening
angles between the quark pairs, as measured in the plane transverse to the 
beam direction.  Flavor excitation is expected to produce both
large and small opening angles, and gluon splitting is expected to produce a 
relatively uniform distribution of opening angles~\cite{Ref:CTEQ,Ref:PYTHIA}. 
%Because the $\jpsi$ and muon are required to have an opening angle less than 
%$\pi/2$, the fraction of events contributed by FC is expected to be suppressed, while GS and FE both contribute to the $\bbbar$ background.

%The basic procedure for the calculation is described in the measurement 
%of the $\bc$ lifetime~\cite{Ref:MarkThesis}. However, we updated many 
%steps of this calculation. Therefore, we describe the full procedure here. 

     The determination of the $\bbbar$ background relies on a PYTHIA MC simulation to generate potential $\bbbar$ background events for the three QCD processes. % FC, FE, and GS. 
%Although we partially trust the kinematic description of these processes given
% by {\sc pythia}, our approach is to 
We constrain the PYTHIA MC simulation with the experimental data 
%In comparing simulated events with experimental data, we 
using the distribution of the opening angle $\dphi$ between the $\jpsi$ and the
 muon in an event.  We select a sample of experimental data called the 
unvertexed-$\jpsiplusmu$-pairs sample as described in 
Sec.~\ref{sec:UnvtxJpsiMu_select} below.  Unvertexed means that there are no requirements that the $\jpsiplusmu$ pairs originate from a common vertex.  From this sample we subtract 
%eliminate %This is a complex procedure requiring the elimination of 
potential signal candidates as well as unvertexed variations of the major 
backgrounds described above.  We fit the $\dphi$ distribution in these data 
with a linear combination of the $\dphi$ distributions of 
PYTHIA-simulated FC, FE, and GS events that are also unvertexed. 
%and from which events that pass our signal criteria are eliminated.  
This procedure allows for a determination of the 
relative fractions of FC, FE, and GS to use in estimating the $\bbbar$ 
background irrespective of the relative fractions that any particular variation
of the PYTHIA parameters might produce.  Using the experimentally 
constrained fractions for the FC, FE, and GS contributions, we calculate the 
$\bbbar$ background by applying the selection requirements for the  
$\jpsiplusmu$-signal sample to the unselected PYTHIA-simulated FC, FE, 
and GS samples.  A valuable cross-check of the background determination 
consists in comparing the sum of all of the backgrounds with the number of 
events in the $\jpsiplusmu$ invariant-mass ranges \lowin and greater than 
6~\gevcc, where the number of events is dominated by background.

%  The $\bbbar$ background calculation is based partially on
%a {\sc pythia} Monte Carlo simulation. With our selection requirements applyed to the {\sc pythia} sample we find that the fractions for the QCD production processes flavor creation (FC), flavor excitation (FE), and gluon splitting (GS) are 25, 55, and 20\%, respectively, for production of $\bptojpsik$ decays and 35,40, and 25\%, respectively, for production of $\jpsi\mu^{+}$ pairs. For the  $\jpsi\mu^{+}$ pairs we do not required that the $\jpsi$ and $\mu^{+}$ be constrained by a common vertex. This Monte Carlo sample is named the ``unvertexed'' $\jpsi\mu^{+}$ pairs. The obvious question is what QCD fractions should be used in determining the $\bbbar$ backgound for $\bctojpsimunu$ decays.  
%A lengthy study led to the conclusion that a more expermimentally driven 
%We chose to constrain the QCD fractions by requiring them to give a good description of the $\dphi$ distributions for the unvertexed $\jpsiplusmu$ pairs in the experimental data.

%%%%%%%%%%%%%%%%%%%%%%%%%%%%%%%%%%%%%%%%%%%%%%%%%%%%%%%%%%%%
\subsubsection{Selecting the unvertexed-$\jpsiplusmu$ pairs} 
\label{sec:UnvtxJpsiMu_select}
%%%%%%%%%%%%%%%%%%%%%%%%%%%%%%%%%%%%%%%%%%%%%%%%%%%%%%%%%%%%
% ++++++++++++++++++++++++++++++++++++++++++++++++++++++++++++++++++++++
\begin{figure*}[tbp]
\centerline{
\makebox{\includegraphics[width=0.33\hsize]{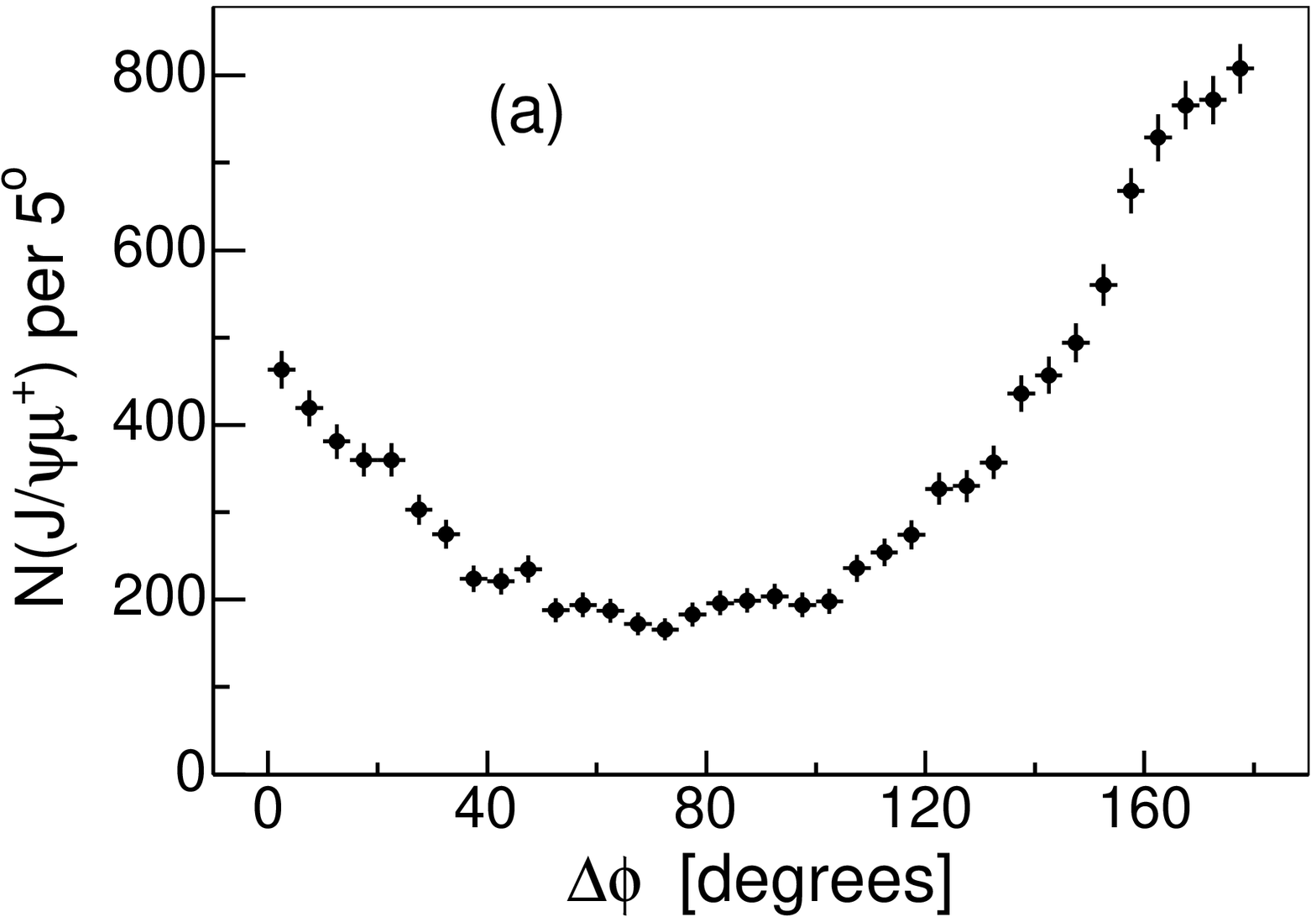}}
%\centerline{
\makebox{\includegraphics[width=0.33\hsize]{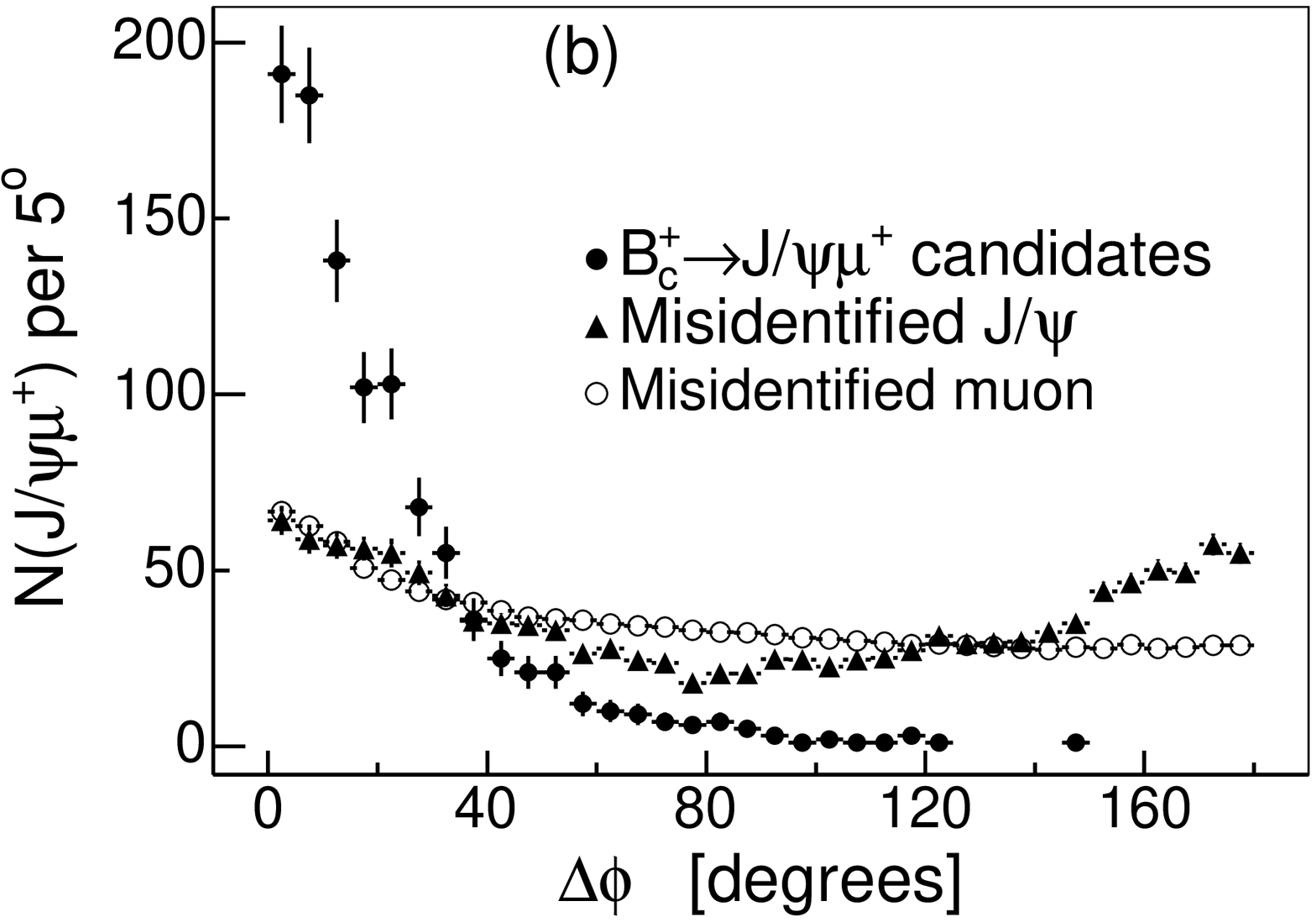}}
%\centerline{
\makebox{\includegraphics[width=0.33\hsize]{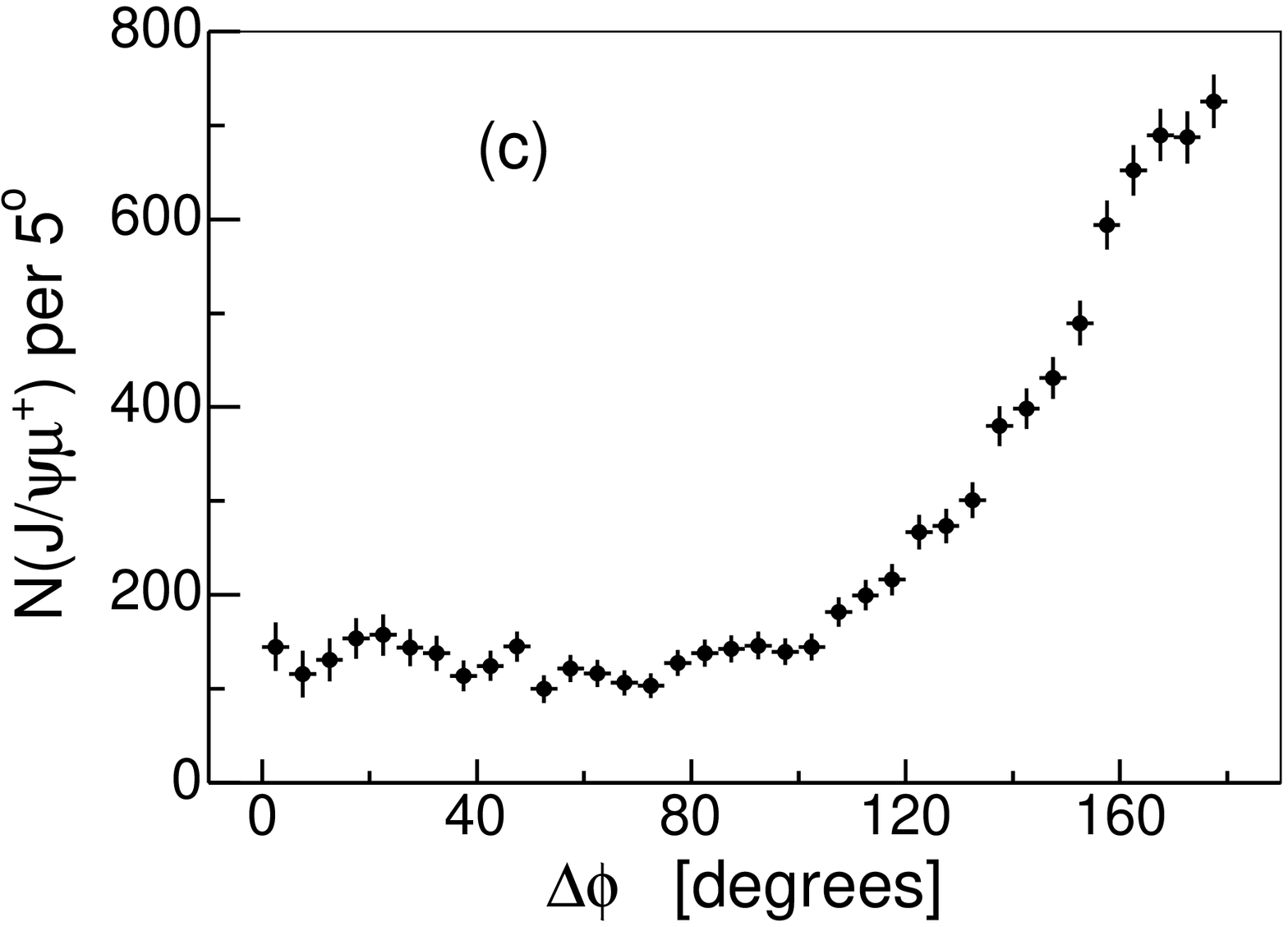}}}
\caption{(a) Distribution versus $\dphi$ of all unvertexed-$\jpsiplusmu$ pairs 
from the experimental data. (b) Three non-$\bbbar$ contributions to (a)
superimposed. (c) Experimental data from (a) with the non-$\bbbar$ 
contributions removed.}
\label{fig:unvtx_nonBB_dphi}
\end{figure*}
% ++++++++++++++++++++++++++++++++++++++++++++++++++++++++++++++++++++++

  The selection requirements for the unvertexed-$\jpsiplusmu$-pairs sample in the data follow the requirements listed in Tables~\ref{tab:jpsi_cuts}--\ref{tab:thirdTrk_cuts} with the modifications: the mass range for the 
$\jpsi$ is reduced from $\pm$50~$\mevcc$ to $\pm$30~$\mevcc$; the decay length 
for the $\jpsi$ is required to be greater than 200~$\mu$m; there is no vertex 
requirement for the trimuon system; and there is no $\dphi$  
requirement between the $\jpsi$ and the third muon.
%In Ref.~\cite{Ref:MarkThesis} the {\sc pythia} tuning was done using 0.36 
%$\fb$ of data. However, in this study we use the complete CDF dataset. 
In the data there may be more than one $p\bar{p}$ interaction 
distributed longitudinally along the interaction region,
%in the longitudinal interaction region 
which has a rms length of about 30~cm. In order 
to restrict the data sample to events in which the $\jpsi$ and 
third muon come from the same $p\bar{p}$ interaction, we require that the $z$ 
separation between the $\jpsi$ and the third muon is less than 2~cm.
  
     The unvertexed-$\jpsiplusmu$ pairs come not only from different 
$b$ hadrons produced in the same $p\bar{p}$ interaction but also from  
non-$\bbbar$ sources: 
%The non-$\bbbar$ contributions are itemized below:
% ++++++++++++++++++++++++++++++++++++++++++++++++++++++++++++++++++++++
\begin{enumerate}
%\itemThe unvertexed-$\jpsiplusmu$ pairs that would pass the vertex probability
%      selection requirement come predominantly from a single $b$ hadron.
\item Single $b$ hadrons contribute to the unvertexed-$\jpsiplusmu$ pairs that 
      would pass the vertex probability requirement. They include the 
      $\bctojpsimux$ event candidates which include the background components 
      having a vertexed $\jpsi$ plus a misidentified muon and 
      misidentified vertexed $\jpsi$ plus a muon.

\item A pion or kaon from an unvertexed
      \jpsihyptrack event is misidentified as a muon.

\item An unvertexed misidentified $\jpsi$ also can be in the 
      unvertexed-$\jpsiplusmu$ system.
\end{enumerate}
% ++++++++++++++++++++++++++++++++++++++++++++++++++++++++++++++++++++++
 
     To produce a pure sample of $\bbbar$ pairs to compare with the 
PYTHIA simulation, it is necessary to estimate the contributions from the
non-$\bbbar$ sources listed above and then subtract them from the selected 
sample of 
unvertexed-$\jpsiplusmu$ pairs shown in Fig.~\ref{fig:unvtx_nonBB_dphi}(a). The 
first non-$\bbbar$ source is identified by applying the vertex probability 
requirement to the unvertexed-$\jpsiplusmu$ pairs, and its $\dphi$ distribution
 is shown in Fig.~\ref{fig:unvtx_nonBB_dphi}(b), labeled as ``$\bctojpsimunu$'' 
candidates.  The essential difference between this $\bctojpsimunu$ sample and 
the signal sample is that the $\dphi$ selection criterion is not applied in 
order to compare the $\bbbar$ data sample with the MC simulation over the 
entire range $\dphi$.  
% Next, we correct for \jpsihyptrack events that do not pass the vertex 
% probability selection requirement but for which the third track is a hadron 
%misidentified as a muon with the procedure described in Sec.~\ref{sec:fake_mu}
% using the unvertexed-\jpsihyptrack sample with the vertexed events in the 
% sample subtracted.  
Next, we estimate the unvertexed misidentified-muon background with the 
procedure described in Sec.~\ref{sec:fake_mu} using the 
unvertexed-\jpsihyptrack sample with the vertexed events subtracted.
The $\dphi$ distribution of unvertexed misidentified-muon background is also 
shown in Fig.~\ref{fig:unvtx_nonBB_dphi}(b).  Finally, the events containing an unvertexed-$\jpsiplusmu$ pair, where the $\jpsi$ is misidentified, are accounted by the method of 
Sec.~\ref{sec:fake_jpsi} using the events from the dimuon mass sidebands of the unvertexed-$\jpsiplusmu$-pairs sample.  The $\dphi$ distribution of
misidentified $\jpsi$ in the unvertexed-$\jpsiplusmu$-pairs sample is also shown in 
Fig.~\ref{fig:unvtx_nonBB_dphi}(b). Subtracting the three non-$\bbbar$ sources shown 
in Fig.~\ref{fig:unvtx_nonBB_dphi}(b) from the unvertexed-$\jpsiplusmu$ pairs in 
Fig.~\ref{fig:unvtx_nonBB_dphi}(a) gives the background-subtracted sample of 
unvertexed-$\jpsiplusmu$ pairs.  This pure $\bbbar$ sample is shown in 
Fig.~\ref{fig:unvtx_nonBB_dphi}c and is used to determine the relative 
fractions of the QCD production processes generated by the PYTHIA 
simulations.

%%%%%%%%%%%%%%%%%%%%%%%%%%%%%%%%%%%%%%%%%%%%%%%%%%%%%%%%%%%%
\subsubsection{Simulated unvertexed-$\jpsiplusmu$ pairs}
 %and $\bptojpsik$ decays} 
\label{sec:unvtx_pythiaJpsiMu}
%%%%%%%%%%%%%%%%%%%%%%%%%%%%%%%%%%%%%%%%%%%%%%%%%%%%%%%%%%%%
% ++++++++++++++++++++++++++++++++++++++++++++++++++++++++++++++++++++++
\begin{figure*}[tbp]
\centerline{
\makebox{\includegraphics[width=0.33\hsize]{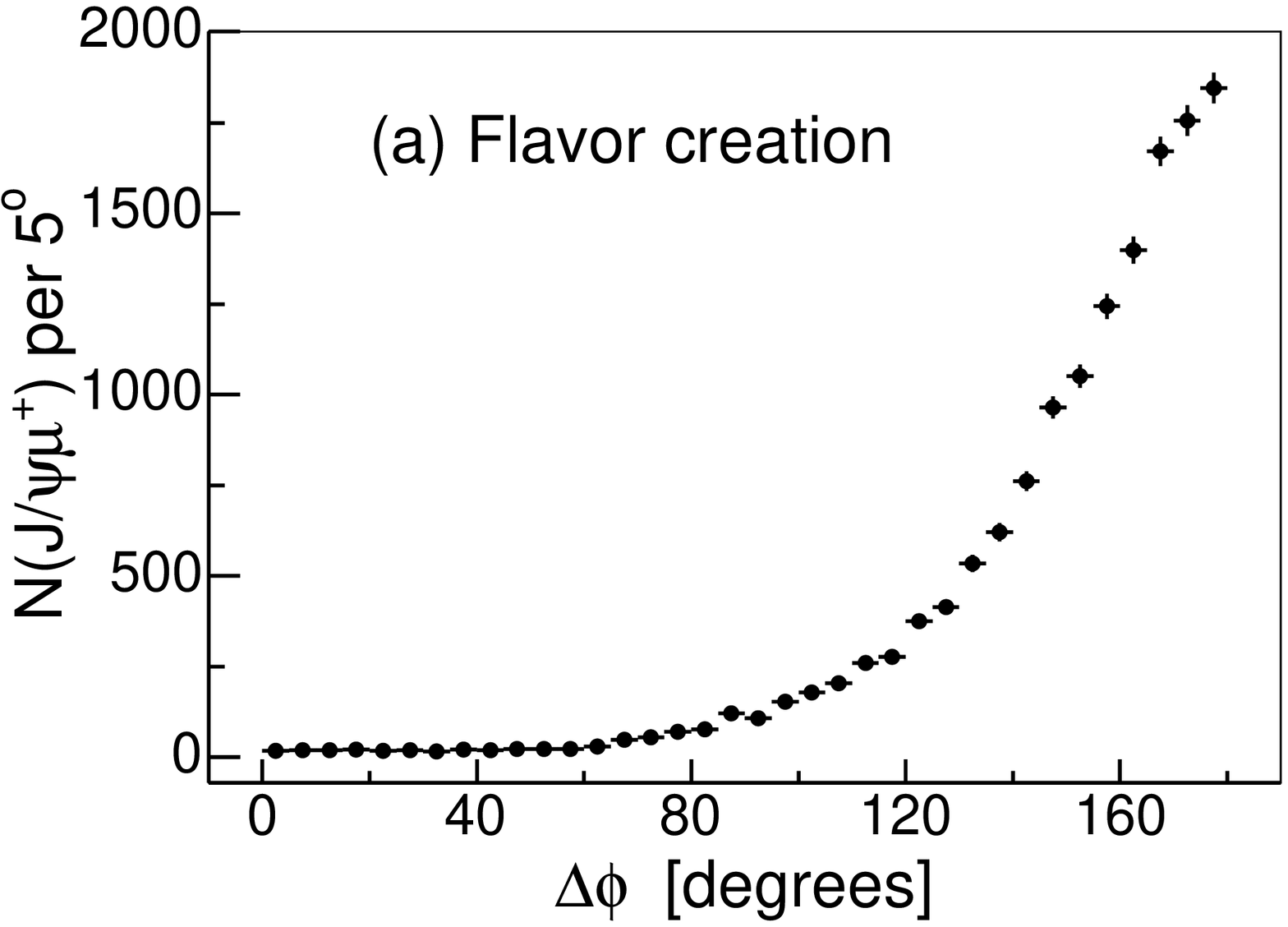}}
%\centerline{
\makebox{\includegraphics[width=0.33\hsize]{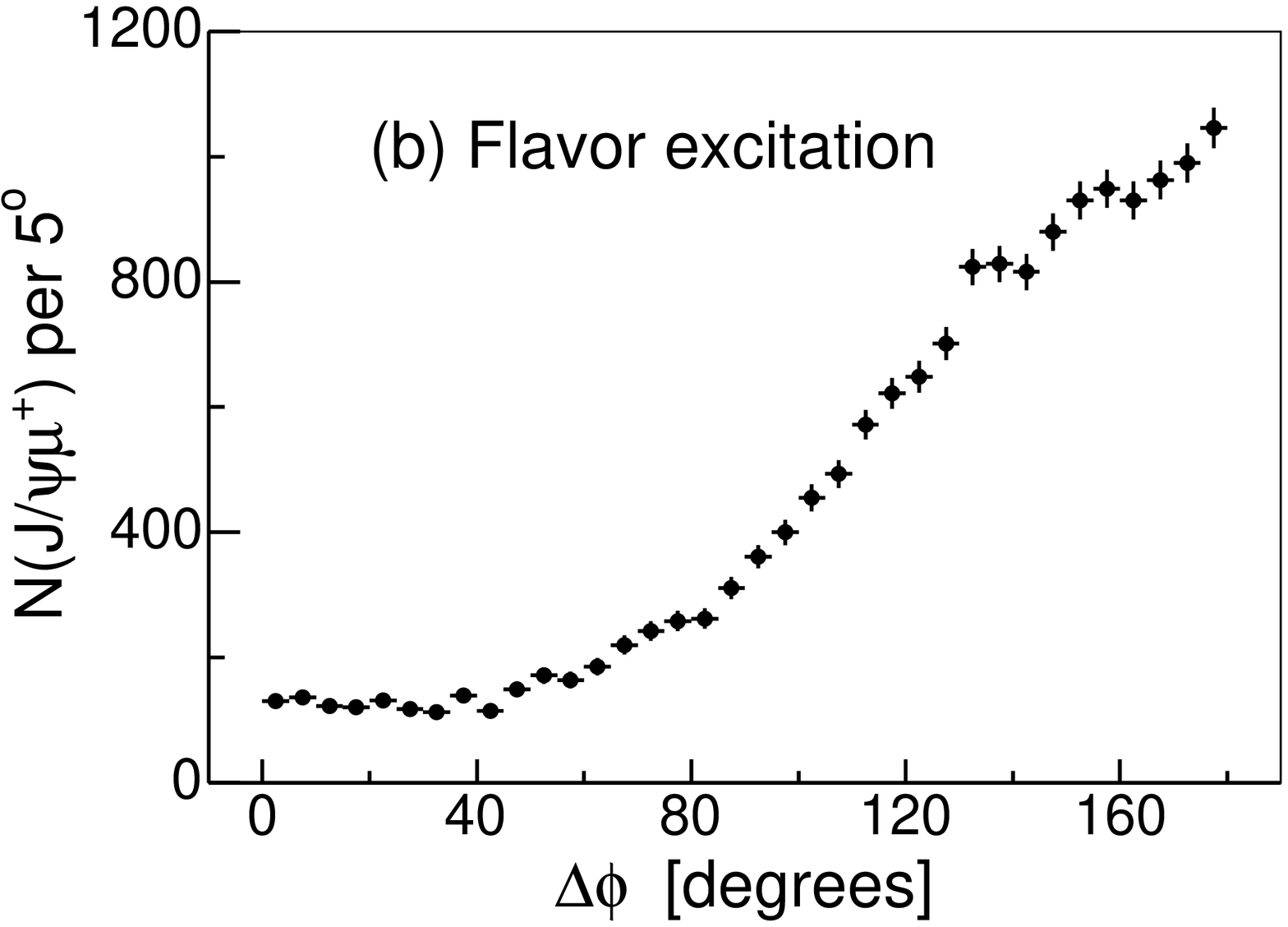}}
%\centerline{
\makebox{\includegraphics[width=0.33\hsize]{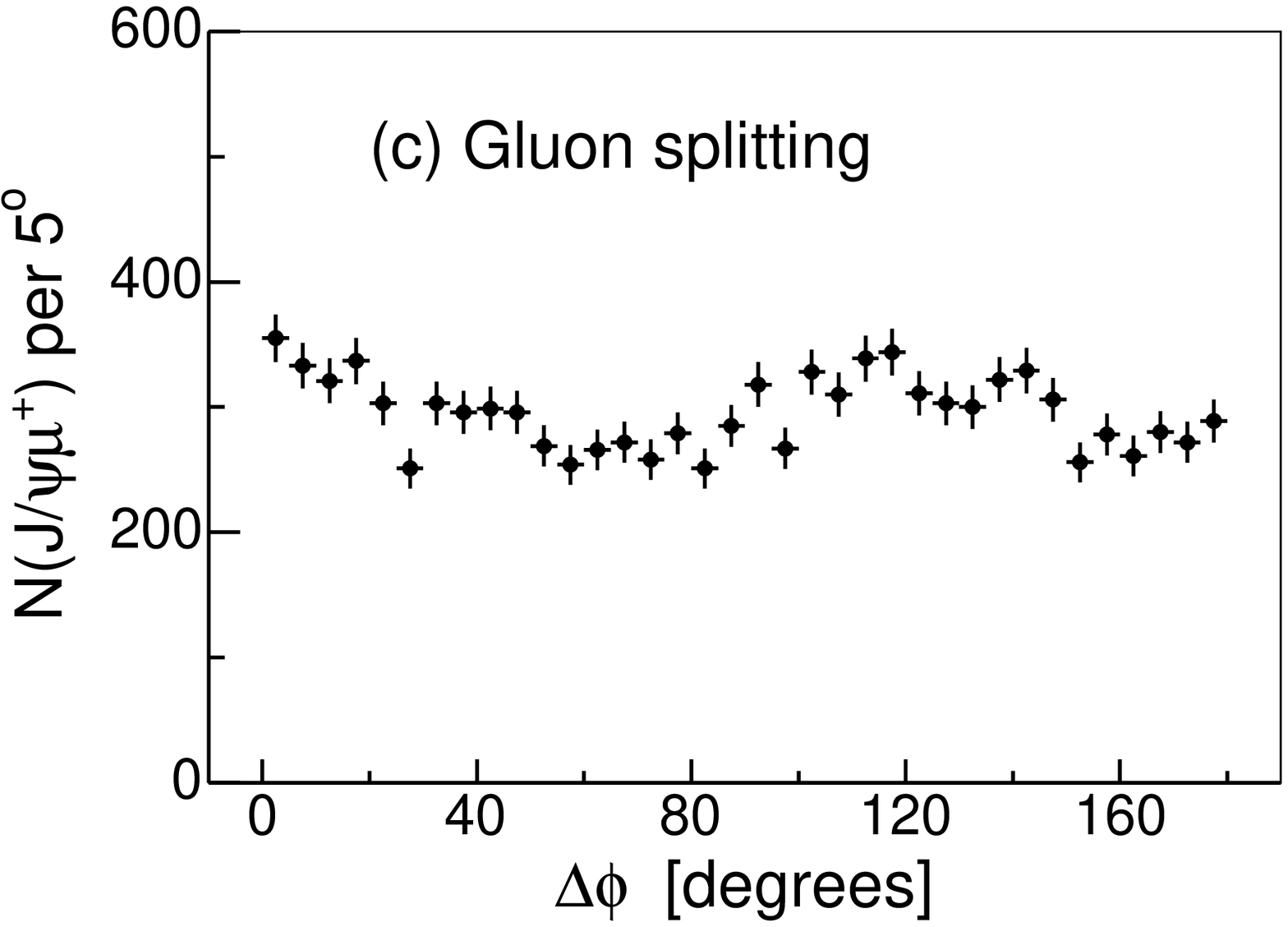}}}
\caption{Distributions of $\dphi$ for the unvertexed-$\jpsiplusmu$ pairs 
simulated from the three QCD production processes: (a) flavor creation, (b) flavor excitation, and (c) gluon splitting. 
%A possible misidentified third muon is rejected using the parent particle 
%identification code. The distributions are fitted with a third order 
%polynomial for FC and  FE and with a second order polynomial for GS
}
\label{fig:unvtx_pythiaDphi}
\end{figure*}
% ++++++++++++++++++++++++++++++++++++++++++++++++++++++++++++++++++++++

%We used following {\sc pythia} settings: MSEL=1, the center of mass energy
%set to 1960 $\gevcc$. Regarding the decay table, we used the best known list 
%of decay table available in 2011 (``master decay table''). There we implement 
%some changes to meet the following requirements:
% ++++++++++++++++++++++++++++++++++++++++++++++++++++++++++++++++++++++
%\begin{itemize}
%\item{The negative $b$, $B^{-}$, $\bar{B^{o}}$, and $\bar{B^{o}_{s}}$ are 
      %allowed to decay according to the master decay table, where the major 
      %sources for the third muons are $\mu^{-}\nu X$, $\jpsi X$ decays, and any
      %other consecutive processes involving the $\jpsi$.}
%\item{The positive $b$, $B^{+}$, $B^{o}$, and $B^{o}_{s}$ are forced to decay 
      %using a user decay table into: $\jpsi X$, $\psi(2S)X$, $\chi_{co}X$,
      %$\chi_{c1}X$, and $\chi_{c2}X$. The decay fractions are proportional to 
      %what is in the master decay table. }
      %\begin{itemize}
      %\item{$\psi(2S)$, $\chi_{co}$, $\chi_{c1}$, and $\chi_{c2}$ from the 
            %positive $b$ are forced to decay into $\jpsi X$ with the branching 
            %ratio's determined by the master decay table.}
      %\item{$\jpsi$ from positive $b$ are forced to decay into $\mu^{+}\mu^{-}$
             %pairs only.}
      %\end{itemize}
%\end{itemize}
% ++++++++++++++++++++++++++++++++++++++++++++++++++++++++++++++++++++++

A PYTHIA sample containing 0.5$\times$10$^{6}$ $\bbbar$ pairs is
generated. Either the $b$ or $\bar{b}$ quark is allowed to decay naturally, 
where the major sources of muons are semileptonic decays of bottom hadrons or 
of their daughter charm hadrons. The $\bar{b}$ or $b$ quark partner is forced 
to decay into a $\jpsi$ or any state which might cascade into a $\jpsi$ meson.
Figures~\ref{fig:unvtx_pythiaDphi}(a)--(c) show the $\dphi$ distributions 
of unvertexed-$\jpsiplusmu$ pairs from the FC, FE, and GS processes, 
respectively. 
%[with all of the events that pass the vertex probability selection 
%requirement removed.] ( statement should be included )
%A possible misidentified third 
%muon is rejected using the parent particle identification code. 
%The $\dphi$ distributions in Fig.~\ref{fig:unvtx_pythiaDphi} are fitted 
%with a third order polynomial for FC and  FE and with a second order 
%polynomial for GS. 
%By viewing Fig.~\ref{fig:unvtx_pythiaDphi} one sees that the polynomial
%functions are not a good fit to  the $\dphi$ distributions, especially the
%GS (c) and the $\chi^2$ are significantly off for all. 

%In comparing these Monte Carlo simulations with the experimental data, we 
%use the Monte Carlo simulated $\dphi$ distributions with any events which pass
% the vertex probability selection requirement removed.

    To normalize the $\bbbar$ background events from the PYTHIA sample to
data, we use the yields of the $\bptojpsik$ decays observed in data. 
%and {\sc pythia}.
In the $\bptojpsik$ decays reconstructed from the PYTHIA simulation 
we  apply all the requirements listed in Tables~\ref{tab:jpsi_cuts}--\ref{tab:thirdTrk_cuts}.  The numbers of $\bptojpsik$ decays produced 
by the three QCD processes are 16 275$\pm$130 (25$\%$ of FC), 35 464$\pm$189 
(55$\%$ of FE), and 12 602$\pm$118 (20$\%$ of GS).

%The invariant mass distributions of 
%reconstructed $\bp$ from $\bptojpsik$ decays produced by the three QCD 
%processes in the {\sc pythia} simulation are shown in 
%Fig.~\ref{fig:bYield_pythia}.  The $\bp$ yield is used to normalization 
%the simulation results to the experimental data.
% ++++++++++++++++++++++++++++++++++++++++++++++++++++++++++++++++++++++
%\begin{figure}[htbp]
%\centerline{
%\makebox{\includegraphics[width=0.33\hsize]{feps/prd_JpsiKMsFC.eps}}
%\makebox{\includegraphics[width=0.33\hsize]{feps/prd_JpsiKMsFE.eps}}
%\makebox{\includegraphics[width=0.33\hsize]{feps/prd_JpsiKMsGS.eps}}}
%\caption{The invariant mass distributions of the $\jpsiplusk$ system from 
%$\bptojpsik$ decays in the {\sc pythia} simulation sample for each QCD 
%production processes: (a) FC, (b) FE, and (c) GS. As 
%in the experimental data the Monte Carlo distributions are fitted with a 
%sum of two Gaussians. The $\bp$ yield is used to normalization the 
%simulation results to the experimental data.}
%\label{fig:bYield_pythia}
%\end{figure}
% ++++++++++++++++++++++++++++++++++++++++++++++++++++++++++++++++++++++

%%%%%%%%%%%%%%%%%%%%%%%%%%%%%%%%%%%%%%%%%%%%%%%%%%%%%%%%%%%%
\subsubsection{Fitting the unvertexed-$\jpsiplusmu$ $\dphi$ 
distribution} 
\label{sec:unvtxJpsiMu_fitData}
%%%%%%%%%%%%%%%%%%%%%%%%%%%%%%%%%%%%%%%%%%%%%%%%%%%%%%%%%%%%
The experimental data shown in Fig.~\ref{fig:unvtx_nonBB_dphi}(c) are fit with 
a linear combination of the the three PYTHIA $\dphi$ distributions shown 
in Figs.~\ref{fig:unvtx_pythiaDphi}(a)--(c). The 
predicted number of $\bbbar$ events for a given $\dphi$ bin is given by
% ++++++++++++++++++++++++++++++++++++++++++++++++++++++++++++++++++++++
\begin{eqnarray}
N_{\bbbar}&=&C(S_{\textit{FC}}N^{\textit{FC}}_{b\bar{b}} + S_{\textit{FE}}N^{\textit{FE}}_{b\bar{b}} + S_{\textit{GS}}N^{\textit{GS}}_{b\bar{b}})\nonumber \\
& &\times \frac{N_{\bp}} {S_{\textit{FC}}N^{\textit{FC}}_{\bp} 
+ S_{\textit{FE}}N^{\textit{FE}}_{\bp} + S_{\textit{GS}}N^{\textit{GS}}_{\bp} }\,.
\label{bbbar_equation}
\end{eqnarray}
% ++++++++++++++++++++++++++++++++++++++++++++++++++++++++++++++++++++++
$C = 0.76 \pm 0.07$ is a correction factor that accounts for the differences 
between the fraction of $b$ quarks fragmenting into $\bp$, the $\bptojpsik$ 
branching fraction, and the known inclusive branching fraction for all $B$ 
hadrons to produce a $\jpsi$ meson~\cite{Ref:PDG3} and the values set in the 
PYTHIA simulation program~\cite{Ref:PYTHIA}. 
In the fit $C$ is constrained by its uncertainty.
% ++++++++++++++++++++++++++++++++++++++++++++++++++++++++++++++++++++++
%\begin{equation}
%C_{norm}\times \frac{F_{pythia}(B^+)\times F_{pythia}(\btojpsik)}{F_{pythia}(H_{b}\to\jpsi X)} = \frac{F_{data}(B^+)\times F_{data}(\btojpsik)}{F_{data}(H_{b}\to\jpsi X)} 
%\label{cnorm_formula}
%\end{equation}
% ++++++++++++++++++++++++++++++++++++++++++++++++++++++++++++++++++++++
%The input numbers to calculate C$_{norm}$ are shown in 
%Table~\ref{cnorm_inputNum}. Using the input numbers from  
%Table~\ref{cnorm_inputNum}, we find that C$_{norm}$ = 0.76$\pm$0.07.
% ++++++++++++++++++++++++++++++++++++++++++++++++++++++++++++++++++++++
%\begin{table}[htbp]
%\begin{center}
%\begin{tabular}{lcc}
%\hline\hline
%                               & PDG 2012                   & {\sc pythia} \\
%\hline
%$\bp$ production fraction      & 0.401$\pm$0.008          & 0.438$\pm$0.0004 \\
%$\btojpsik$ branching fraction & (1.016$\pm$0.033)10$^{-3}$ & set 0.106     \\
%$H_{b}\to\jpsi X$ production fraction & 0.0116$\pm$0.0010   & set 1         \\
%\hline\hline
%\end{tabular}
%\end{center}
%\caption{The input quantities to calculate an expected $C_{norm}$ value.}
%\label{cnorm_inputNum}
%\end{table}
% ++++++++++++++++++++++++++++++++++++++++++++++++++++++++++++++++++++++
The parameters $S_{\textit{FC}}$, $S_{\textit{FE}}$, and $S_{\textit{GS}}$ are the scale 
factors for the different QCD production processes in PYTHIA.  The fit 
allows the scale factors to float subject to the constraint that their sum must
equal three. The numbers of PYTHIA events in a given $\dphi$ bin as
shown in Figs.~\ref{fig:unvtx_pythiaDphi}(a)--(c) are $N^{\textit{FC}}_{\bbbar}$,
$N^{\textit{FE}}_{\bbbar}$, and $N^{\textit{GS}}_{\bbbar}$, respectively. 
The total number of $\bptojpsik$ decays in the data shown in 
Fig.~\ref{fig:threeTrk_mass}(b) is $N_{\bp}$. The numbers of $\bptojpsik$  decays
produced by the three QCD processes in PYTHIA are $N^{\textit{FC}}_{\bp}$, 
$N^{\textit{FE}}_{\bp}$, and $N^{\textit{GS}}_{\bp}$, respectively.  The last term in 
Eq.~(\ref{bbbar_equation}) along with $C$ normalizes the three  
PYTHIA samples to the experimental data. 
  
\begin{figure}[tbp]
\centerline{
%\makebox{\includegraphics[width=0.5\hsize]{feps/prd_fcAndFE_dPhiPsiMu.eps}}
\makebox{\includegraphics[width=1.0\hsize]{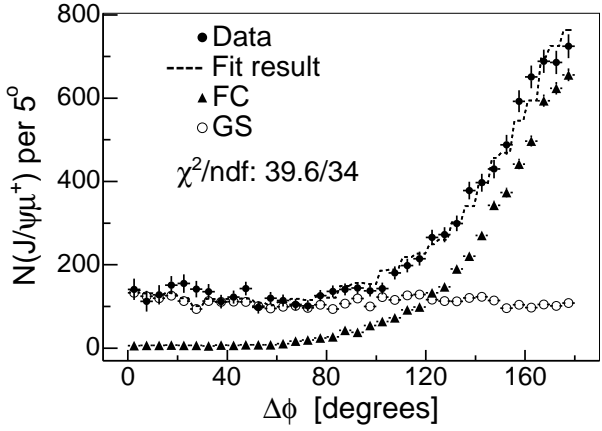}}}
\caption{Fit of the $\dphi$ distribution of the unvertexed-$\jpsiplusmu$ data 
for the combination of FC plus GS.}
\label{fig:unvtx_FitDphi}
\end{figure}

%The fit done by the method of least squares:
% ++++++++++++++++++++++++++++++++++++++++++++++++++++++++++++++++++++++
%\begin{equation}
%\chi^2 = \sum{\frac{(N^{data}_{\bbbar}-N^{pred}_{\bbbar})^2}{\sigma^2}},
%\label{chisquare}
%%\end{equation}
% ++++++++++++++++++++++++++++++++++++++++++++++++++++++++++++++++++++++
%where, $N^{data}_{\bbbar}$ is the number of experimental events for given 
%$\dphi$ bin from Fig.~\ref{unvtx_nonBB_dphi} (right), $N^{pred}_{\bbbar}$ is
%prediction described by Eq.~(\ref{bbbar_equation}) and $\sigma^2$ is 
%statistical uncertainty of data and {\sc pythia} samples
%calculated as $\sigma^2$ = $\sigma^2_{data}$+$K_{FC}\sigma^2_{FC}$+
%$K_{FE}\sigma^2_{FE}$+ $K_{GS}\sigma^2_{GS}$. The coefficients $K_{FC,FE,GS}$
%are to account the scale factors for each production mechanism and $\bp$ 
%related normalization.
%Figure~\ref{unvtx_FitDphi3} shows the fit results with the three QCD production
%processes: FC+FE+GS.  
     The result of the fit is given in Table~\ref{dphi_fitParams3}.
% ++++++++++++++++++++++++++++++++++++++++++++++++++++++++++++++++++++++
%\begin{figure}[htbp]
%\centerline{
%\epsfxsize3.2in\epsffile{bb_eps/pytV2_dPhiUnvPsiMu_FEfree_noTemplates.eps}
%\epsfxsize4.5in\epsffile{bb_eps/pytV2_dPhiUnvPsiMu_FEfree_noTmpl_v2.eps}}
%\caption{Fits of the $\dphi$ distribution of the unvertexed $\jpsiplusmu$
%data with the three QCD production mechanisms.}
%\label{unvtx_FitDphi3}
%\end{figure}
% ++++++++++++++++++++++++++++++++++++++++++++++++++++++++++++++++++++++
% ++++++++++++++++++++++++++++++++++++++++++++++++++++++++++++++++++++++
\begin{table}[tbp]
\caption{Results of the least-squares fit of the $\dphi$ distribution of the 
unvertexed-$\jpsiplusmu$ data with the three QCD production processes.}  
%The expected value for C$_{norm}$ is 0.76$\pm$0.07.  A contribution from 
%flavor excitation is rejected by the fit.}
\begin{center}
\begin{tabular}{lc}
\hline\hline
                & FC+FE+GS            \\
\hline
$C$             & 0.70$\pm$0.03       \\
$S_{\textit{FC}}$    & 3-$S_{FE}$-$S_{GS}$ \\
$S_{\textit{FE}}$    & -0.11$\pm$0.10      \\
$S_{\textit{GS}}$    &  1.60$\pm$0.07      \\
$\chi^{2}$/ndf  & 38.5/33             \\
\hline\hline
\end{tabular}
\end{center}
\label{dphi_fitParams3}
\end{table}
% ++++++++++++++++++++++++++++++++++++++++++++++++++++++++++++++++++++++
  The least-squares fit disfavors a contribution from the FE process by 
returning $S_{\mathit FE}$ of --0.11$\pm$0.10.  A linear combination of 
FC and GS terms gives a reasonable least-squares fit to the data.  The fitting 
function for the FC plus GS combination is shown in 
Eq.~(\ref{bbbar_equationFCandGS}), 
% ++++++++++++++++++++++++++++++++++++++++++++++++++++++++++++++++++++++
\begin{eqnarray}
N_{\bbbar}&=& C(S_{\textit{FC}}N^{\textit{FC}}_{b\bar{b}} + S_{\textit{GS}}N^{\textit{GS}}_{b\bar{b}})\nonumber \\
& &\times \frac{N_{\bp}} {S_{\textit{FC}}N^{\textit{FC}}_{\bp} + S_{\textit{GS}}N^{\textit{GS}}_{\bp} }
\,,
\label{bbbar_equationFCandGS}
\end{eqnarray}
% ++++++++++++++++++++++++++++++++++++++++++++++++++++++++++++++++++++++
where the sum of $S_{\textit{FC}}+S_{\textit{GS}}=2$.
Numerical results from the fit shown in Fig.~\ref{fig:unvtx_FitDphi} are as 
follows: $C = 0.73\pm0.01$, $S_{\textit{GS}} = 1.02\pm0.03$, 
$S_{\textit{FC}} = 2-S_{GS}$. 
%given in Table~\ref{tab:dphi_fitParams}. 
The factors $S_{\textit{FC}}$, $S_{\textit{GS}}$, and $C$ together with 
Eq.~(\ref{bbbar_equationFCandGS}) are used in Sec.~\ref{sec:bbbar_result_sys} 
to calculate the number of $\bbbar$ background events.
% ++++++++++++++++++++++++++++++++++++++++++++++++++++++++++++++++++++++
% ++++++++++++++++++++++++++++++++++++++++++++++++++++++++++++++++++++++

% ++++++++++++++++++++++++++++++++++++++++++++++++++++++++++++++++++++++
%\begin{table}[tbp]
%\caption{Fit results for the data shown in Fig.~\ref{fig:unvtx_FitDphi}.}
%The expected value for C$_{norm}$ is 0.76$\pm$0.07. The fit parameters shown 
%are used to calculate the number of $\bbbar$ background events.}
%\begin{center}
%\begin{tabular}{lcc}
%\hline\hline
%               & FC+GS       \\
%\hline
%C$_{norm}$      & 0.73$\pm$0.01 \\
%$S_{FC}$        & 2-$S_{GS}$    \\
%$S_{FE}$        & not used      \\
%$S_{GS}$        & 1.02$\pm$0.03 \\
%$\chi^{2}$/ndf  & 39.6/34       \\
%\hline\hline
%\end{tabular}
%\end{center}
%\label{tab:dphi_fitParams}
%\end{table}
% ++++++++++++++++++++++++++++++++++++++++++++++++++++++++++++++++++++++
% ++++++++++++++++++++++++++++++++++++++++++++++++++++++++++++++++++++++
\begin{table*}[tbp]
\caption{Systematic uncertainties in the number of $\bbbar$ background events 
in the $\jpsiplusmu$ mass ranges \lowin, \sigwin, and greater than 6~$\gevc$.}
\begin{center}
\begin{tabular}{lcccc}
\hline\hline
$\bbbar$ sys               & \lowin & \sigwin & \hiwin   \\
\hline
$S_{\textit{FE}}$ = 0.0 or 0.1  & -0.3 & -4.9 & -3.0  \\
Misidentified muon increased      & -0.1 & -1.5 & -0.9  \\
Misidentified muon reduced        & +0.2 & +2.7 & +1.7  \\
\hline
 Total                     & $\pm$0.4 & $\pm$5.8 & $\pm$3.6 \\
\hline\hline
\end{tabular}
\end{center}
\label{tab:bbbarSyst_items}
\end{table*}
% ++++++++++++++++++++++++++++++++++++++++++++++++++++++++++++++++++++++
% ++++++++++++++++++++++++++++++++++++++++++++++++++++++++++++++++++++++
\begin{table*}[tbp]
\caption{Expected numbers of $\bbbar$ background events in the signal 
region. The uncertainties are statistical only and their sources include the 
sizes of the trimuon systems, the number of $\bp$ events, and the statistical 
uncertainty of the scale factors. The value of $C$ returned by the 
fit is 0.73$\pm$0.01, while the expected one is 0.76$\pm$0.07.}
\begin{center}
\begin{tabular}{lcccc}
\hline\hline
$\bbbar$ background    & $N_{\bbbar}$(MC) & $S_{i}$ & $N_{\bp}$ (MC) & $N_{\bbbar}$  \\
\hline
FC & 36.5  & $2-S_{GS}$     & $16275\pm 130$ & $12.9\pm 4.1$          \\
FE & 185   & 0              & $35464\pm 189$ & 0        \\
GS & 443.5 & $1.02\pm 0.03$ & $12602\pm 118$ & $165.7\pm 11.7$         \\
\hline
Total                  &         & -            &    -           & $178.6\pm 12.4$            \\
\hline\hline
\end{tabular}
\end{center}
\label{tab:bbbar_results}
\end{table*}
% ++++++++++++++++++++++++++++++++++++++++++++++++++++++++++++++++++++++

     One source of systematic uncertainty in the determination of the $\bbbar$ background arises from the choice to force the contribution of FE to be zero.  
%The fit results using all three QCD components given in 
%Table~\ref{dphi_fitParams3} indicate that the value $S_{FE}=-0.11\pm0.10$ is 
% unphysical since the smallest possible physical value is zero. 
%As an estimate of the systematic uncertainty introduced by choosing 
% $S_{FE}=0$, we use the difference 
We estimate the corresponding systematic uncertainty using the difference
between the predicted number of $\bbbar$ events for 
the two values 0 and 0.1 for $S_{\textit{FE}}$.  A second source is introduced by 
the uncertainty in the estimate of the unvertexed misidentified-muon component 
of the $\dphi$ distribution of unvertexed-$\jpsiplusmu$ events. The 
misidentified-muon component is removed prior to fitting the PYTHIA 
predictions to the data; hence, its uncertainty propagates into the 
determination of the $\bbbar$ background. To determine this systematic 
uncertainty, the $\dphi$ distribution of unvertexed-$\jpsiplusmu$ pairs shown 
in Fig.~\ref{fig:unvtx_nonBB_dphi}(c) is increased and decreased by the amount 
of the lower and upper values of the unvertexed misidentified-muon systematic 
uncertainty, respectively. The scale factors are refit for these two cases and 
the change in the predicted $\bbbar$ background is determined. 
The systematic uncertainties from these two sources are summarized in
Table~\ref{tab:bbbarSyst_items}.  The total systematic uncertainty is 
calculated by adding the results from the three rows in quadrature.

\subsubsection{Results for the $\bbbar$ background} 
\label{sec:bbbar_result_sys}
%%%%%%%%%%%%%%%%%%%%%%%%%%%%%%%%%%%%%%%%%%%%%%%%%%%%%%%%%%%%
% ++++++++++++++++++++++++++++++++++++++++++++++++++++++++++++++++++++++
\begin{figure}[tbp]
\centerline{
\makebox{\includegraphics[width=1.0\hsize]{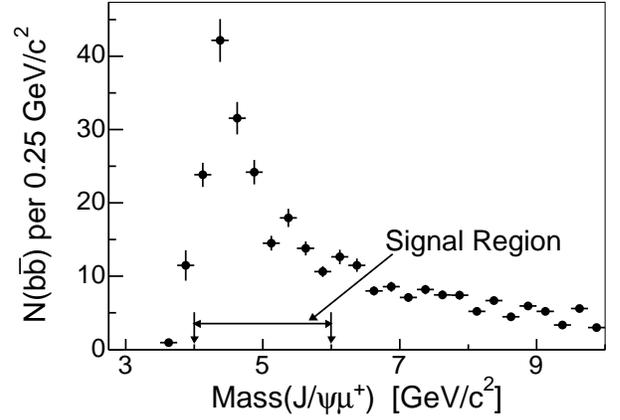}}}
\caption{Invariant-mass distribution of the $\bbbar$ background determined from a PYTHIA MC simulation constrained by the experimental data. The
error bars represent the statistical uncertainties. }
\label{fig:bbbar_jpsiMuMass}
\end{figure}
% ++++++++++++++++++++++++++++++++++++++++++++++++++++++++++++++++++++++
% ++++++++++++++++++++++++++++++++++++++++++++++++++++++++++++++++++++++
\begin{table*}[Htbp]
\caption{Total background for $\bctojpsimux$ decays in three 
invariant-mass ranges. The doubly misidentified contribution is
subtracted from the total to avoid double counting.  
%The statistical uncertainty is given first, 
%then the systematic uncertainty.  
Entries with no 
statistical uncertainties listed represent determinations for which the 
statistical uncertainty is negligible compared with the systematic 
uncertainty. Enries with no systematic uncertainties are estimated to have negligible systematic uncertainties compared with the statistical errors. 
%Since the misidentified-$\jpsi$ background is calculated using the dimuon sidebands near the $\jpsi$ invariant mass, the uncertainty is statistical only.
}
\begin{center}
\begin{tabular}{lcccc}
\hline\hline
$\bctojpsimux$ background & \lowin    & \sigwin       & \hiwin   \\
\hline
Misidentified $\jpsi$  & $11.5\pm2.4$(stat) & $96.5\pm6.9$(stat) & $25.0\pm3.5$(stat) \\
Misidentified muon     & $86.7^{+2.4}_{-4.2}$(syst) & $344.4^{+9.6}_{-16.5}$(syst) & $32.1^{+0.9}_{-1.5}$(syst)        \\
Doubly misidentified   & $5.1^{+0.1}_{-0.2}$(syst) & $19.0^{+0.5}_{-0.9}$(syst) & $5.2^{+0.1}_{-0.3}$(syst)        \\
$\bbbar$ background & $12.4\pm2.4$(stat) & $178.6\pm12.4$(stat) & $110.4\pm10.7$(stat)\\
                    & $\pm0.4$(syst) & $\pm5.8$(syst) & $\pm3.6$(syst) \\   
\hline
Total misidentified+$\bbbar$ bg.  & $105.5\pm3.4$(stat) & $600.5\pm14.2$(stat) & $162.3\pm11.3$(stat)\\
                    & $^{+2.4}_{-4.2}$(syst) & $^{+11.2}_{-17.5}$(syst) & $^{+3.7}_{-3.9}$(syst) \\
\hline\hline
\end{tabular}
\end{center}
\label{tab:total_background}  
\end{table*}
% ++++++++++++++++++++++++++++++++++++++++++++++++++++++++++++++++++++++
% ++++++++++++++++++++++++++++++++++++++++++++++++++++++++++++++++++++++
\begin{table*}[tbp]
\caption{$\bctojpsimux$ candidates and background subtractions from Table~\ref{tab:total_background}.}
%\begin{small}
\begin{center}
\begin{tabular}{lccc}
\hline\hline
                            & \lowin      & \sigwin      & \hiwin      \\
\hline
$N(\bctojpsimux)$, reconstr.& $132\pm 11.5$ & $1370\pm 37.0$ & $208\pm 14.4$ \\
Sum of misidentified+$\bbbar$ bg.  & $105.5^{+4.2}_{-5.4}$ & $600.5^{+18.1}_{-22.5}$ & $162.3^{+11.9}_{-12.0}$\\
\hline
$N_{\textrm{obs}}$  & $26.5^{+12.2}_{-12.7}$ & $769.5^{+41.2}_{-43.3}$ & $45.7\pm18.7$ \\
\hline\hline
\end{tabular}
\end{center}
\label{tab:sumBcObserved}
%\end{small}
\end{table*}
% ++++++++++++++++++++++++++++++++++++++++++++++++++++++++++++++++++++++

     Having determined the correct scale factors to use for the PYTHIA 
simulation of the QCD $\bbbar$ processes, the $\bbbar$ background in the 
$\jpsiplusmu$ event sample is calculated using 
Eq.~(\ref{bbbar_equationFCandGS}).  The number of FC and GS events from the 
PYTHIA simulation is determined by requiring that all the simulated 
$\jpsiplusmu$ events satisfy \textit{all} the requirements listed in 
Tables~\ref{tab:jpsi_cuts}--\ref{tab:thirdTrk_cuts} to reconstruct 
the $\bctojpsimunu$ decay.  
%More specifically, this means the vertex probability and $\dphi$ selection 
%requirements for the $\jpsiplusmu$ event candidates.  
In addition to the $\bc$ selection requirements for the PYTHIA sample, 
we require that the third muon does not originate from 
a pion or kaon and that it originates from a different particle than the 
$\jpsi$ originates from. Other than the $\jpsiplusmu$ events, the quantities 
needed for the calculation are the QCD scale factors and $C$, 
%given in Table~\ref{tab:dphi_fitParams}, 
the number of $\bp$ mesons in the data shown in Fig.~\ref{fig:threeTrk_mass}(b), 
and the numbers of $\bptojpsik$  decays produced by the QCD 
processes in the PYTHIA simulation given in 
Sec.~\ref{sec:unvtx_pythiaJpsiMu}. 
%shown in Fig.~\ref{fig:bYield_pythia}. 
A summary of the input quantities and the results for the $\bbbar$ background
in the signal region is given in Table~\ref{tab:bbbar_results}.  The second 
column gives the numbers of $\jpsiplusmu$ events simulated by PYTHIA 
passing the $\bc$ selection requirements after contributions
from the dimuon sideband region are subtracted.
%the $\bc$ selection requirements
%The number of the $\jpsiplusmu$ {\sc pythia} simulated events passing 
%the $\bc$ selection requirements is given in 
%Table~\ref{tab:pythiaVtxJpsiMu_yield}. 
% ++++++++++++++++++++++++++++++++++++++++++++++++++++++++++++++++++++++
%\begin{table}[htbp]
%\caption{The number of the $\jpsiplusmu$ {\sc pythia} simulated events passing
%the $\bc$ selection requirements for the signal and sideband dimuon invariant 
%mass regions. Each number represents the $\jpsiplusmu$ events for the signal
%mass region, \sigwin.}
%\begin{center}
%\begin{tabular}{lccc}
%\hline\hline
%              & Flavor Creation & Flavor Excitation & Gluon Splitting \\
%\hline
%$\jpsi$ Signal        & 54      & 230               & 471             \\
%$\jpsi$ Sideband      & 17.5    & 45                & 27.5            \\
%\hline
%Sideband subtracted   & 36.5    & 185               & 443.5           \\
%\hline\hline
%\end{tabular}
%\end{center}
%\label{tab:pythiaVtxJpsiMu_yield}
%\end{table}
% ++++++++++++++++++++++++++++++++++++++++++++++++++++++++++++++++++++++
%Thus, in the FC process 36.5 events are within the signal mass window, in 
%the FE there are 185 events, and in the GS, 443.5 events. The final results 
%for the $\bbbar$ background in the signal region are summarized in 
%Table~\ref{tab:bbbar_results}. 
The uncertainty in the $\bbbar$ background 
is due to several sources.  There are statistical uncertainties in the yields
of the four simulated samples $N^{\textit{FC}}_{\bbbar}$, $N^{\textit{GS}}_{\bbbar}$, 
$N^{\textit{FC}}_{\bp}$, and $N^{\textit{GS}}_{\bp}$, and in 
the determination of the $\bptojpsik$ sample in the experimental data. 
Finally, there are correlated uncertainties in the parameters  $C$, 
$S_{\textit{FC}}$, and $S_{\textit{GS}}$ that are determined by the fit to the $\dphi$ 
distribution in the unvertexed-$\jpsiplusmu$ sample.  The resulting 
invariant-mass distribution of the $\bbbar$ background is shown in 
Fig.~\ref{fig:bbbar_jpsiMuMass}.
 
     The total $\bbbar$ background event yields in the invariant-mass ranges 
3--4~GeV/$c^2$, 4--6~GeV/$c^2$, and greater than 6~\gevcc are 
$12.4\pm2.4$(stat)$\pm0.4$(syst), $178.6\pm12.4$(stat)$\pm5.8$(syst), and 
$110.4\pm10.7$(stat)$\pm3.6$(syst), respectively.

%%%%%%%%%%%%%%%%%%%%%%%%%%%%%%%%%%%%%%%%%%%%%%%%%%%%%%%%%%%%
\subsection{Total background} 
\label{sec:tot_bkg}
%%%%%%%%%%%%%%%%%%%%%%%%%%%%%%%%%%%%%%%%%%%%%%%%%%%%%%%%%%%%

     The backgrounds to the $\bctojpsimux$ decays discussed above are 
summarized in Table~\ref{tab:total_background} with their statistical and 
systematic uncertainties.  The misidentified-$\jpsi$ 
background, misidentified-muon background, and $\bbbar$ background are
included.  The doubly misidentified background contribution is
subtracted %from the total 
to avoid double counting. Entries with no 
statistical uncertainties listed represent determinations for which the 
statistical uncertainty is negligible compared with the systematic 
uncertainty.  The misidentified-$\jpsi$ background is calculated using 
the dimuon sidebands near the $\jpsi$ invariant mass.  Since there are no systematic uncertainties that are significant, the uncertainty is only
statistical.

The number $N_{\textrm{obs}}$ of $\bctojpsimux$ signal candidates 
%above the backgrounds 
%described above 
is presented in Table~\ref{tab:sumBcObserved}. The statistical and systematic 
uncertainties are combined in quadrature.
The top row in Table~\ref{tab:sumBcObserved} reports the 
number of reconstructed $\bctojpsimux$ candidates shown in Fig.~\ref{fig:threeTrk_mass}a. The quantity $N_{\textrm{obs}}$ is used to calculate the final
$\bctojpsimunu$ yield.

%%%%%%%%%%%%%%%%%%%%%%%%%%%%%%%%%%%%%%%%%%%%%%%%%%%%%%%%%%%%
\section{Contributions to \boldmath{$\protect\bctojpsimux$} 
from other \boldmath{$\protect\bc$} decays} 
\label{sec:odm}
%%%%%%%%%%%%%%%%%%%%%%%%%%%%%%%%%%%%%%%%%%%%%%%%%%%%%%%%%%%%
% ++++++++++++++++++++++++++++++++++++++++++++++++++++++++++++++++++++++
\begin{table*}[Htbp]
\caption{$\bc$ decay modes and their BF$_1$ from the theoretical predictions of Kiselev~\cite{Ref:Kiselev}. 
The BF$_2$ column represents other decays and associated branching 
fractions necessary to reach the trimuon system. The ``product BF 
Kiselev'' represents the product BF$_1$BF$_2$ for the Kiselev 
predictions~\cite{Ref:Kiselev}, and the sum is normalized to 1. 
The ``Ivanov'' column represents a similar sum based on the theoretical 
predictions of Ivanov~\cite{Ref:Ivanov}.}
\begin{center}
\begin{tabular}{lcccll}
\hline\hline
$\bc$ decay mode &  BF$_1$ pred & Secondary decay mode & BF$_2$ & Product BF \\
&&&& Kiselev & Ivanov \\
\hline
$\jpsiplusmunu$      & 0.01900 & None & & 0.8424 & 0.8872 \\
$\psi(2S)\mu^+\nu$   & 0.00094 & $\psi(2S)\to J/\psi$+... & 0.595   & 0.0248 & 0.0017 \\
$B^{0}_{s}\mu^+\nu$  & 0.04030 & $B^{0}_{s}\to J/\psi$+... & 0.0137 & 0.0245 & 0.0065 \\
$B^{*0}_{s}\mu^+\nu$ & 0.05060 & $B^{*0}_{s}\to J/\psi$+... & 0.0137 & 0.0307 & 0.0139 \\
$B^{0}\mu^+\nu$      & 0.00340 & $B^{0}\to J/\psi$+... & 0.0109	  & 0.0016 & 0.0003 \\
$B^{*0}\mu^+\nu$     & 0.00580 & $B^{*0}\to J/\psi$+... & 0.0109    & 0.0028 & 0.0003 \\
$\jpsi\tau\nu$       & 0.00480 & $\tau\to\mu$+... & 0.178 	  & 0.0378 & 0.0373 \\
$\psi(2S)\tau\nu$    & 0.00008 & $\psi(2S)\to J/\psi$+... \\
&& $\tau\to\mu$+... &  0.595*0.178 			  & 0.0004 & 0.0000 \\
$\jpsi D^{+}_{s}$    & 0.00170 & $D^{+}_{s}\to\mu$+... & 0.0864	  & 0.0065 & 0.0126 \\
$\jpsi \dst_{s}$     & 0.00670 & $D^{*+}_{s}\to\mu$+... & 0.0864    & 0.0257 & 0.0359 \\
$\jpsi D^{+}$        & 0.00009 & $D^{+}\to\mu$+... & 0.168	  & 0.0007 & 0.0011 \\
$\jpsi \dst$         & 0.00028 & $D^{*+}\to\mu$+... & 0.168	  & 0.0021 & 0.0032 \\
\hline\hline
\end{tabular}
\end{center}
\label{tab:branch_frac}
\end{table*}
% ++++++++++++++++++++++++++++++++++++++++++++++++++++++++++++++++++++++
% ++++++++++++++++++++++++++++++++++++++++++++++++++++++++++++++++++++++
\begin{table*}[Htbp]
\caption{Trimuon survival fractions for the various decay modes 
using the product of branching fractions based on the 
predictions of Kiselev {\cal {B}}(K)~\cite{Ref:Kiselev}. The event fractions 
for each decay 
are determined from the MC simulation with the number of surviving 
events shown at the bottom of each column.  The fractions in each column add 
to 1.0.}
\begin{center}
\begin{tabular}{lcccc}
\hline\hline
$\bc$ decay mode & {\cal {B}}(K) & \lowin  & \sigwin & \hiwin  \\
\hline
$\jpsiplusmunu$      & 0.8424 & 0.9007 & 0.9612 & 1.0 \\
\hline
$\psi(2S)\mu^+\nu$   & 0.0248 & 0.0251 & 0.0200 & 0   \\
$B^{0}_{s}\mu^+\nu$  & 0.0245 & 0.0114 & 0.0001 & 0   \\
$B^{*0}_{s}\mu^+\nu$ & 0.0307 & 0.0160 & 0       & 0   \\
$B^{0}\mu^+\nu$      & 0.0016 & 0       & 0       & 0   \\
$B^{*0}\mu^+\nu$     & 0.0028 & 0.0011 & 0       & 0   \\
$\jpsi\tau^+\nu$     & 0.0378 & 0.0411 & 0.0110 & 0   \\
$\psi(2S)\tau^+\nu$  & 0.0004 & 0.0011 & 0.0001 & 0   \\
$\jpsi D^{+}_{s}$    & 0.0065 & 0       & 0.0017 & 0   \\
$\jpsi \dst_{s}$     & 0.0257 & 0.0034 & 0.0056 & 0   \\
$\jpsi D^{+}$        & 0.0007 & 0       & 0.0001 & 0   \\
$\jpsi \dst$         & 0.0021 & 0       & 0.0003 & 0   \\
\hline
Total 3$\mu$ events  &        &  876    & 28342   & 1301\\
\hline\hline
\end{tabular}
\end{center}
\label{tab:otherDM_list}
\end{table*}
% ++++++++++++++++++++++++++++++++++++++++++++++++++++++++++++++++++++++
  
     After subtracting backgrounds, the trimuon sample still contains contributions from other $\bc$ decay
modes, in addition to the decay $\bctojpsimunu$.  For example, a $\bc$
might decay into a $\psi(2S) \mu^+ \nu$ state, followed by the $\psi(2S)$
decay into a $\jpsi\, \pi^+ \pi^-$ final state.  Another example is a $\bc$ 
decay into $\jpsi \,\tau^+ \nu$ state followed by the $\tau$ decay into a muon and two neutrinos.
The fraction of these events that meets the selection requirements is small 
but nonzero.
    
    We consider a set of $\bc$ decay modes taken from the 
theoretical predictions of Kiselev~\cite{Ref:Kiselev}. 
Table~\ref{tab:branch_frac} shows the list of the $\bc$ 
decay modes and their branching fractions  used in the MC 
simulation.  Another set of theoretical $\bc$ decay modes that is sufficiently 
complete to allow an estimate of the number of events in our signal sample from other $\bc$ decay modes is given by Ivanov and collaborators~\cite{Ref:Ivanov}.  The  difference in the estimate of the mumber of events from other decays modes from these two bodies of work is used to estimate the systematic uncertainty in this correction.  The correction is small, approximately 30 events, but the two sets of branching-fraction predictions differ by approximately 50\% of the correction.  Using BGENERATOR~\cite{Ref:jpsi_inclusive}, we generate 
%  $2.2~\times~10^{7}$ $\bc$ decays including 
$\bctojpsimunu$ decays and eleven other decay modes that can yield trimuon 
events. The fraction of these events that meets the selection requirements is reported in Table~\ref{tab:otherDM_list}. 
% ++++++++++++++++++++++++++++++++++++++++++++++++++++++++++++++++++++++
\begin{table*}[tbp]
\caption{Final numbers of $\bctojpsimunu$, $N_{\bc}$.  The statistical and 
systematic errors are combined. The last row presents the number of simulated 
$\bctojpsimunu$ events in the three mass regions. They are scaled so 
that the number in the signal region is consistent with the experimental 
data.  The MC sample's  %is sufficiently large that its 
statistical uncertainties are small compared with the statistical uncertainties
 in the experimental data.}
\begin{small}
\begin{center}
\begin{tabular}{lccc}
\hline\hline
Mass range ($\gevcc$)      & \lowin      & \sigwin      & \hiwin      \\
\hline
$N_{\textrm{obs}}$  & $26.5^{+12.2}_{-12.7}$ & $769.5^{+41.2}_{-43.3}$ & $45.7\pm18.7$ \\
Other decay modes &  $2.6\pm1.9$ & $30.0\pm16.4$         & 0             \\
\hline			    
$N_{\bc}$         & $23.9^{+12.3}_{-12.8}$ & $739.5^{+44.3}_{-46.3}$ & $45.7\pm18.7$ \\
$N(\bctojpsimunu)$, MC &$22.8\pm0.6$ & $739.5$%\pm3.7$ 
& $27.6\pm0.6$ \\
\hline\hline
\end{tabular}
\end{center}
\label{tab:bc_bkgSubtracted}
\end{small}
\end{table*}
% ++++++++++++++++++++++++++++++++++++++++++++++++++++++++++++++++++++++
    
     Our method uses the number $N_{\textrm{obs}}$ of observed $\bc$ candidates in 
the data as shown in Table~\ref{tab:sumBcObserved} after all other backgrounds 
have been subtracted except for the other decay modes. In the signal region 
4--6~GeV/$c^2$, we observe $N_{\textrm{obs}} = 769.5$ events. The number of 
$\bctojpsimunu$ events in the data is given by 
$N_{\bc}=N_{\textrm{obs}}-N_{\textrm{other}}$ where $N_{\textrm{other}}$ is the number due 
to other decay modes. This can be rewritten as 
$$N_{\textrm{other}}=N_{\textrm{obs}}\left(1-\frac{N_{\bc}}{N_{\textrm{obs}}}\right)\,.$$
The fraction $N_{\bc}$/$N_{\textrm{obs}}$ equals 0.961
%$$\frac{N_{\bc}}{N_{\textrm{obs}}=0.961$$ 
for the signal region 4--6~GeV/$c^2$ and is given in 
Table~\ref{tab:otherDM_list}.  
%For the signal region 4-6~GeV/$c^2$, it is $$\frac{N_{\bc}}{N_{\textrm{obs}}=0.961\,.$$  study shows that
%the surviving fractions of the other decay modes combined with respect to 
%the $\bctojpsimunu$ channel are: 9.9\% in the $\lowin$ mass range, 3.9\% in
%the signal mass range, and no contribution in the $\hiwin$ mass range. 
Thus, $N_{\textrm{other}}$ is %769.5$\times (1-0.961) = 
30.0$\pm1.6$(stat) events in the signal 
region and %26.5$\times (1-0.901) = 
2.6$\pm1.2$(stat) events in the \lowin mass 
range. %Note that using the %surviving fractions of the other decay modes based on the
%Ivanov predictions (BF(I))~\cite{Ref:Ivanov} give for $N_{other}$ %769.5$\times$(1-0.974)= 
%20.0 events in the signal mass range.  This difference indicates that we can expect the systematic uncertainty in $N_{other}$ to be of the order of 10 events (see Sec.~\ref{sec:bc_bkg_sys}).
The difference between the Kiselev and Ivanov predictions for the $\bctojpsimunu$ branching fraction is 9\%~\cite{Ref:Kiselev,Ref:Ivanov}. This results in a systematic uncertainty of $\pm16.3$ events in $N_{\bc}$.
%at the level of 2.2$\%$ of $N_{B^+_c}$ or 16.3 events.

%%%%%%%%%%%%%%%%%%%%%%%%%%%%%%%%%%%%%%%%%%%%%%%%%%%%%%%%%%%%
\section{\boldmath{$\protect\bc$} Signal} 
\label{sec:bc_excess}
%%%%%%%%%%%%%%%%%%%%%%%%%%%%%%%%%%%%%%%%%%%%%%%%%%%%%%%%%%%%
% ++++++++++++++++++++++++++++++++++++++++++++++++++++++++++++++++++++++
\begin{figure}[tbp]
\centerline{
\makebox{\includegraphics[width=1.0\hsize]{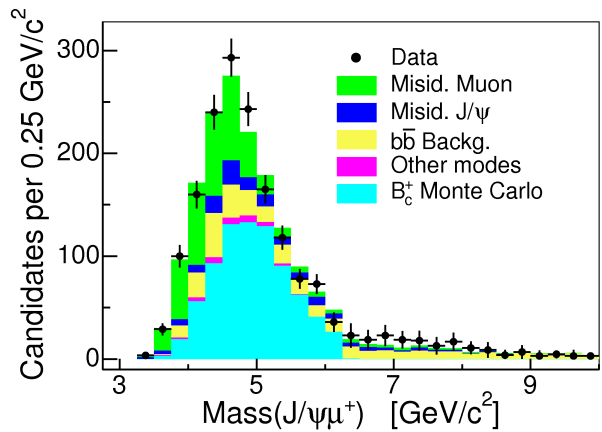}}}
\caption{Invariant-mass distribution of the $\bctojpsimu$ candidate 
events using the full CDF Run II data sample with a MC simulated signal 
sample and the calculated backgrounds superimposed.  Details of the contributions are described in the main text.
%``Misid. Muon'' is the misidentified-muon background corrected for the doubly misidentified background, while ``Other modes'' indicates the contribution from the the other decay modes. ``$B_{c}$ Monte Carlo'' stands for the simulated $\bctojpsimunu$ decays.  The simulated sample size is normalized to the number of signal events in the signal region after subtracting background plus other decay modes.  
The error bars are the statistical uncertainties on the data and background predictions combined.}
\label{fig:bc_bgen_bkgSuperimposed}
\end{figure}
% ++++++++++++++++++++++++++++++++++++++++++++++++++++++++++++++++++++++

%The $N_{\textrm{obs}}$ from Table~\ref{tab:sumBcObserved} is used as an input to calculate the other decay modes contribution. 
     The estimated number of events from other decay modes that contribute to 
the $\bctojpsimux$ signal and sidebands, observed $N_{\textrm{obs}}$ and the final 
number $N_{\bc}$ of $\bctojpsimunu$, are shown in 
Table~\ref{tab:bc_bkgSubtracted}.  The statistical and systematic 
uncertainties are combined. 
The result for $N_{\bc}$ in the \lowin and greater than 6~\gevcc mass regions 
compared with the number of simulated $\bctojpsimunu$ events in these regions 
yields an important cross-check on the overall size of the experimental 
backgrounds in the 4--6 $\gevcc$ signal region.  The \lowin and greater than 
6~\gevcc mass regions are populated predominantly by background, while the 
signal region has 740 $\bctojpsimunu$ decays and 630 background events 
including the other decay modes.  By normalizing a Monte Carlo sample of 
$\bctojpsimunu$ events to the measured number of events after background 
subtraction in the signal region, we predict the expected number of 
$\bctojpsimunu$ decays in the \lowin and greater than 6~\gevcc mass regions.  
%If these numbers are consistent with the observed numbers in these mass 
%regions, it gives confidence in the technique. %that we correctly calculated 
%the sum of backgrounds plus other decay modes in these regions.  
From Table~\ref{tab:bc_bkgSubtracted}, we expect 23 $\bctojpsimunu$ decays
and observe 24$\pm$12 in the \lowin mass region.  In the greater than 6~\gevcc 
region, we expect 28 and observe 46$\pm$19.  This gives confidence in the 
calculation of the sum of background yield plus other decay modes in the 
signal region. 

     The invariant-mass distribution of the $\jpsiplusmu$ events
is shown in Fig.~\ref{fig:bc_bgen_bkgSuperimposed} with 
simulated signal and backgrounds superimposed. ``Misid. muon''
is the misidentified-muon background corrected for the doubly
misidentified background, while ``other modes'' indicates
the contribution from the other decay modes. ``$B_{c}$ Monte
Carlo'' stands for simulated $\bctojpsimunu$ decays.  The
simulated sample size is normalized to the number of signal events in the 
signal region after subtracting background and other decay modes.  
%It is to be noted again that 
After accounting for the small $\bctojpsimunu$ signal component in the \lowin and greater than 6~\gevcc mass regions, we correctly model the background in these regions.

%%%%%%%%%%%%%%%%%%%%%%%%%%%%%%%%%%%%%%%%%%%%%%%%%%%%%%%%%%%%
\section{Relative efficiency of \boldmath{$\protect\bptojpsik$} 
to \boldmath{$\protect\bctojpsimunu$}} 
\label{sec:rel_eff}
%%%%%%%%%%%%%%%%%%%%%%%%%%%%%%%%%%%%%%%%%%%%%%%%%%%%%%%%%%%%

To determine $\ratio$, we need to determine the efficiencies used in Eq.~(\ref{relcross}). These efficiencies are collected together into $\erel=\epsilon_{\bp}/(\epsilon_{\bc}\times\epsilon_{\mu})$.  
%The ratio of the production cross sections times branching fractions of
%the $\bctojpsimunu$ relative to the production cross section times branching fraction $\bptojpsik$ can be written as
% ++++++++++++++++++++++++++++++++++++++++++++++++++++++++++++++++++++++
%\begin{equation}
%\frac{\sigbrbc}{\sigbrb} = \frac{N_{\bc}}{N_{\bp}}\times\erel,
%\label{Xsec_formula}
%\end{equation}
% ++++++++++++++++++++++++++++++++++++++++++++++++++++++++++++++++++++++
%where $N_{\bc}\equiv N(\bctojpsimunu)$ is found in
%Table~\ref{tab:bc_bkgSubtracted} within the signal mass region \sigwin 
%and $N_{\bp}$ is obtained from the fit to Fig.~\ref{fig:threeTrk_mass}b as described in Sec.~\ref{sec:three_trk}. 
%The relative efficiency $\erel$ is defined as
% ++++++++++++++++++++++++++++++++++++++++++++++++++++++++++++++++++++++
%\begin{equation}
%    \erel = \frac{\epsilon_{\bp}}{\epsilon_{\bc}}\frac{1}{\epsilon_{\mu}}\,,
%\label{erel_formula}
%\end{equation}
% ++++++++++++++++++++++++++++++++++++++++++++++++++++++++++++++++++++++
The efficiencies $\epsilon_{\bp}$ and $\epsilon_{\bc}$ are the geometrical acceptances for $\bptojpsik$ and $\bctojpsimunu$ decays, respectively, in the CDF II detector corrected for effects discussed below, and $\epsilon_{\mu}$ is the third-muon detection efficiency in the CMU and CMP 
detectors. The ratio $\epsilon_{\bp}/\epsilon_{\bc}$
%  $$\frac{\epsilon_{\bp}}{\epsilon_{\bc}}$$
includes a small correction for the relative trigger efficiency between kaons 
and muons. The muon identification efficiency for CMUP muons is 
0.962$\pm$0.007(stat)$\pm$0.021(syst)~\cite{Ref:CMUP_eff}. 
%This efficiency is normalized to the Monte Carlo simulation. 
%Fig.~\ref{fig:cmupEff_vsPt} shows the data to Monte Carlo normalized 
%CMUP efficiency as a function of $\pt$. 
% ++++++++++++++++++++++++++++++++++++++++++++++++++++++++++++++++++++++
%\begin{figure}[htbp]
%\centerline{
%\makebox{\includegraphics[width=0.5\hsize]{feps/prd_CMUPEff_vsPT.eps}}}
%\caption{The data to Monte Carlo normalized CMUP efficiency as function 
%of $\pt$ from Ref.~\cite{Ref:CMUP_eff}}
%\label{fig:cmupEff_vsPt}
%\end{figure}
% ++++++++++++++++++++++++++++++++++++++++++++++++++++++++++++++++++++++
Because the effects due to multiple Coulomb scattering and the stopping of muons in the absorber at low $\pt$ are
modeled accurately by the simulation, the normalized efficiency of the CMUP 
%can be taken as flat 
is uniform over the $\pt$ range greater than 3~$\gevc$.
%as seen in Fig.~\ref{fig:cmupEff_vsPt}.

We determine the efficiencies $\epsilon_{\bp}$ and $\epsilon_{\bc}$ with 
MC simulations.  Knowledge of the transverse momentum %  production 
spectra for the $\bp$ and $\bc$ is essential to determine the relative
efficiency correctly.  
%As most other efficiencies that may be
%incorrectly modeled in the simulation are expected to cancel in the
%ratio of the efficiencies, accurate knowledge of the $\bp$ and $\bc$
%spectra, which do not cancel, are very important to the relative cross
%section measurement.  
In order to determine the relative efficiency,
we use the generated samples of $\bptojpsik$, $\bctojpsimunu$, and
$\bcst\to\bc+\gamma$ decays.  All data from the MC events are passed
through the full detector and trigger simulation.  The events 
that meet the dimuon 
%which survive the di-muon 
trigger requirements are processed in the same way as experimental data.

%%%%%%%%%%%%%%%%%%%%%%%%%%%%%%%%%%%%%%%%%%%%%%%%%%%%%%%%%%%%
\subsection[$\bp$ and $\bc$ $\pt$ spectra]{\boldmath{$\bp$} and \boldmath{$\bc$} \boldmath{$\pt$} spectra}
\label{sec:bc_spectrum}
%%%%%%%%%%%%%%%%%%%%%%%%%%%%%%%%%%%%%%%%%%%%%%%%%%%%%%%%%%%%
   
     The $\bptojpsik$ acceptance calculation is based on the FONLL spectrum~\cite{Ref:FONLL}, where FONLL stands for fixed-order plus next-to-leading logs.  As the FONLL spectrum shows some discrepancies in the low-$\pt$ region with respect to the data, a corrected FONLL spectrum is used. 
%The FONLL spectrum shows some discrepancies in the low $\pt$ region with 
% respect to the our data. Thus, a corrected FONLL spectrum is used to 
% determine the $\bptojpsik$ acceptance.
   
\begin{figure}[Htbp]
\centerline{
\makebox{\includegraphics[width=1.0\hsize]{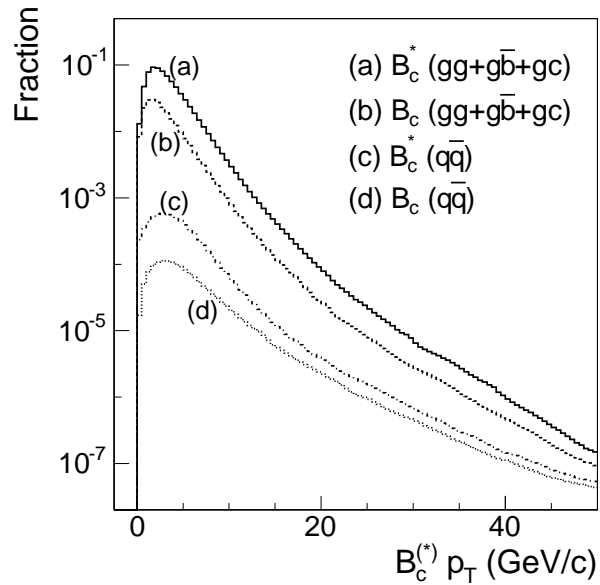}}} 
\caption{Spectra for $\bc$ and $\bcst$ due to various production processes are shown.  The processes are scaled to reflect the weight used in composing the final spectrum.}
\label{fig:bc_bcstSpec}
\end{figure}

     In generating the $\bctojpsimunu$ MC sample, we follow the  
theoretical work on $\bc$ production, the general-mass
variable-flavor-number (GMVFN) model of Chang 
\textit{et al.}~\cite{Ref:BcLatest_Chang}, 
which has the following advantages: it includes $\bc$ and $\bcst$ spectra; 
it includes production via the interactions of gluons and heavy sea quarks, 
$gb$ and $gc$, as well as pure $gg$ fusion; and it includes a small contribution from $q\bar{q}$ production.  Figure~\ref{fig:bc_bcstSpec} 
%Figure~\ref{fig:bcSpec_oldVSnew} illustrates the differences between
%what we have used in CDF previously to describe $\bc$
%production~\cite{Ref:BcExclusive}, which included ground state $\bc$
%decays only, and the updated spectrum.  For comparison, we also show
%the spectrum obtained from the $q\bar{q}$ and $gg$ processes only.
%One can see from the figure that the updated spectrum is softer than
%the previous one.  Figure~\ref{fig:bc_bcstSpec} shows the contribution
%to the updated spectrum from the different production processes for
%$\bc$ and $\bcst$.
shows that the $\bc$ and $\bcst$ spectra are similar, but the $\bc$ produced in $\bcst\to\bc\gamma$ decays is softer than that produced directly.
% ++++++++++++++++++++++++++++++++++++++++++++++++++++++++++++++++++++++
% ++++++++++++++++++++++++++++++++++++++++++++++++++++++++++++++++++++++
% ++++++++++++++++++++++++++++++++++++++++++++++++++++++++++++++++++++++
%\begin{figure}[htbp]
%  \centerline{
%    \includegraphics[width=0.6\hsize]{feps/prd_bcPtSpectraPrev_fig9.eps}
%  }
%  \caption{The $\bc$ $\pt$ spectrum (a) previously used in
%    CDF~\cite{Ref:BcExclusive} is compared with the updated spectrum (b)
%    based on the latest theoretical
%    predictions~\cite{Ref:BcLatest_Chang}.  The updated spectrum (c) with
%    only $gg$ and $q\bar{q}$ processes is shown for comparison.}
%\label{fig:bcSpec_oldVSnew}
%\end{figure}
% ++++++++++++++++++++++++++++++++++++++++++++++++++++++++++++++++++++++
The composition of the $\bc$ spectrum used in this measurement makes
use of the $\bc$ and $\bcst$ cross sections given in Tables I--II of
Ref.~\cite{Ref:BcLatest_Chang}.  According to this calculation,
made for Tevatron energy 1.96~TeV using $\pt(\bc)> 4~\gevc$
and rapidity $|y|< 0.6$, the total production cross
sections for the $\bc$ and $\bcst$ mesons are 0.7 and 2.3 nb,
respectively.  In Table~\ref{tab:bcFrac_Xsect} we present both the combined 
contributions of $gg+g\bar{b}+gc$ and $q\bar{q}$ to $\bc$ and $\bcst$ 
production.
%same
%predictions in the form of the $\bc$ and $\bcst$ production fractions
%relative to the total $\bc$ cross section.

% ++++++++++++++++++++++++++++++++++++++++++++++++++++++++++++++++++++++
\begin{table}[tbp]
\caption{Cross section fractions for $\bc$ and $\bcst$ based
on calculations from Ref.~\cite{Ref:BcLatest_Chang}, where
``$gg+g\bar{b}+gc$'' represents the combined contributions from the $gg$
fusion, $g\bar{b}$ and $gc$ production subprocesses, and $q\bar{q}$
represents the quark-antiquark production mechanism.}
\begin{center}
\begin{tabular}{lccc}
\hline\hline
Production fractions &  & $gg+g\bar{b}+gc$ & $q\bar{q}$  %& Fractions of total $\sigma$                     
\\
\hline
$\bc$                & & 0.994            & 0.006       %& $\frac{\sigma(\bc)}{\sigma(\bc+\bcst)}=0.237$   
\\
$\bcst$              & & 0.991            & 0.009       %& $\frac{\sigma(\bcst)}{\sigma(\bc+\bcst)}=0.763$ 
\\
\hline\hline
\end{tabular}
\end{center}
\label{tab:bcFrac_Xsect}
\end{table}
% ++++++++++++++++++++++++++++++++++++++++++++++++++++++++++++++++++++++
%The new spectrum is composed of 23.7$\%$ $\bc$ and 76.3$\%$ $\bcst$. The $\bc$ is produced $99.4\%$ of the time from $gg+g\bar{b}+gc$ and $0.6\%$ of the time from the $q\bar{q}$ annihilation process.  The $\bcst$ is produced $99.1\%$ of the time from gluon related mechanisms and $0.9\%$ of the time from the $q\bar{q}$ production process. 
The authors of Ref.~\cite{Ref:BcLatest_Chang} provided the $\pt$ and rapidity distributions for both $\bc$ and $\bcst$ mesons from the various production mechanisms.

     In the MC simulation, we assign the $\bcst$ mass to be
$M_{\bc} + 0.076~\gevcc$ based on the theoretically predicted value
from Baldicchi and Prosperi~\cite{Ref:BcStar_mass}. In this work the
authors predict a range of $\bcst$ masses varying with the model
used. We use the highest of the predicted $\bcst$ masses in order to
assign a conservative systematic uncertainty on the amount of $\bcst$ production relative to $\bc$ production. 
%discussed in Section~\ref{sec:sys}.  
The mass difference between the $\bcst$ and
the $\bc$ is too small for $\pi^0$ production.  Consequently, the
$\bcst$ are assumed to decay exclusively to the $\bc \gamma$ final state.  
%The $\bcst$ is assumed to decay promptly (electromagnetically) and the lifetime is set to zero.  
%This assumption is reasonable, since the $\bcst$ decays to the $\bc$ electromagnetically.

%%%%%%%%%%%%%%%%%%%%%%%%%%%%%%%%%%%%%%%%%%%%%%%%%%%%%%%%%%%%
\subsection[Comparison of MC $\bp$ and 
$\bc$ $\pt$ spectra with data]{Comparison of MC \boldmath{$\bp$} and \boldmath{$\bc$} \boldmath{$\pt$} spectra with data}
\label{sec:bp_spectrum}
%%%%%%%%%%%%%%%%%%%%%%%%%%%%%%%%%%%%%%%%%%%%%%%%%%%%%%%%%%%%

\begin{figure}[Htbp]
\centerline{
\makebox{\includegraphics[width=1.0\hsize]{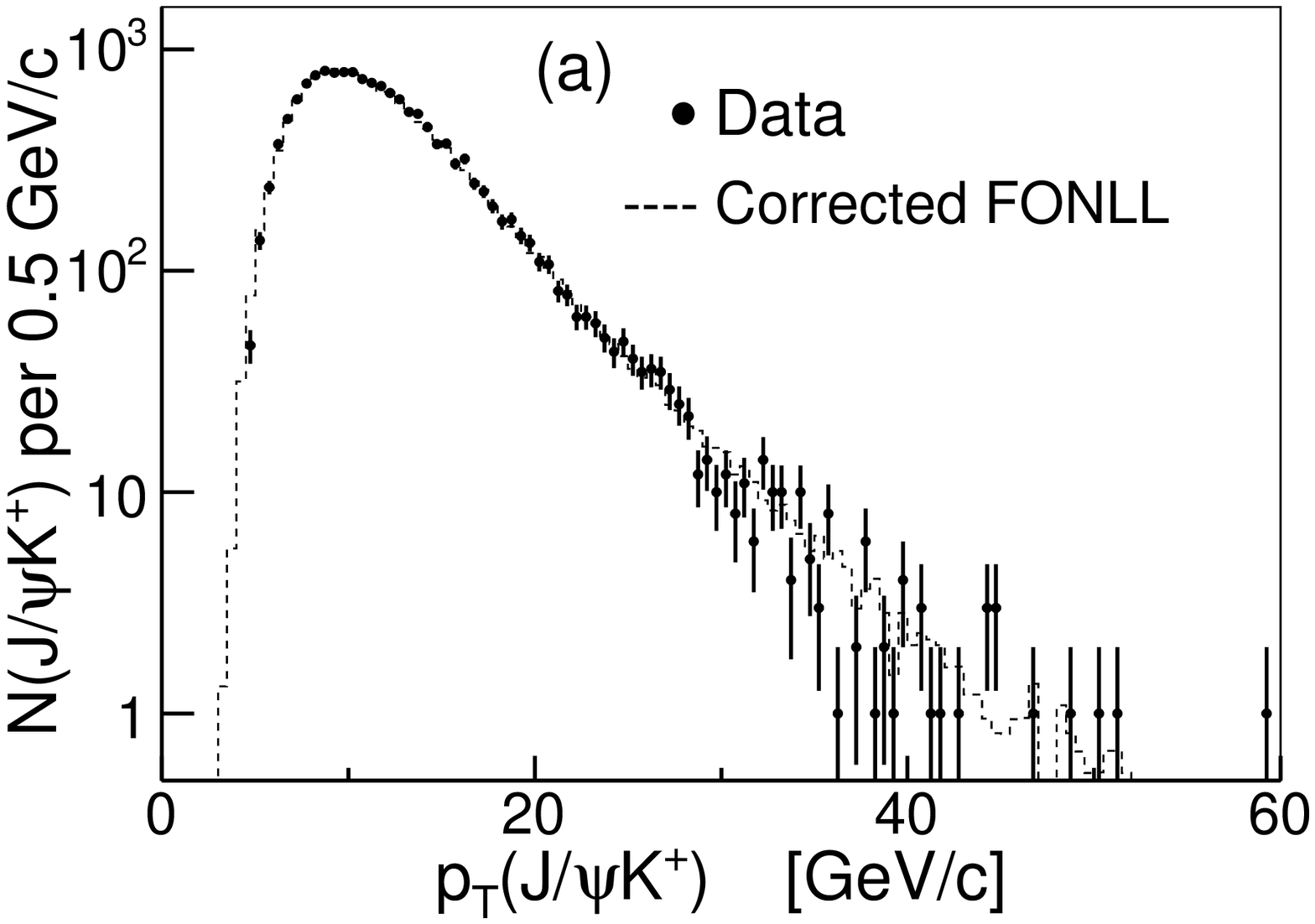}}}
\centerline{
\makebox{\includegraphics[width=1.0\hsize]{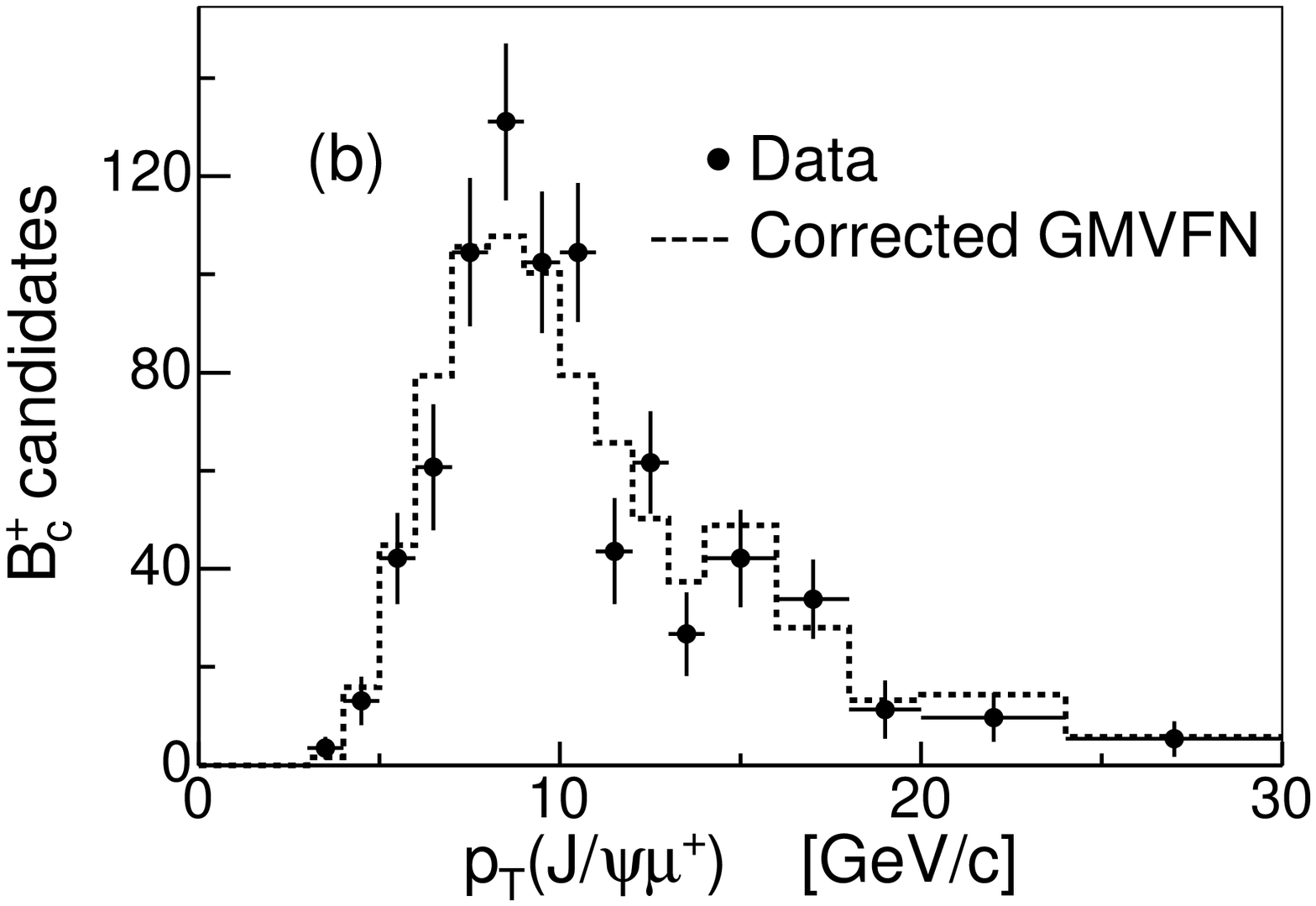}}}
\caption{%The $p_{T}$ spectra 
Transverse momenta distributions for (a) the $\jpsiplusk$ and (b) the $\jpsiplusmu$ samples.  Both data plots are background subtracted, and the 
theoretically predicted spectra are corrected using data.}
\label{fig:ptSpec_bcAndBp}
\end{figure}

The $\bctojpsimunu$ and $\bptojpsik$ samples generated using the corrected
$\pt$ spectra are 
%  The MC generated samples using the corrected theoretically 
%  predicted $\pt$ spectra of the $\bctojpsimunu$ and the $\bptojpsik$ are 
compared with data in Fig.~\ref{fig:ptSpec_bcAndBp}, where the same 
selection requirements are applied to data and simulation. %  Both the 
Experimental data and simulated distributions are selected with the 
requirement that the invariant-mass value should lie within the signal 
mass region \sigwin for the $\bc$ and within $\pm$50 $\mevcc$ of the $\bp$ 
mass for the $\bptojpsik$ decays.  Both $\pt$ 
distributions for data are background subtracted.  The simulated distributions are normalized to the data distributions.
% ++++++++++++++++++++++++++++++++++++++++++++++++++++++++++++++++++++++
% ++++++++++++++++++++++++++++++++++++++++++++++++++++++++++++++++++++++
%%%%%%%%%%%%%%%%%%%%%%%%%%%%%%%%%%%%%%%%%%%%%%%%%%%%%%%%%%%%
\subsection{Results for the relative efficiency}
\label{sec:epsilon_rel}
%%%%%%%%%%%%%%%%%%%%%%%%%%%%%%%%%%%%%%%%%%%%%%%%%%%%%%%%%%%%

In calculating $\erel$, we first determine $\epsilon_{\bc}$ and
$\epsilon_{\bp}$ separately and then calculate the ratio
$\erel=\epsilon_{\bp}/(\epsilon_{\bc}\times\epsilon_{\mu})$ for
$p_T(B)>\, 6$ $\gevc$, where $B$ is the $\bc$ ($\bp$) for 
the $\epsilon_{\bc}$ ($\epsilon_{\bp}$) calculations.  
For $\epsilon_{\bc}$ and $\epsilon_{\bp}$, both the generated and
reconstructed events are determined from a sample with a
generator-level requirement of $\pt(B) > 6~\gevc$.  
The generator-level $\bc$ or $\bp$ events that satisfy $\pt(B)> 6~\gevc$ 
 and $|y|<1.0$ are counted in this sample as the generated events, 
while all events  are passed through the detector and trigger 
simulation with all the analysis selection criteria applied.  Finally, a 
requirement that the $\pt$ be greater than 6~$\gevc$ is applied to the reconstructed $\jpsiplusmu$ in the $\bc$ case and to the reconstructed $\jpsiplusk$ in the $\bp$ case.  For the reconstructed events there is no requirement made on the rapidity.  In both cases 
$\epsilon_{\bc (\bp)}$ is the ratio of reconstructed to generated 
events. 
%The reconstructed events are not a proper subset of the generated events, since reconstruction may push events above or below the $\pt$ threshold. This procedure takes into account detector smearing effects.  It also corrects (?) $\epsilon_{\bc}$ for the effect of the missing neutrino in the $\bc$ decay.  For example, if the $\pt(\jpsiplusmu)$ cut is raised or lowered while holding the $p_T$ cut on the generated events constant, $\epsilon_{\bc}$ will be changed to compensate(?).  The presence of a missing neutrino in the $\bc$ decay is one out of two effects in explaining why $\epsilon_{\bc}$ is smaller than $\epsilon_{\bp}$. Second contribution to this effect is a fact that the predicted $\pt(\bp)$ spectrum from Ref.~\cite{Ref:FONLL} is harder than the $\pt(\bc)$ from Ref.~\cite{Ref:BcLatest_Chang}.

     In the acceptance calculation there is a small correction (approximately 3.4\% in the value of $\erel$) for the fact that
XFT efficiencies in data are different for kaons and muons. Assuming that muons and pions are similar, the model, based on data, parametrizes the XFT efficiency for kaons and pions as a function of $1/\pt$ relative to the same efficiencies as estimated in the MC simulation for the acceptance~\cite{Ref:KarenThesis}.  The muon efficiency $\epsilon_{\mu}$ %given in Eq.~(\ref{erel_formula}) 
depends on the CMU and CMP muon detectors alone and is not included in 
these results.  
%We used from~\cite{Ref:KarenThesis} the XFT efficiency measured in data to that measured in Monte Carlo simulation for kaons and pions as function of inverse $\pt$. We assign the weight for each an event in Monte Carlo as w$_{XFT}$ = a$_0$ + a$_1$/$\pt$, where a$_0$ and a$_1$ are the line parameters known for given run number. Reference~\cite{Ref:KarenThesis} provides us with 5 sets of parameters covering all run numbers. Here we assume that the muon XFT efficiency is approximately equal to the pion XFT efficiency. 

\begin{figure}[Htbp]
\centerline{
\makebox{\includegraphics[width=1.0\hsize]{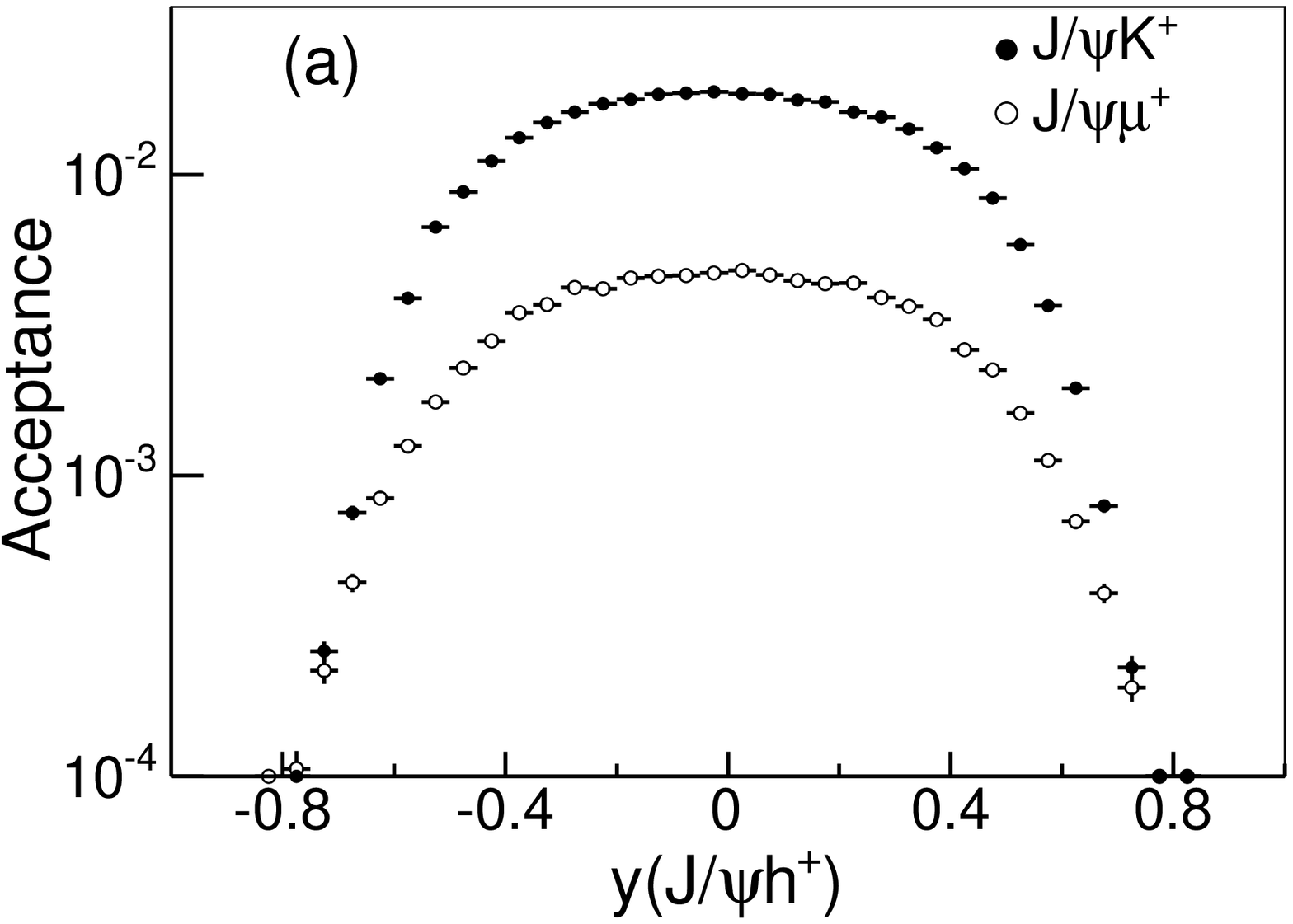}}}
\centerline{
\makebox{\includegraphics[width=1.0\hsize]{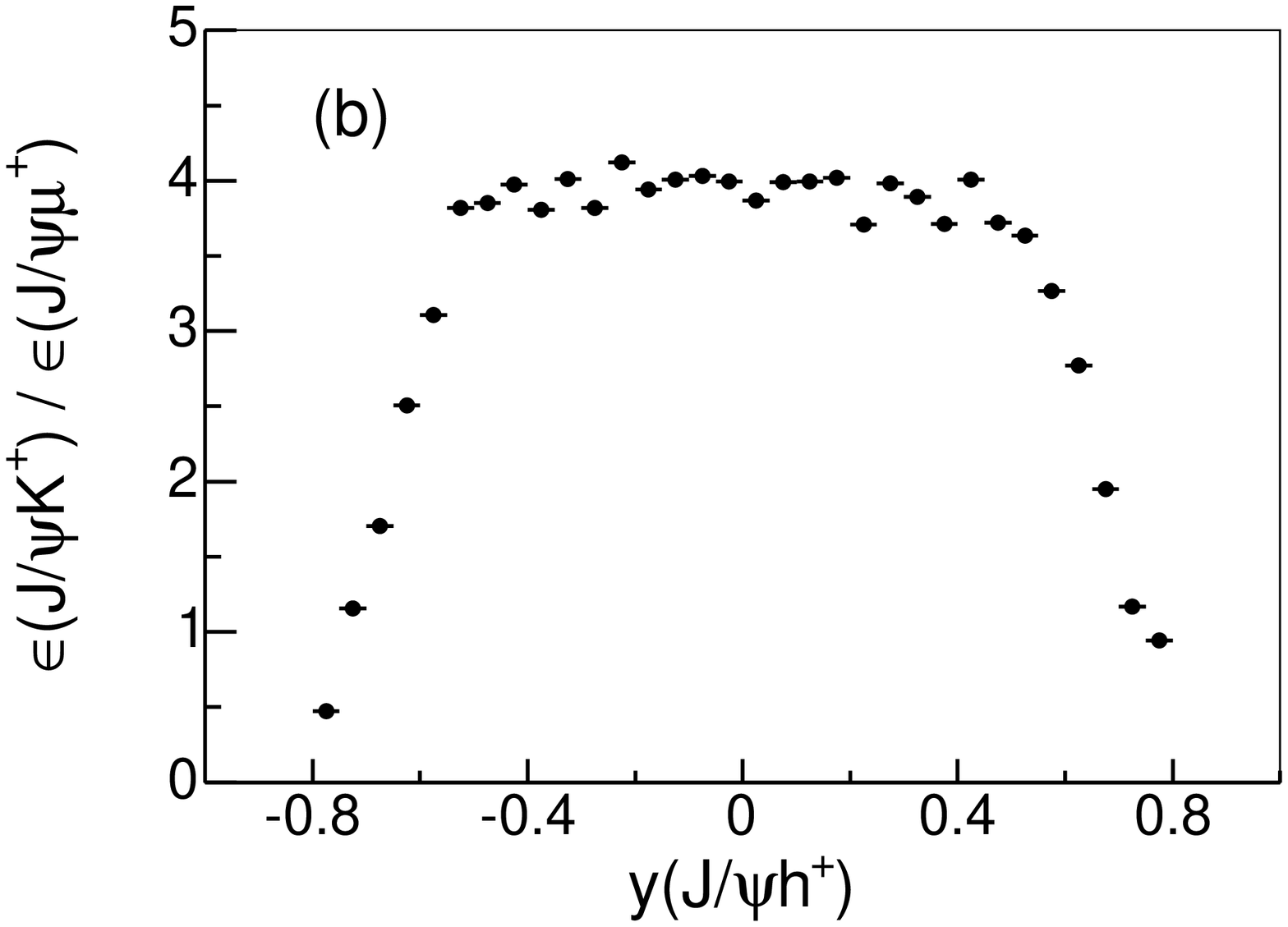}}}
\caption{(a) Comparison of the acceptances for the $\jpsiplusmu$ and 
$\jpsiplusk$ systems and (b) the $\epsilon_{\bp}/\epsilon_{\bc}$ ratio as a
function of the rapidity.}
\label{fig:acceptVSrapidity}
\end{figure}

     The results for the acceptances of $\bctojpsimunu$ decays for the 
various $\bc$ production mechanisms as discussed in 
Sec.~\ref{sec:bc_spectrum} are shown in Table~\ref{tab:acc_bc_gmvfn}.
% ++++++++++++++++++++++++++++++++++++++++++++++++++++++++++++++++++++++
\begin{table*}[Htbp]
\caption{$\bc$ acceptance for different production mechanisms.}
\begin{center}
\begin{tabular}{lcccc}
\hline\hline  
                      & \multicolumn{2}{c}{$gg+g\bar{b}+gc$}    & \multicolumn{2}{c}{$q+\bar{q}$}    \\
Production process    & $\bc$            & $\bcst\to\bc\gamma$ & $\bc$       & $\bcst\to\bc\gamma$ \\
\hline
%Generated events      & 22 095 939       & 22 095 939           & 22 095 939    & 22 095 939              \\
%Signal events         & 39624.7          & 38081.1              & 75564.2     & 55764.9              \\
%\hline
$\epsilon_{\bc}$ (\%) & 0.179$\pm$0.001  & 0.172$\pm$0.001      & 0.342$\pm$0.001 & 0.252$\pm$0.001   \\
\hline\hline
\end{tabular}
\end{center}
\label{tab:acc_bc_gmvfn}
\end{table*}
% ++++++++++++++++++++++++++++++++++++++++++++++++++++++++++++++++++++++
Using the production cross-section fractions for $\bc$ and $\bcst$ given in Table~\ref{tab:bcFrac_Xsect} combined with the predicted production 
cross sections for $\bc$ and $\bcst$ of 0.7 and 2.3 nb, respectively, a 
weighted average of the acceptances is calculated to determine the total 
acceptance $\epsilon_{\bc}$ for $\bctojpsimunu$ presented in 
Table~\ref{tab:acc}. The acceptance $\epsilon_{\bp}$ for $\bptojpsik$ is also 
shown in Table~\ref{tab:acc} and its calculation is simpler because there is 
only one production spectrum involved in its determination.  Both results are 
for $\pt(B)>6~\gevc$.
% ++++++++++++++++++++++++++++++++++++++++++++++++++++++++++++++++++++++
\begin{table}[Htbp]
\caption{Acceptances of $\bc$ and $\bp$ for $\pt>6~\gevc$. Small corrections for different XFT track efficiencies for muons and kaons are applied.}
\begin{center}
\begin{tabular}{lcc}
\hline\hline
                       & $\bptojpsik$        & $\bctojpsimunu$     \\
\hline
%Generated events       & 22 095 939          & 22 095 939            \\
%Signal events          & 152076              & 38619.5             \\
%\hline
$\epsilon_{\bp, \bc}$ (\%)  & $0.688\pm 0.002$  & $0.175\pm 0.001$  \\
\hline\hline
\end{tabular}
\end{center}
\label{tab:acc}
\end{table}
% ++++++++++++++++++++++++++++++++++++++++++++++++++++++++++++++++++++++ 
Comparisons of the acceptances for the $\jpsiplusmu$ and $\jpsiplusk$ 
systems and the $\epsilon_{\bp}/\epsilon_{\bc}$ ratio as a function of the 
rapidity are shown in Fig.~\ref{fig:acceptVSrapidity}.
% ++++++++++++++++++++++++++++++++++++++++++++++++++++++++++++++++++++++
% ++++++++++++++++++++++++++++++++++++++++++++++++++++++++++++++++++++++    
     Using %Eq.~(\ref{erel_formula}), where
$\epsilon_{\bp}$ and  $\epsilon_{\bc}$ from Table~\ref{tab:acc} 
and $\epsilon_{\mu}$ from the opening of Sec.~\ref{sec:rel_eff}, the value of $\erel$ is
% ++++++++++++++++++++++++++++++++++++++++++++++++++++++++++++++++++++++
\begin{equation}
\erel = 4.093\pm0.038\textrm{(stat)}.
%= \frac{0.6883\times10^{-2}}{0.1748\times10^{-2}\times0.962} 
\label{result_erel}
\end{equation}
% ++++++++++++++++++++++++++++++++++++++++++++++++++++++++++++++++++++++

     Using Eq.~(\ref{relcross}), $N_{\bc}$ from 
Table~\ref{tab:bc_bkgSubtracted}, $N_{\bp}$ from 
Fig.~\ref{fig:threeTrk_mass}(b), and $\erel$, we find
%%%%%%%%%%%%%%%%%%%%%%%%%%%%%%%%%%%%%%%%%%%%%%%%%%%%
\begin{equation}
{\cal R} = 0.211\pm0.012\textrm{(stat)}.
\label{result_ratio}
\end{equation}
\subsection{Systematic uncertainties for the relative efficiency}
\label{sec:rel_eff_sys}
%%%%%%%%%%%%%%%%%%%%%%%%%%%%%%%%%%%%%%%%%%%%%%%%%%%%%%%%%%%%

We consider the systematic uncertainty associated with the prediction of the
relative efficiency due to knowledge of the $\bc$ lifetime, the $\bc$
production spectrum, the $\bp$ production spectrum,  the difference
between the $K$ and $\mu$ tracking efficiencies in the XFT, and the muon identification efficiency for CMUP muons. The total systematic uncertainty in $\erel$  is summarized in Table~\ref{tab:erel_sys}.  The individual systematic uncertainties are discussed below.

% ++++++++++++++++++++++++++++++++++++++++++++++++++++++++++++++++++++++
\begin{table}[tbp]
\caption{Systematic uncertainty assigned to $\erel$.}
\begin{center}
\begin{tabular}{lc}
\hline\hline
Source        & Systematic uncertainty \\
\hline 
$\bc$ lifetime            & $^{+0.134}_{-0.147}$    \\
$\bc$ spectrum            & $^{+0.356}_{-0.303}$    \\
$\bp$ spectrum            & $\pm0.055$              \\
XFT efficiency            & $\pm0.070$              \\
CMUP muon efficiency           & $^{+0.092}_{-0.087}$    \\ 
\hline
Total                     & $^{+0.401}_{-0.359}$    \\
\hline\hline 
\end{tabular} 
\end{center} 
\label{tab:erel_sys}
\end{table}
% ++++++++++++++++++++++++++++++++++++++++++++++++++++++++++++++++++++++

%%%%%%%%%%%%%%%%%%%%%%%%%%%%%%%%%%%%%%%%%%%%%%%%%%%%%%%%%%%%
\subsubsection{Systematic uncertainty from the $\bc$ lifetime}
\label{sec:bc_life_sys}
%%%%%%%%%%%%%%%%%%%%%%%%%%%%%%%%%%%%%%%%%%%%%%%%%%%%%%%%%%%%

The systematic uncertainty for $\erel$ due to the uncertainty in the $\bc$ lifetime is estimated by varying the
$\bc$ lifetime in MC simulations by one standard deviation %$\pm 9~\mu$m
relative to the current world average value~\cite{Ref:PDG3}.
%, $\ct(\bc)=137~\mu$m. 
%To determine the systematic uncertainty, we generate two 
%$\bc$ Monte Carlo samples using {\sc BGenerator} with $\bc$~lifetimes, 
%$\ct(\bc)$, of $128~\mu$m and $146~\mu$m. For the $\bc$ lifetime systematic 
%uncertainty we find: 
The systematic uncertainty is $\Delta\erel$ = $^{+0.134}_{-0.147}$.

%%%%%%%%%%%%%%%%%%%%%%%%%%%%%%%%%%%%%%%%%%%%%%%%%%%%%%%%%%%%
\subsubsection{Systematic uncertainty from the $\bc$ and $\bp$ production spectra}
\label{sec:bc_bp_spec_sys}
%%%%%%%%%%%%%%%%%%%%%%%%%%%%%%%%%%%%%%%%%%%%%%%%%%%%%%%%%%%%

\begin{figure}[tbp]
\centerline{
\makebox{\includegraphics[width=1.0\hsize]{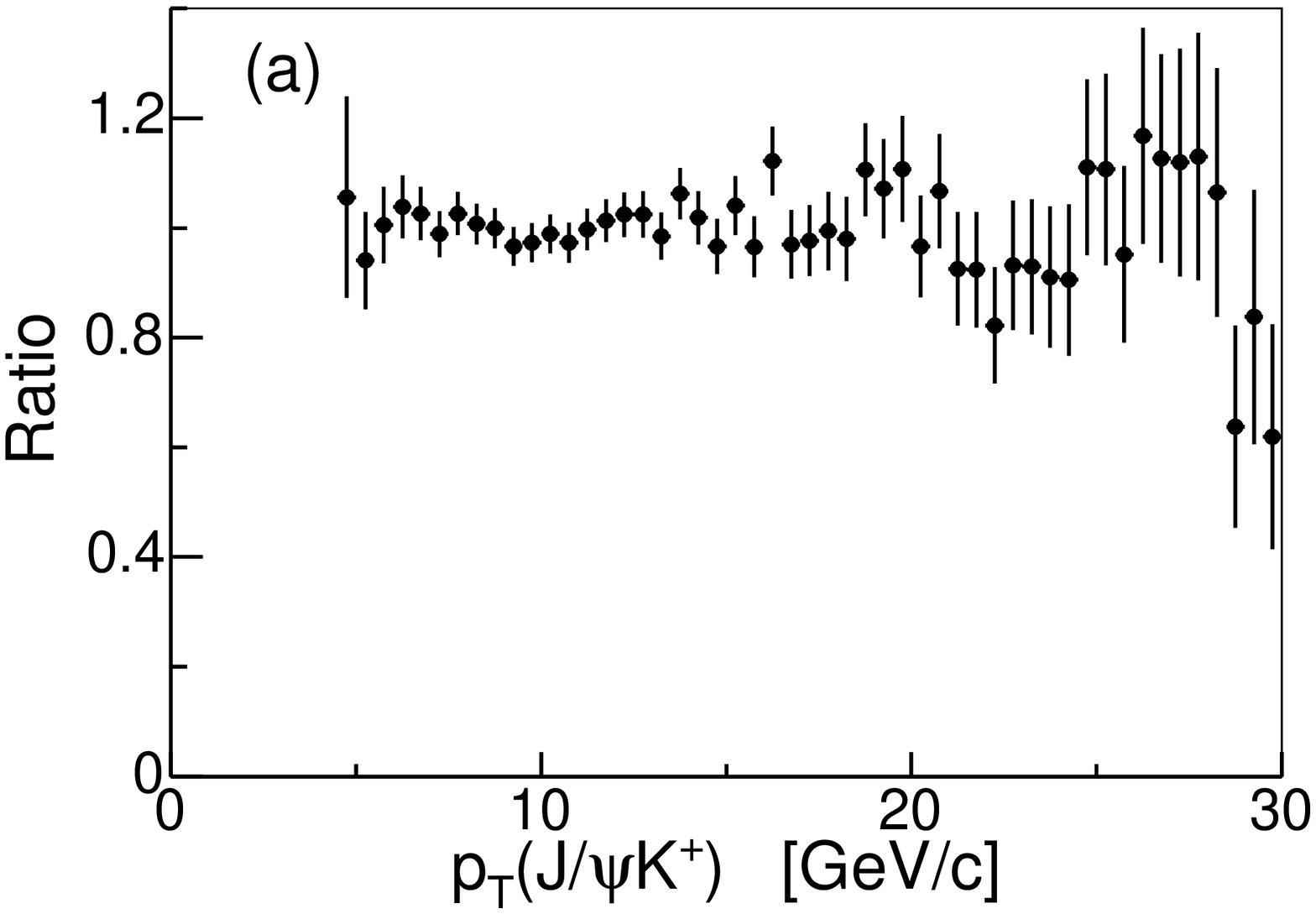}}}
\centerline{
\makebox{\includegraphics[width=1.0\hsize]{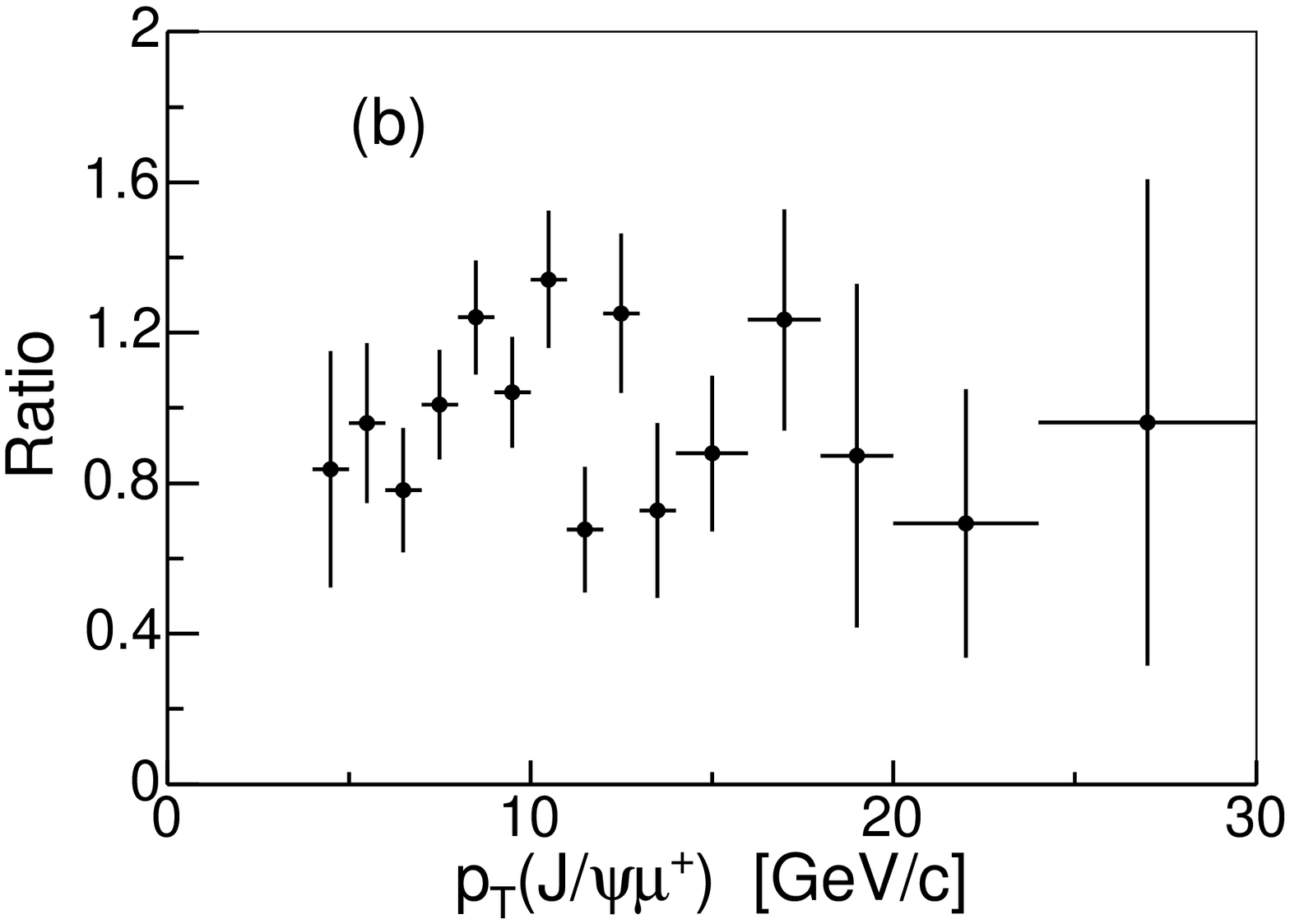}}}
\caption{Ratio of the data to the MC simulation (a) versus 
$\pt(\jpsiplusk)$ for the $\bp$ and (b) versus $p_T(\jpsiplusmu)$ for the $\bc$. 
%Both data plots have been background subtracted. 
Both theoretically predicted spectra are corrected using data.}
\label{fig:BcBpToPredRatio_vsPt}
\end{figure}

The systematic uncertainty associated with the calculations of the $\bc$ and 
$\bp$ production spectra is derived by
%  For the $\bc$ and $\bp$ production-spectrum systematic-uncertainty 
%  calculations, the systematic uncertainty is derived by 
comparing the bin-by-bin $p_T$ spectrum given by the data directly with that of simulated 
events %   in the detector 
produced using the corrected theoretical production spectra 
(see Fig.~\ref{fig:ptSpec_bcAndBp}). The ratios of the data to the MC 
simulation versus $\pt(\jpsiplusk)$ for the $\bp$ mesons and versus 
$p_T(\jpsiplusmu)$ for the $\bc$ mesons are shown in 
Fig.~\ref{fig:BcBpToPredRatio_vsPt}.
% ++++++++++++++++++++++++++++++++++++++++++++++++++++++++++++++++++++++
% ++++++++++++++++++++++++++++++++++++++++++++++++++++++++++++++++++++++

The data to corrected-MC ratio plots (Fig.~\ref{fig:BcBpToPredRatio_vsPt})
for both cases are used to estimate an average ratio for $p_T(B)>$ 6 $\gevc$, 
$\bar{R} = \sum{(w_i\times R_i)}/\sum{w_i}\,,$ 
where \textit{i} is the bin number, $R_i$ is the ratio in bin \textit{i}, 
$w_i$ = 1/$\sigma^2_i$ and $\sigma_{\bar{R}}$ = 
$\sqrt{\sum{\sigma^2_i}/[n(n-1)]}$.
%   $$w_i = \frac{1}{\sigma^2_i}\,,$$ and $$\sigma_{\bar{R}} = 
%   \sqrt{\frac{\sum{\sigma^2}}{n(n-1)}}\,.$$ 
Thus, we find $\bar{R}(\bc)$ = $1.00\pm0.08$ and $\bar{R}(\bp)$ = 
$0.999\pm0.013$. We assign systematic uncertainties of 8\% and 1.3\% for the 
$\bc$ and $\bp$ ratios, respectively. The $\erel$ systematic uncertainties are
$^{+0.356}_{-0.303}$ for the $\bc$ and $\pm0.055$ for the $\bp$ spectra,
respectively.

%%%%%%%%%%%%%%%%%%%%%%%%%%%%%%%%%%%%%%%%%%%%%%%%%%%%%%%%%%%%
\subsubsection{Differences in the efficiency of kaons and muons in the XFT simulation}
\label{sec:xft_sys}
%%%%%%%%%%%%%%%%%%%%%%%%%%%%%%%%%%%%%%%%%%%%%%%%%%%%%%%%%%%%

A small source of systematic uncertainty arises from the different XFT
efficiencies for kaons and muons due to the different $\dedx$ characteristics of these
particles in the COT.  The difference in ionization gives different single-hit efficiencies for kaons and muons that result in different XFT efficiencies as functions of $\pt$. 
%This difference exists due to the fixed number of hits required by the XFT in the COT, which produces different transverse momentum distributions for kaons and muons. 
These differences are not
modeled in the simulations.  % we use to determine the relative efficiency.  
We model this systematic uncertainty by weighting the
MC simulation to reproduce kaon and pion transverse-momentum
distributions with and without the XFT efficiencies determined from
data~\cite{Ref:KarenThesis}.     %   We find that 
The $\erel$ difference between using and not using the XFT correction is 0.14.
Comparison of the MC simulation with experimental data gives a systematic 
uncertainty of 50\% of the correction or $\pm0.07$.

%%%%%%%%%%%%%%%%%%%%%%%%%%%%%%%%%%%%%%%%%%%%%%%%%%%%%%%%%%%%
\subsubsection{Muon identification efficiency}
\label{sec:cmup_sys}
%%%%%%%%%%%%%%%%%%%%%%%%%%%%%%%%%%%%%%%%%%%%%%%%%%%%%%%%%%%%

We use Ref.~\cite{Ref:CMUP_eff} for the muon identification efficiency for CMUP muons and its 
systematic uncertainty to calculate the contribution to the uncertainty in
$\erel$. The measured systematic uncertainty for the detection efficiency of CMUP muons is about 2.2\%.  It yields a systematic uncertainty for $\erel$ 
of $\Delta\erel$ = $^{+0.092}_{-0.087}$.

%%%%%%%%%%%%%%%%%%%%%%%%%%%%%%%%%%%%%%%%%%%%%%%%%%%%%%%%%%%%
%\subsection{Total systematic uncertainty}
%\label{sec:total_sys}
%%%%%%%%%%%%%%%%%%%%%%%%%%%%%%%%%%%%%%%%%%%%%%%%%%%%%%%%%%%%

%%%%%%%%%%%%%%%%%%%%%%%%%%%%%%%%%%%%%%%%%%%%%%%%%%%%%%%%%%%%
\section{Results and Conclusions} 
\label{sec:results}
%%%%%%%%%%%%%%%%%%%%%%%%%%%%%%%%%%%%%%%%%%%%%%%%%%%%%%%%%%%%
% ++++++++++++++++++++++++++++++++++++++++++++++++++++++++++++++++++++++
\begin{table}[tbp]
\caption{Summary of values and uncertainties used in the measurement of 
$\ratio$ 
%the ratio of the production cross sections times branching fractions of $\bctojpsimunu$ to $\bptojpsik$. The result is given 
for $p_T> 6~\gevc$ and $|y|<0.6$.}
\begin{center}
\begin{tabular}{lll}
\hline\hline
Quantity           & Value                                               \\
\hline
%$N(\bctojpsimunu)$ & $739.5\pm 39.6$ (stat) $^{+19.8}_{-23.9}$ (syst)     \\
$N(\bctojpsimunu)$ & $740\pm 45$ (stat+syst)    \\
$N(\bptojpsik)$    & $14~338\pm 125$ (stat)                               \\
%$\erel$            & $4.093\pm 0.038$ (stat)  $^{+0.401}_{-0.359}$ (syst) \\
$\erel$            & $4.09\pm 0.04$ (stat) $^{+0.40}_{-0.36}$ (syst) \\
\hline
$\ratio$ & $0.211\pm 0.012$(stat)$^{+0.021}_{-0.020}$ (syst)\\
\hline\hline
\end{tabular}
\end{center}
\label{tab:cross_section}
\end{table}
% ++++++++++++++++++++++++++++++++++++++++++++++++++++++++++++++++++++++  
% ++++++++++++++++++++++++++++++++++++++++++++++++++++++++++++++++++++++
\begin{table}[tbp]
\caption{Systematic uncertainties for $\ratio$.}
\begin{center}
\begin{tabular}{lc}
\hline\hline
Source                 & Systematic uncertainty   \\
\hline	
$\bc$ background       & $^{+0.0057}_{-0.0068}$   \\
$\erel$                & $^{+0.0207}_{-0.0185}$   \\ 
\hline
Total                  & $^{+0.0214}_{-0.0197}$   \\
\hline\hline 
\end{tabular} 
\end{center} 
\label{total_systematics}
\end{table}
% ++++++++++++++++++++++++++++++++++++++++++++++++++++++++++++++++++++++
% ++++++++++++++++++++++++++++++++++++++++++++++++++++++++++++++++++++++
\begin{table*}[Htbp]
\caption{Branching-fraction predictions for the decay $\bctojpsimunu$.}
\begin{center}
\begin{tabular}{lccccccccccccc}
\hline\hline
               & \multicolumn{13}{c}{Branching-fraction predictions in \%}  \\
\hline
Reference  & \cite{Ref:Kiselev} & \cite{Ref:Huang2008} & \cite{Ref:Nobes2000} & \cite{Ref:Ebert2003} & \cite{Ref:Ivanov} & \cite{Ref:Chang1994} & \cite{Ref:Hernandez2006} & \cite{Ref:Anisimov1999} & \cite{Ref:Wang2009} & \cite{Ref:Ke2014} & \cite{Ref:Colangelo2000} & \cite{Ref:El-Hady2000} & \cite{Ref:Qiao2013}   \\
Prediction & 1.9 & 2.37 & 1.44 & 1.21 & 2.07 & 2.35 & 1.5 & 1.2 & 1.49 & 1.15 & 1.47 & 2.01 & 6.7 \\	
\hline\hline 
\end{tabular} 
\end{center} 
\label{semileptonic_predictions}
\end{table*}
% ++++++++++++++++++++++++++++++++++++++++++++++++++++++++++++++++++++++
   
     The result of the measurement of $\ratio$ based on the complete CDF Run\,II data set, which corresponds to an integrated luminosity of 8.7 fb,$^{-1}$ is
% ++++++++++++++++++++++++++++++++++++++++++++++++++++++++++++++++++++++
\begin{equation}
  {\cal R} = 0.211\pm 0.012~\textrm{(stat)} ^{+0.021}_{-0.020}~\textrm{(syst)}  
\end{equation}
% ++++++++++++++++++++++++++++++++++++++++++++++++++++++++++++++++++++++
for $\pt(\bc)> 6~\gevc$ and $|y|<0.6$.  The numbers of $\bctojpsimunu$ and
 $\bptojpsik$ decays, and 
the relative efficiency between the two, are summarized in 
Table~\ref{tab:cross_section}.
%which includes all uncertainties in the measurement.
The total systematic uncertainties for the ratio $\ratio$
are summarized in Table~\ref{total_systematics}.
         
     The result ${\cal R} = 0.211 \pm 0.024$ 
can be compared to the Run\,I measurement from CDF~\cite{Ref:RunIa,*Ref:RunIb}, 
${\cal R} = 0.13 \pm 0.06$ based on a sample corresponding to 0.11~fb$^{-1}$ of 
integrated luminosity at $\sqrt{s} = 1.8$~TeV.

     Using theoretical predictions for ${\cal B}(\bctojpsimunu)$ and independent measurements for ${\cal B}(\bptojpsik)$ and $\sigma(\bp)$, we calculate the total $\bc$ cross section.  The measured quantities are ${\cal B}(\bptojpsik) = (1.027\pm 0.031)\times
10^{-3}$~\cite{Ref:PDG3} and $\sigma(\bp) = 2.78\pm 0.24~\mu$b for
$\pt(\bp) > 6~\gevc$ and $|y|<1$~\cite{Ref:BpSpectrum}. Assuming 
that the observed value of $\ratio$ for $|y|<0.6$ approximates the value for 
$|y|<1$, we find
%our measured value of $\ratio$ is also valid for $|y|<1$, we find
% ++++++++++++++++++++++++++++++++++++++++++++++++++++++++++++++++++++++
\begin{widetext}
\begin{equation}
  \sigbrbc = 0.602\pm 0.034~\textrm{(stat)} ^{+0.060}_{-0.063}~\textrm{(syst)} \pm 0.055~\textrm{(other)}~\textrm{nb}  
\label{bcsigmabf}
\end{equation}
\end{widetext}
% ++++++++++++++++++++++++++++++++++++++++++++++++++++++++++++++++++++++
for $\pt(\bc)> 6~\gevc$ and $|y|<1$.  In Eq.~(\ref{bcsigmabf}) the statistical and systematic uncertainties are from the measurement of $\ratio$ and other is the combined experimental uncertainty in the measurements of ${\cal B}(\bptojpsik)$ and $\sigma(\bp)$.  Combining the uncertainties in quadrature gives $\sigbrbc = 0.60\pm 0.09~\textrm{nb}$.   
     To extract the total $\bc$ production cross section from this result, it is necessary to consider the predictions for the branching fraction for the semileptonic decay $\bctojpsimunu$.  Table~\ref{semileptonic_predictions} summarizes the many predictions.
The approaches to the calculation of this semileptonic branching fraction include: QCD sum rules \cite{Ref:Kiselev, Ref:Huang2008}, relativistic constituent-quark models \cite{Ref:Nobes2000, Ref:Ebert2003,  Ref:Ivanov}, a quark model using the Bethe-Salpeter equation \cite{Ref:Chang1994}, a nonrelativistic constituent-quark model \cite{Ref:Hernandez2006}, covariant-light-front quark models \cite{Ref:Anisimov1999, Ref:Wang2009, Ref:Ke2014}, QCD relativistic-potential models \cite{Ref:Colangelo2000, Ref:El-Hady2000}, and nonrelativistic QCD \cite{Ref:Qiao2013}.
With the exception of Ref.~\cite{Ref:Qiao2013}, all of the %other 
theoretical results shown in Table~\ref{semileptonic_predictions} predict the branching fraction ${\cal B}(\bctojpsimunu)$ in the range 1.15--2.37~\%.  Using this selection of theoretical predictions, we find the total $\bc$ cross section to
be in the range 25$\pm$4 to 52$\pm$8~nb 
for $\pt(\bc)> 6~\gevc$ and $|y|<1$, where the uncertainties reflect only the experimental uncertainties of the measurements used in the calculation.  The result is a measure of the 
combined cross section for production to the ground state plus any excited $\bc$ state that 
%eventually decays 
cascades into the ground state prior to its weak-interaction decay. 
%  flavor-changing decay. 
    
     This result is %significantly 
higher than the theoretical prediction of Chang \textit{et al.},~\cite{Ref:BcLatest_Chang,Ref:Chang_ExcitedBc}, which estimates the sum of the production cross sections to $\bc$ and $\bcst$, 
$\sigma(\bc+\bcst)$, to be 5~nb for $\sqrt{s}=1.96$~TeV, $\pt >
4~\gevc$, and $|y|<1$.  Similarly, Ref.~\cite{Ref:Saleev} reports $\sigma(\bc+\bcst)=7.4\pm 5.4$~nb for $\sqrt{s}=1.8$~TeV, $\pt> 6~\gevc$, and 
$|y|<1$.  If we consider the prediction ${\cal B}(\bctojpsimunu)=6.7^{+2.5}_{-1.4}$\% given in Ref.~\cite{Ref:Qiao2013}, then our result for the $\bc$ production cross section is 9.0$^{+3.6}_{-2.3}$~nb (theoretical uncertainty included), in reasonable agreement with the predictions of Refs.~\cite{Ref:BcLatest_Chang,Ref:Chang_ExcitedBc,Ref:Saleev}.
      
     If the branching fraction ${\cal B}(\bctojpsimunu)$ is in the approximate range 1.2--2.4\% as given by 12 of the 13 predictions in Table~\ref{semileptonic_predictions}, then there is a discrepancy between the theoretical $\bc$ production cross section and the estimate made from the experimental results presented here.  This discrepancy would be mitigated if the production cross section to $\bc$ states higher in mass than the $\bcst$ were also large.  Therefore, it would be very useful to have a new prediction of the $\bc$ production cross section at the exact kinematic values of this experimental result that takes into account all production to excited $\bc$ states that cascade to the ground state.  The discrepancy would also disappear if ${\cal B}(\bctojpsimunu)$ is approximately 7\% as predicted by Ref.~\cite{Ref:Qiao2013}.

\begin{acknowledgments}
     We thank the Fermilab staff and the technical staffs of the participating institutions for their vital contributions. This work was supported by the U.S. Department of Energy and National Science Foundation; the Italian Istituto Nazionale di Fisica Nucleare; the Ministry of Education, Culture, Sports, Science and Technology of Japan; the Natural Sciences and Engineering Research Council of Canada; the National Science Council of the Republic of China; the Swiss National Science Foundation; the A.P. Sloan Foundation; the Bundesministerium f\"ur Bildung und Forschung, Germany; the Korean World Class University Program, the National Research Foundation of Korea; the Science and Technology Facilities Council and the Royal Society, United Kingdom; the Russian Foundation for Basic Research; the Ministerio de Ciencia e Innovaci\'{o}n, and Programa Consolider-Ingenio 2010, Spain; the Slovak R\&D Agency; the Academy of Finland; the Australian Research Council (ARC); and the EU community Marie Curie Fellowship Contract No. 302103. 
\end{acknowledgments}

\bibliography{bcXsec_8fb_prd_gpv5}    % Produces the bibliography via BibTeX.

%%%%%%%%%%%%%%%%%%%%%%%%%%%%%%%%%%%%%%%%%%%%%%%%%%%%%%%%%%%%

\end{document}